 \documentclass[useAMS]{mn2e}
\usepackage{psfig}

\usepackage{times}
\def\lsim{ \lower .75ex\hbox{$\sim$} \llap{\raise .27ex \hbox{$<$}} }
\def\gsim{ \lower .75ex \hbox{$\sim$} \llap{\raise .27ex \hbox{$>$}} }

\title[BL Lac objects  of {\it Fermi}] 
{TeV BL Lac objects at the dawn of the {\it Fermi} era}

\author[Tavecchio et al.]
{F. Tavecchio$^1$\thanks{E--mail: fabrizio.tavecchio@brera.inaf.it}, 
G. Ghisellini$^1$, G. Ghirlanda$^1$, L. Foschini$^1$, L. Maraschi$^2$  \\
$^1$ INAF -- Osservatorio Astronomico di Brera, via E. Bianchi 46, I--23807
Merate, Italy\\
$^2$ INAF -- Osservatorio Astronomico di Brera, via Brera 28, I--20121 Milano, Italy\\
}

\begin{document}


\maketitle

\begin{abstract} 
We reconsider the emission properties of the BL Lac objects emitting in the high-energy $\gamma$-ray band exploiting the new information in the MeV-GeV band obtained by the  Large Area Telescope (LAT) onboard the {\it Fermi Gamma-Ray Space Telescope} in its first three months of operation. To this aim we construct the spectral energy distribution of all the BL Lacs revealed by LAT and of the known TeV BL Lacs not detected by LAT, also including data from the {\it Swift} satellite, and model them with a simple one-zone leptonic model.
The analysis shows that the BL Lacs detected by LAT (being or not already detected in the TeV band) share similar physical parameters. While some of the TeV BL Lacs not revealed by LAT have spectral energy distributions and physical parameters very similar to the LAT BL Lacs, a group of objects displays peculiar properties (larger electron energies and smaller magnetic fields) suggesting different physical conditions in the emission region. Finally, we discuss possible criteria to effectively select good new candidates for the Cherenkov telescopes among the LAT sources, presenting a list of predicted fluxes in the very high-energy band calculated including the effect of the absorption by the extragalactic background light.
\end{abstract}
 
\begin{keywords} radiation mechanisms: non-thermal --- $\gamma$--rays: theory ---$\gamma$--rays: observations -- BL Lac objects: general
\end{keywords}

\section{Introduction}
 
 The gamma-ray extragalactic sky at high ($>100$ MeV) and very high ($>100$ GeV) energies is dominated by blazars, associated to radio-loud active galactic nuclei with a relativistic jet closely oriented to the Earth. The resulting relativistic amplification of the non-thermal jet emission makes blazars extreme objects, with apparent luminosities exceeding in the most powerful sources $10^{48}$ erg/s and variability timescales as short as minutes (e.g. Aharonian et al. 2007a, Albert et al. 2007a).
 
Blazars are divided in two classes: i) flat spectrum radio quasars (FSRQs) show broad emission lines typical of quasars in their optical spectra and are the most powerful sources; ii) BL Lac objects, instead, display rather weak (or event absent) emission lines (with, by definition, equivalent width less than 5\AA) and have smaller luminosities. 
For both classes,  the Spectral Energy Distribution (SED) is characterized by two broad peaks, generally interpreted as due to synchrotron (the low energy peak) and inverse Compton emission (at high energy; an hadronic origin for the high energy peak is supposed in the hadronic models, e.g. Muecke et al. 2003). It is widely accepted that the SEDs of blazars follow a sequence with the luminosity (Fossati et al. 1998, Donato et al. 2001): the powerful sources are characterized by SEDs whose synchrotron and IC peaks are in the submm/IR and MeV band, respectively, while the low-power BL Lacs show the maxima of their components in the optical/UV or even X-ray band and in the GeV-TeV region. Ghisellini et al. (1998) proposed that the sequence could be related to the balance between acceleration and cooling acting on the electrons emitting most of the power: electrons in the jet of FSRQs are characterized by strong radiative losses, and thus cannot reach large energies. In BL Lacs, instead, the small cooling allows the electrons to be accelerated to very high energy, determining high frequencies for the synchrotron and IC emission components (see Ghisellini \& Tavecchio 2008a, 2009 for a refined view).
 
Blazars emitting in the TeV band are still a small group but their number is fastly increasing (see De Angelis et al. 2008 and Aharonian et al. 2008a for recent reviews)\footnote{updates at {
http://www.mppmu.mpg.de/$\sim$rwagner/sources}}. The interest for these sources is driven by the possibility to get interesting clues on the acceleration processes of charged particles in relativistic flows and the possible use of their TeV emission to characterized the poorly known extragalactic background light (EBL; e.g. Stecker et al. 1992, Stanev \& Franceschini 1998, Mazin \& Raue 2007). The favorite candidates for the TeV detection are blazars peaking at the highest energies, called high frequency peaked BL Lacs (HBLs, Giommi \& Padovani 1994), showing the peak of the synchrotron component in the UV-X--ray band and the IC peak close to 100 GeV. For this reason most of the observation time has been devoted to observe HBLs (one of the most used list of candidates was presented in Costamante \& Ghisellini 2002, hereafter CG02). However, the new generation of Cherenkov telescopes now operating (HESS, MAGIC, VERITAS) have enough sensibility to observe other, less bright in the TeV band, blazars. Indeed, a handful of IBL (intermediate BL Lacs) and LBLs (low frequency peak BL Lac objects) belong to the present group ot TeV BL Lacs. One of the most critical points in the search for new TeV blazars is that the present Cherenkov telescopes have a quite limited field of view and cannot provide a full-sky survey. 
For this reason the selection of candidates has to rely on a variety of indirect methods. The simplest criterion (e.g. Stecker et al. 1996) is to choose the HBL whose synchrotron component  peaks in the X-ray band assuming that the X-ray flux is a proxy for the flux of the TeV emission. A more refined method is based on two parameters, the X-ray and the radio flux (CG02). This method has been quite successful in helping to select the targets, though not all the candidates have yet been detected.
 
Recently, the {\it Fermi} collaboration released  a catalogue of AGNs (with galactic latitude $|b|>10$ deg) detected with high significance (10$\sigma$ above 100 MeV) by the Large Area Telescope (LAT) in the first three months of operation (Abdo et al. 2009a, A09 hereafter). Excluding two radiogalaxies (Cen A and NGC 1275), the remaining 104 sources are all blazars (58 FSRQs, 42 BL Lacs and 4 with uncertain classification\footnote{Note that one of the sources classified as FSRQ is a radio-loud Narrow Line Seyfert 1 galaxy (Abdo  et al. 2009b, Foschini et al. 2009).}). One interesting feature of this catalogue is the large number of BL Lacs. For comparison, the previous EGRET list of BL Lacs detected in the 100 MeV -- 10 GeV band (Hartman et al. 1999) comprised only 14 BL Lacs (out a total of 60 blazars). The knowledge of the GeV part of the spectrum is rather important, allowing to better determine the shape and the position of the the peak of the emission of the TeV BL Lacs, when located between the bands covered by LAT and by the Cherenkov telescopes (see e.g. Aharonian et al. 2009). Moreover, the large number of BL Lacs, together with the good description of the spectrum, is a useful tool to find new TeV candidates. 

The SEDs of the 23 most powerful blazars  of the A09 list (with $\gamma$-ray luminosity exceeding $10^{48}$ erg s$^{-1}$) are presented and discussed in Ghisellini, Tavecchio \& Ghirlanda (2009). The emission models for the SEDs of all the A09 sources are presented in Ghisellini et al. (2009a).
In this paper we focus our attention on the BL Lacs reported in the A09 list, including also the BL Lacs detected in the TeV band but not by LAT.  We also discussed possible criteria to select, among the LAT BL Lacs, the best candidates to be detected by Cherenkov Telescopes.  

We use $H_{\rm 0}\rm =70\; km\; s^{-1}\; Mpc^{-1} $, $\Omega_{\Lambda}=0.7$, $\Omega_{\rm M} = 0.3$.

\begin{table}
\centering
\begin{tabular}{llllll}
\hline
\hline
Name  &Alias &$z$  &{\it S?} &E?  & TeV?\\
\hline
00311--1938    &KUV       &0.610   &Y &  &  \\
0048--097      & PKS        & --         & Y &   & \\
0109+22        &S2            &--          &Y &UL  &\\ 
0118--272      &PKS         &0.559    &  &UL  &\\
0133+388      & B3          & --          &   &    &\\
0141+268$^n$      & TXS        &--           &   &  &\\
0219+428      &3C66A     &0.444     &Y &Y  & Y\\
0301--243      &PKS       &0.260      &Y &UL  &\\
0447--439      &PKS       &0.107     &Y   &    &\\
0502+675       &1ES       &0.314$^a$     &Y   &    &\\
0712+5033     &GB6      & --           &Y    & ? &\\
0716+714       &TXS       &0.26       &Y &Y   & Y\\
0735+178       &PKS       &0.424     &Y &Y   &\\
0814+425       &OJ 425   &0.53      &Y &UL  &\\
0851+202       &OJ 287    &0.306   &Y &UL  & \\ 
1011+496       &1ES       &0.212     &Y &    &Y\\ 
1050.7+4946    &MS        &0.140   &Y &    &\\
1054+2210$^n$    &CLASS   & --          &      &   &\\
10586+5628     &RX        &0.143    &  &    &\\   
1101+384       &Mkn 421   &0.031   &Y &Y  &Y\\ 
1215+303       &B2        &0.13         &  &      &\\
1219+285       &W Comae    &0.102   &Y &Y  &Y\\
1250+532$^n$       & TXS        & --         &      &    & \\
1424+240       &PKS         & $>0.67^{S}$        & Y &     &Y\\
1514--241      &Ap Lib     &0.048      &Y &Y  & \\ 
15429+6129  & RXS       & --           & Y   &   & \\
1553+11        &PG        &0.36$^u$  &Y &   &Y \\ 
1652+398       &Mkn 501   &0.0336 &Y &Y  &Y \\ 
1717+177       &PKS       &0.137     &Y &   & \\
1749+096       &OT 081    &0.322   &Y &UL & \\
1959+650       &1ES       &0.047     &Y &   &Y \\ 
2005--489      &PKS       &0.071     &Y &Y  &Y \\ 
2136--428      & MH       & --            & Y & UL & \\ 
2155--304      &PKS       &0.116     &Y &Y  &Y \\
2200+420       &BL Lac    &0.069   &Y &Y  &Y \\
2322+396      & B3          & --          & Y &  & \\
\hline
\hline 
\end{tabular}
\vskip 0.4 true cm
\caption{The  {\it Fermi} BL Lac objects  in the A09 list. We have excluded 6 sources, defined as BL Lacs based on the small equivalent width of the lines, but showing broad emission lines typical of quasars (see Ghisellini et al. 2009a).
In the last 3 columns we indicate if there are {\it Swift} observations; if the source was 
detected by EGRET (UL stands for an upper limit given
by Fichtel et al. 1994) and if it is a TeV source.
$u$: redshift uncertain;
$S$: lower limit from Sbarufatti et al. (2005);
$n$: not studied in this paper due to lack of multiwavelength data;
$a$: redshift from NED. A09 reports $z=0.416$.
}
\label{sample}
\end{table}

\begin{table}
\centering
\begin{tabular}{llll} 
\hline  
\hline
Name       &  Alias &  $z$    & {\it S}? \\
\hline 
0152+017 & RGB   &  0.080   & Y \\
0229+200 & 1ES    &   0.140   & Y \\
0347--121 &  1ES    &  0.188   & Y\\
0548--322 &  PKS  &   0.069  &   Y  \\
0710+591   & RGB   &  0.125  & Y \\
0806+524 &  1ES    &  0.138    &  Y \\
1101--232 &  1ES     &   0.186   &   \\
1133+704  &Mkn180   &  0.046  &   Y\\
1218+304 &  1ES      & 0.182  &  \\
1426+428   &  H    &   0.129   &  Y \\
2344+514$^*$ & 1ES &   0.044  &  \\
2356--309   &  H    &   0.165  &   \\
\hline
\hline
\end{tabular}
\caption{The known TeV emitting BL Lacs not in the A09 list. $^*$ $|b|<10$ deg.
}
\label{tevnolat}
\end{table}

\section{BL Lac objects in the Fermi/LAT}

The BL Lacs reported in A09 are reported in Table \ref{sample}, where we also report whether there are {\it Swift} observations, whether the source was detected by EGRET (or there is an upper limit, Fichtel et al. 1994) and whether the source has been detected in the TeV band. We exclude from this list 6 sources (AO 0235+164, PKS 0332-403, PKS 0426-380, PKS 0537-441, PKS 1057-79, S5 1803+784), classified as BL Lac in A09, but showing broad emission lines with luminosities typical of quasars (see Ghisellini et al. 2009a for details). Moreover, for three sources reported in Table \ref{sample} (0141+268, 1054+2210 and 1250+532) the paucity of the data prevents even a rough description of the SED and we excluded them in the following analysis.
Therefore the total number of LAT BL Lacs considered here is 33. This list contains 12 BL Lacs detected in the TeV band, exactly half of the known TeV BL Lacs\footnote{Note, however, that 1ES 2344+514 lies at a galactic latitude $|b|<10 $ deg, not covered by the A09 list.}.

\begin{figure}
\hskip -1.3 cm
\vskip -0.8cm
\psfig{figure=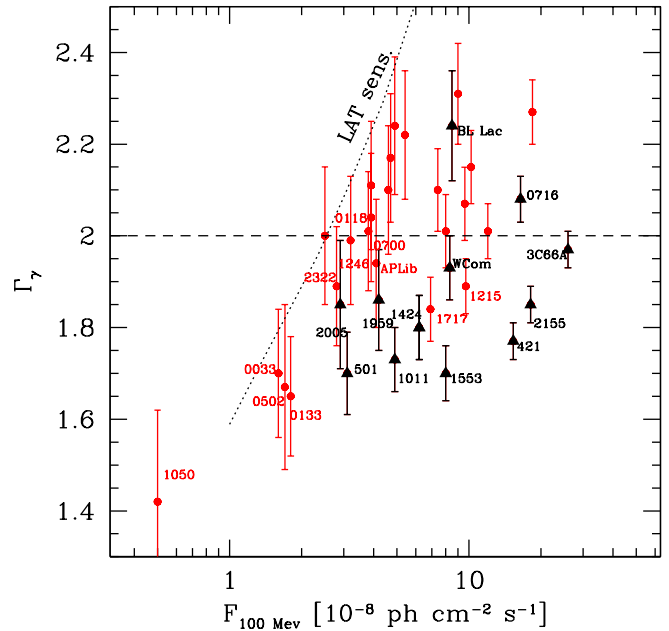,width=10cm,height=10cm}
\vskip -0.3 cm
\caption{Gamma-ray photon index versus the integral flux above 100 MeV for the BL Lac objects detected by {\it Fermi}/LAT (A09). Black triangles refer to BL Lacs detected in the TeV band (see Table \ref{sample}). TeV detected sources belong to the hardest sources, while the distribution of fluxes is very similar for both groups. The dashed line reports the limit sensitivity for the sources reported in A09. Note that the LAT sensitivity increases with the hardness of the spectra.}
\label{histo}
\end{figure}

A09 report the spectral parameters (flux and slope) obtained with a power law model fit to the average LAT spectrum obtained during the first three months.  The spectral characteristics of the TeV (black  triangles) and non-TeV BL Lacs (red circles) are compared in Fig. \ref{histo}, in which we report the photon index and the flux above 100 MeV resulting from the power law fit. 
Clearly, TeV sources preferentially lay on the lower part of the plot,  indicating a harder GeV spectrum than that of the non TeV ones. This is not surprising, since most of the BL Lacs detected so far at TeV energies are HBLs with the high-energy peak of the SED above 100 GeV, implying a hard spectrum in the LAT band. Only BL Lac itself and S5 0716+71 (and, marginally, 3C66A and W Comae) have a soft photon index. Occasionally, BL Lac displays a very hard GeV spectrum (e.g. Nandikotkur et al. 2007). The hardest (and faintest) source is MS 1050.7+4946 ($z=0.140$), with a photon index $\Gamma_{\gamma}=1.4\pm0.2$. TeV and non-TeV sources share the same distribution of fluxes. This fact ensures that the difference in the distributions in slope is an intrinsic feature and not the result of a selection effect related to the higher sensitivity of LAT for harder sources (dotted line in Fig.\ref{histo}, see also A09).

In Table \ref{tevnolat} we report the 12 TeV BL Lacs not present in the A09 list. Neither of these sources was detected by EGRET. In the following we will also consider these sources for the modeling of the SED.

\section{Analysis of the {\it Swift} data}

Several of the BL Lac in A09 (and of the non-LAT TeV sources) have been observed by {\it Swift}.

{\it Swift}/XRT (Burrows et al. 2005) and {\it Swift}/UVOT (Roming et al. 2005) data have been reduced using the standard procedure (described in, e.g., Ghisellini et al. 2007), with the latest version of the \texttt{HEASOFT} (6.6.3) and \texttt{CALDB} (2009/5/6) packages\footnote{http://heasarc.gsfc.nasa.gov/docs/software/lheasoft/}. XRT data have been fitted with \texttt{XSPEC12}. In most cases a single power law model absorbed with a column density fixed to the Galactic value (taken from Kalberla et al. 2005) provides a good fit of the data. In few cases a broken power law (again with a Galactic column density) is required. Results are shown in Table \ref{xrt}.
UVOT magnitudes have been corrected for Galactic absorption using the values of Schlegel et al. (1998) for the V, B and U filters and the formulae by Pei (1992) for the UV filters and converted into fluxes following Poole et al. (2008). Results are reported in Table \ref{uvot}.

\section{Spectral Energy Distributions}

We assembled the SEDs of the BL Lacs in the A09 and of the 12 TeV BL Lac not included there
 using historical (mainly from the NASA Extragalactic Database\footnote{http://nedwww.ipac.caltech.edu/} and the BZCAT/ASDS archive\footnote{http://www.asdc.asi.it/bzcat/}), {\it Fermi}/LAT from A09 (when available) and, if available, {\it Swift} data taken in (or close to) the period of the {\it Fermi} observations (August 4 -- October 30, 2008). When a large amount of data are available for a given source (for instance the well studied Mkn 421, Mkn 501 and PKS 2155-304) we choose to plot the most representative datasets, to show the possible range of variations in the different bands. For clarity, we report only the TeV spectra already corrected for the effect of the absorption through the interaction with the EBL. Absorption is taken into account considering only one of the models discussed in literature, the "lowSFR" model of Kneiske et al. (2004), based on a low estimate of the EBL close to values recently derived (e.g. Aharonian et al. 2006).

For the sources detected by LAT we plot the result of the fit with a power law model to the average spectrum (lower bow tie) using the parameters reported in A09. To show the variability we also report the spectrum (upper bow-tie) obtained by using the weekly averaged peak flux provided by A09 and assuming the same slope of the average spectrum. This last assumption is probably violated in some of the sources, since the spectrum likely changes with the flux (possibly in a rather complex fashion, see e.g. Aharonian et al. 2009).  As already discussed, the LAT spectra for HBLs are generally hard, falling onto the raising part of the high-energy peak. In IBL/LBL sources the LAT spectra rather probe the peak of the emission. We warn the reader that, as also discussed in A09, not always the power law model provides a good fit of the data over the entire LAT band. In particular, the slope is often dictated by the low energy bins, characterized by the smallest errors. In general, we expect that the GeV spectra of HBLs soften approaching the peak of the bump. A clear example is given in Aharonian et al. (2009) for PKS 2155-304. For IBLs and LBLs the situation could be even more complex, since the LAT band is likely centered on the high-energy peak. For the TeV BL Lacs not present in A09 we report an upper limit in the GeV band calculated using the sensitivity curve reported in A09 (dotted line in Fig.\ref{histo}).

For the sources detected by LAT it is interesting to compare the LAT and the TeV spectra. In general, for the sources with a good coverage at TeV energies (especially Mkn 421, Mkn 501, PKS 2155-304, 1ES 1959+650), the LAT spectrum agrees very well with TeV spectra taken at low level. For the sources with a limited number of observations/detections in the TeV band, the LAT data seems to connect rather smoothly with the available TeV spectrum. 

Finally, another interesting point to consider concerns the variability in the GeV band. The great majority of these BL Lac are known to be extremely variable in the TeV band, showing variations of factors as large as 10 on timescales as short as hours or even minutes (e.g. Aharonian et al. 2007, Albert et al. 2007a). On the other hand, variability in the LAT band seems to be more moderate (less than a factor of 2 in flux), as shown by the peak flux spectra. However, variability on timescale shorter than one week is probably diluted and thus not visible in these data.

Our aim is mainly to reproduce "average" SEDs of the sources, to investigate these BL Lacs as a population. For this reason, in the case of the sources detected by LAT we choose to calculate the model using the LAT spectrum as a guideline to reproduce the high-energy peak. Indeed, the 3-months average LAT spectrum is a more reliable indicator of the average state than the fastly-evolving TeV spectrum. This is also reinforced by the fact that published TeV spectra are often biased, being taken during high levels of the emission. LAT data, instead, are more representative of the average state of the sources. As noted above, in most cases the model  is not very distant from the observed TeV spectra at low levels. In these cases we try to locate the position of the high-energy peak between the LAT and the (soft) TeV spectrum. For the sources not detected by LAT we model the SED using the TeV spectrum.

When possible we fix the synchrotron component by using {\it Swift} data (providing a good coverage of the crucial optical-UV and X-ray band) taken during the three months covered by the LAT observations (or close to the TeV observations for the sources not detected by LAT). When {\it Swift} data exactly covering the period of high-energy observations are not available we choose to use, if available, the data closest in time. In most cases they have been obtained few months after or before the observations of LAT or TeV telescopes. However, we are aware that, given the extreme variability displayed by these sources, especially at X-ray frequencies, these SEDs should be taken with some cautions. For instance, for two cases (PKS 2155-304, 3C66A) there are more than one {\it Swift} observations in the period covered by the LAT showing important variations of the X-ray emission, both in flux and slope. In these cases we calculated two models, corresponding to the extremes of the X-ray flux. 
For PG 1553+113 and RGB J0152+017 there are no {\it Swift} observations close in time with LAT. However, we find archival observations showing huge variations in the X-rays. For PG 1553+113, for which no X-ray observation nearly simultaneous with the $\gamma$-ray data exits, we produced two different models using the two extreme X-ray spectra. For RGB J0152+017  we use the X-ray spectrum (taken with {\it XMM} and {\it RXTE}, Aharonian et al. 2008) close in time with the TeV observations. In the case of Mrk180, for which we have three {\it Swift} observations (one in spring 2008, two just after the LAT observing period) showing huge variability we model the low level X-ray spectrum
and the MAGIC spectrum. Finally, for the interesting case of 1426+428 we present two models, one referring to the average HEGRA and CAT spectrum (Aharonian et al. 2003a) and one reproducing  the recent simultaneous {\it Swift} and MAGIC data (Leonardo et al. 2009).

\section{Modeling of the SEDs}

We use the one-zone synchrotron self-Compton model fully described in Maraschi \& Tavecchio (2003). Briefly, the emitting region is a sphere with radius $R$ with a tangled and uniform magnetic field $B$. The relativistic effects are described by the Doppler factor $\delta$. The (purely phenomenological) distribution of the emitting relativistic electrons is described by a broken power law model with normalization $K$ and indices $n_1$ from $\gamma_{\rm min}$ to $\gamma_{\rm b}$ and $n_2$ above the break up to $\gamma_{\rm max}$. The model includes the full Klein-Nishina cross-section for the calculation of the inverse Compton spectrum, particularly important for the emission above the GeV band (e.g. Tavecchio et al. 1998). 

We recall that with one-zone models, such as that adopted here, one can not reproduce the emission at the longest wavelengths, since the emission is self absorbed below the millimeter band. That part of the SED is due to outer regions of the jet, not important for the modeling of the high-energy emission. Multi zone emission models, such as those discussed in the framework of structured jet models (Georganopoulos \& Kazanas 2003, Ghisellini et al. 2005), could be required for the detailed modeling of the SED even at $\gamma$-ray energies (e.g. Anderhub et al. 2009). However, with the available non-simultaneous data the one-zone model can satisfactorily reproduce most of the SEDs.

\begin{figure}
\hskip -0.3 cm
\vskip -1.1cm
\psfig{figure=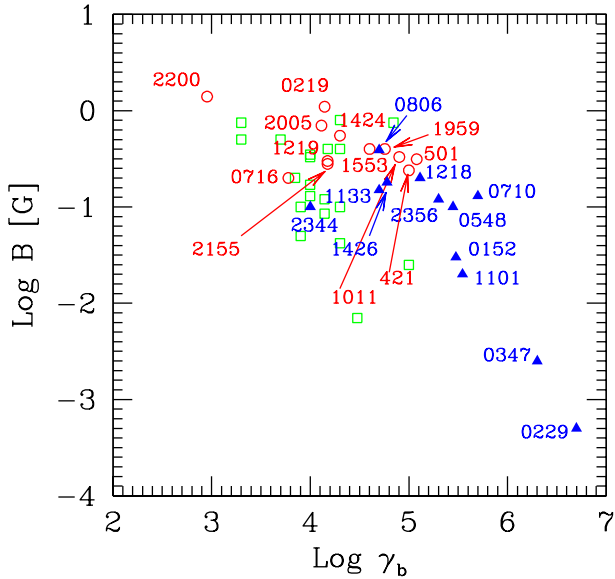,width=10cm,height=10cm}
\vskip -0.7 cm
\caption{Magnetic field versus the break Lorentz factor $\gamma _{\rm b}$ for the BL Lacs modeled with the one-zone synchrotron-SSC model. Red open circles (blue filled triangles) show the values for the known TeV sources detected (not detected) by LAT. The two populations are clearly divided, with the sources not detected by LAT populating the region with small magnetic field and large electron Lorentz factor. Open green squares are for the LAT BL Lacs not detected in the TeV band.
}
\label{fitparam}
\end{figure}

\begin{figure*}
\vskip -1 cm
\psfig{figure=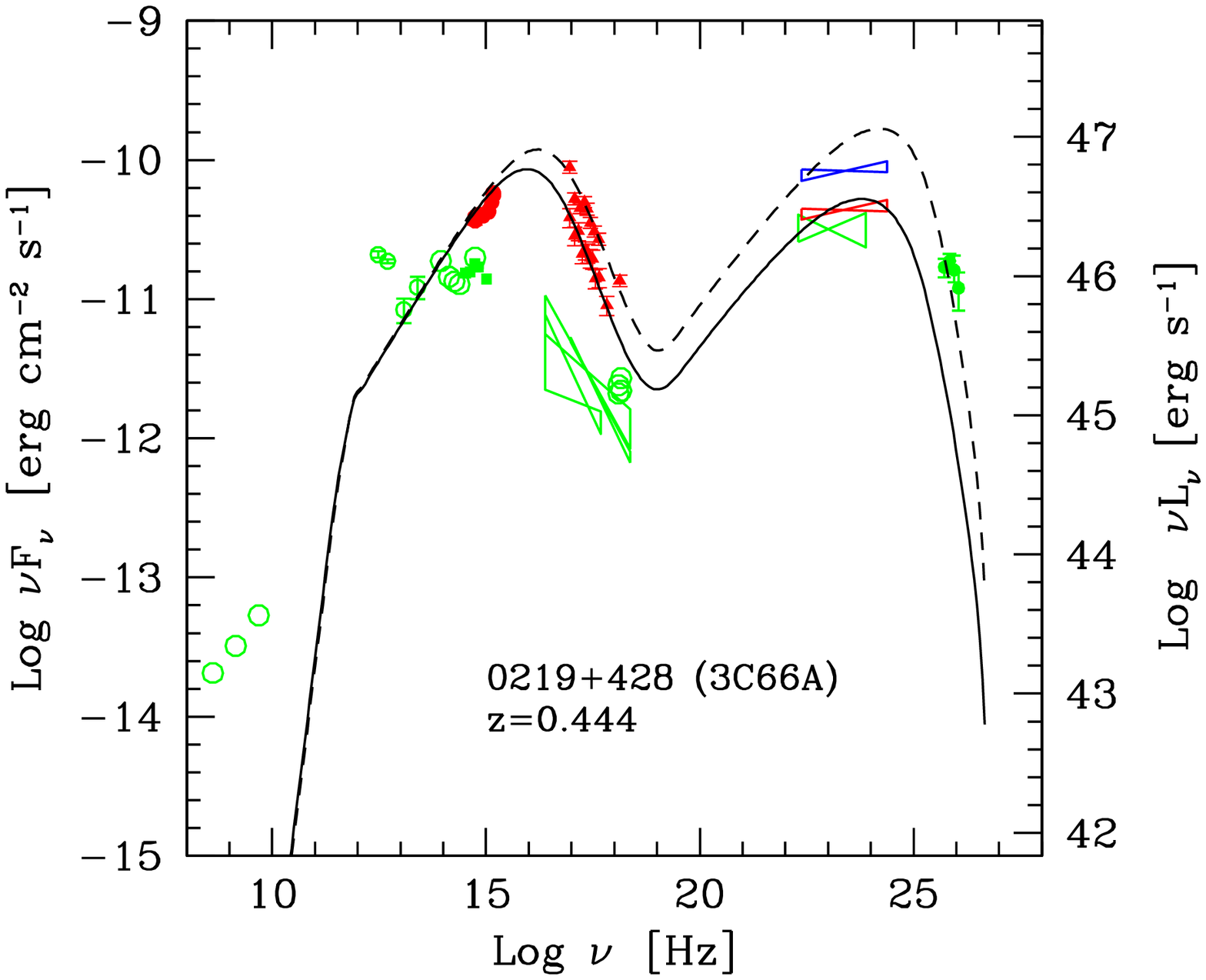,width=7cm,height=7cm}
\vskip -1.5 cm
\psfig{figure=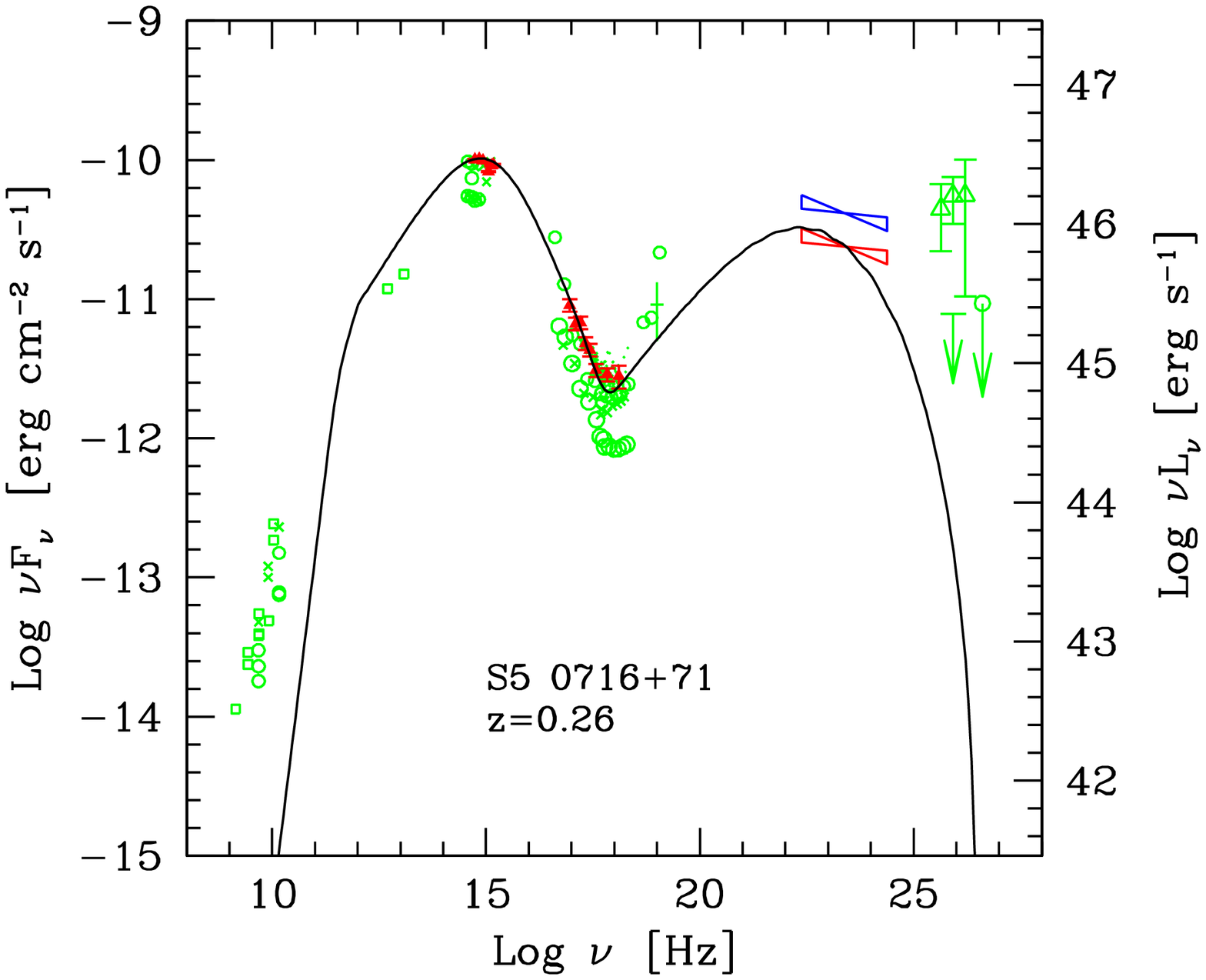,width=7cm,height=7cm}
\vskip -1.5 cm
\psfig{figure=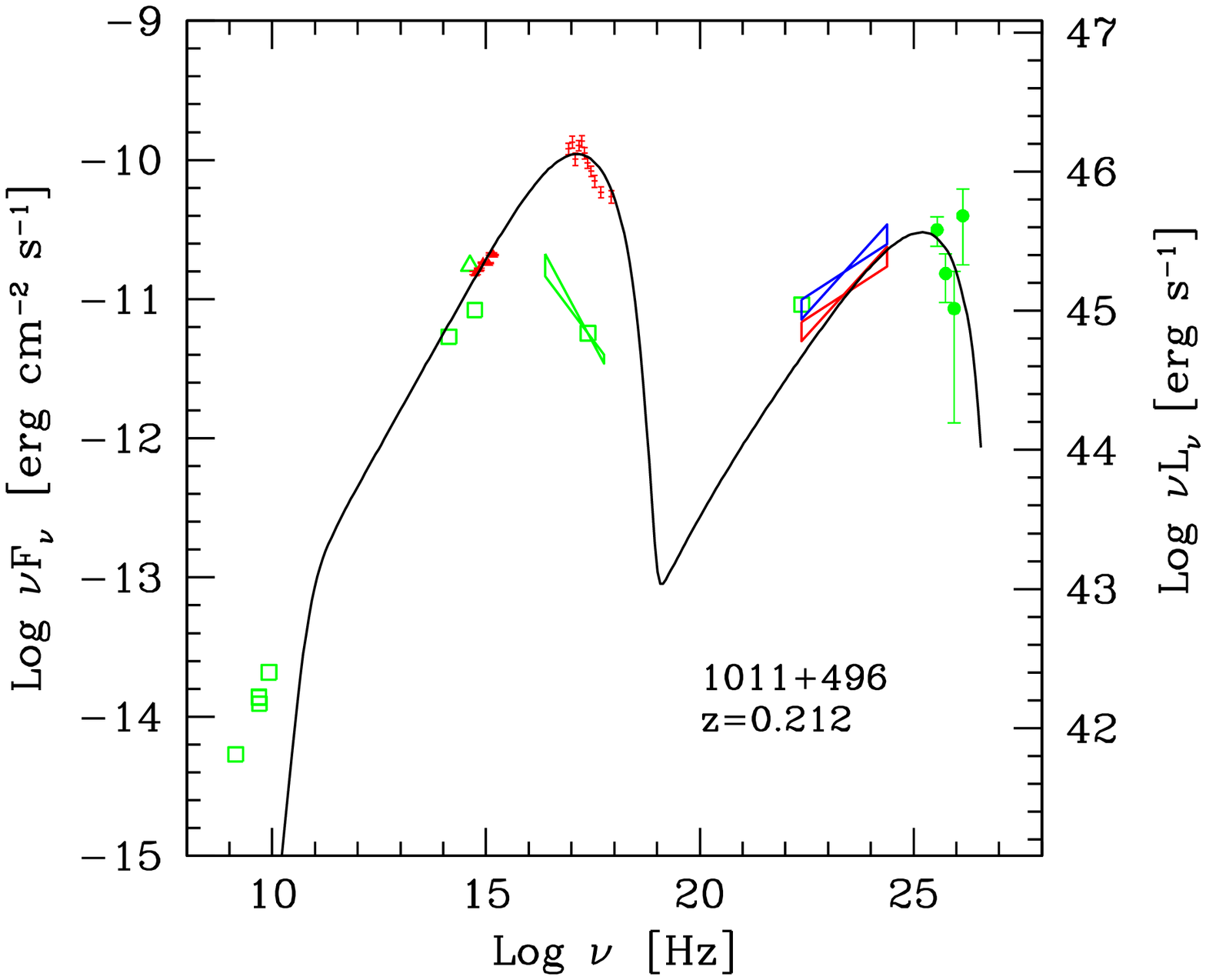,width=7cm,height=7cm}
\vskip -1.5 cm
\psfig{figure=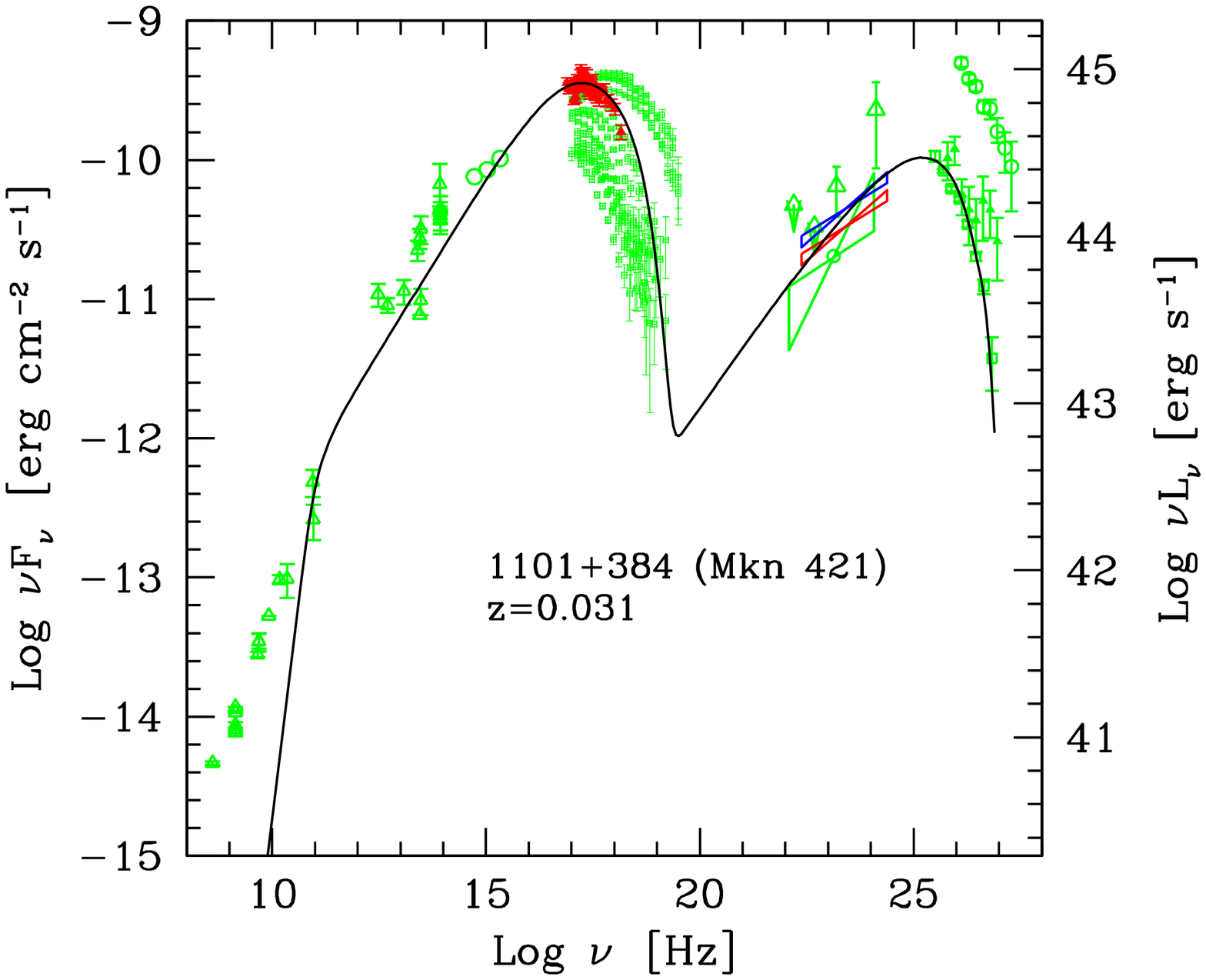,width=7cm,height=7cm}
\vskip -23.6 cm
\hskip 7 cm
\psfig{figure=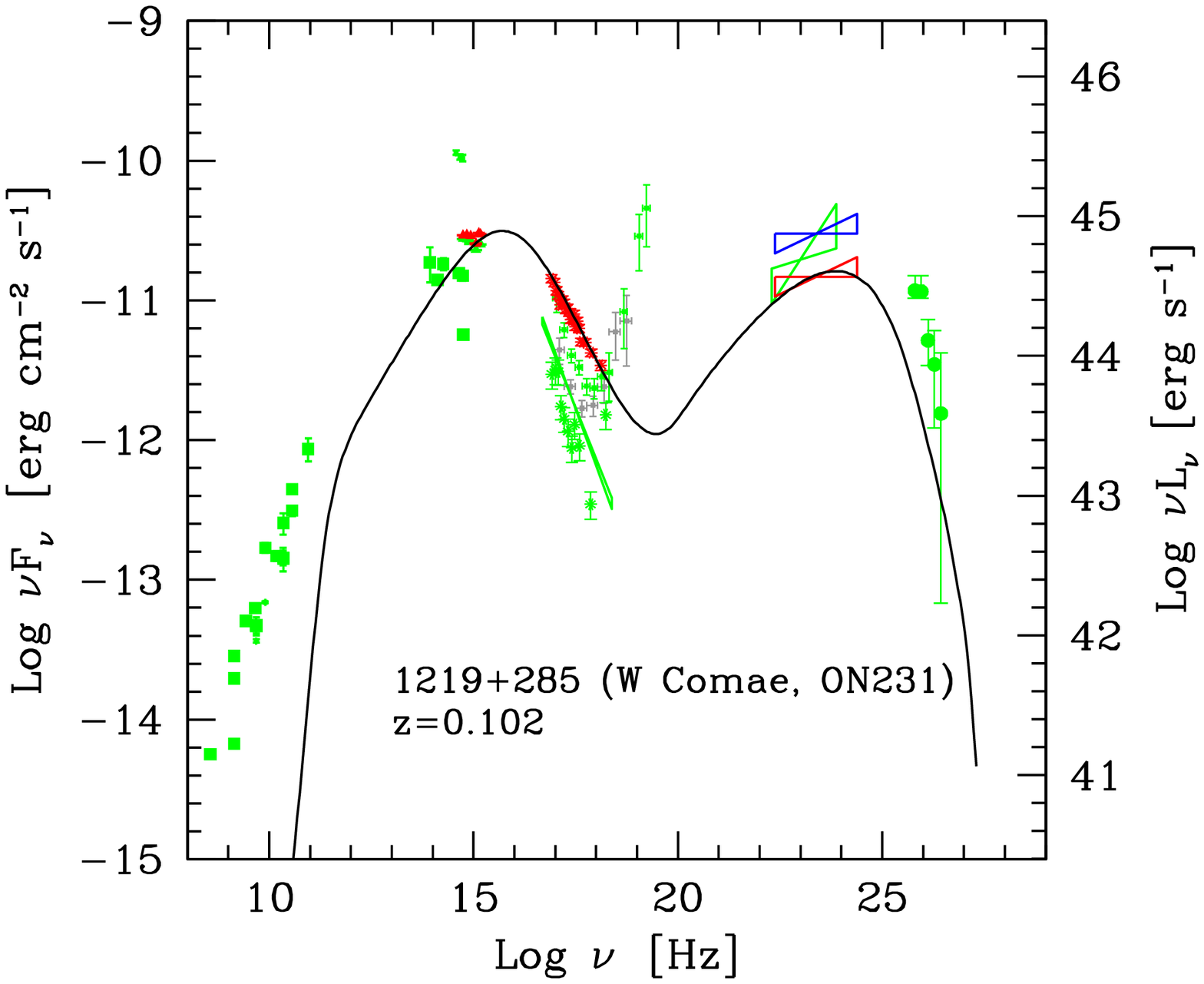,width=7cm,height=7cm}
\vskip -1.5 cm
\hskip 7 cm
\psfig{figure=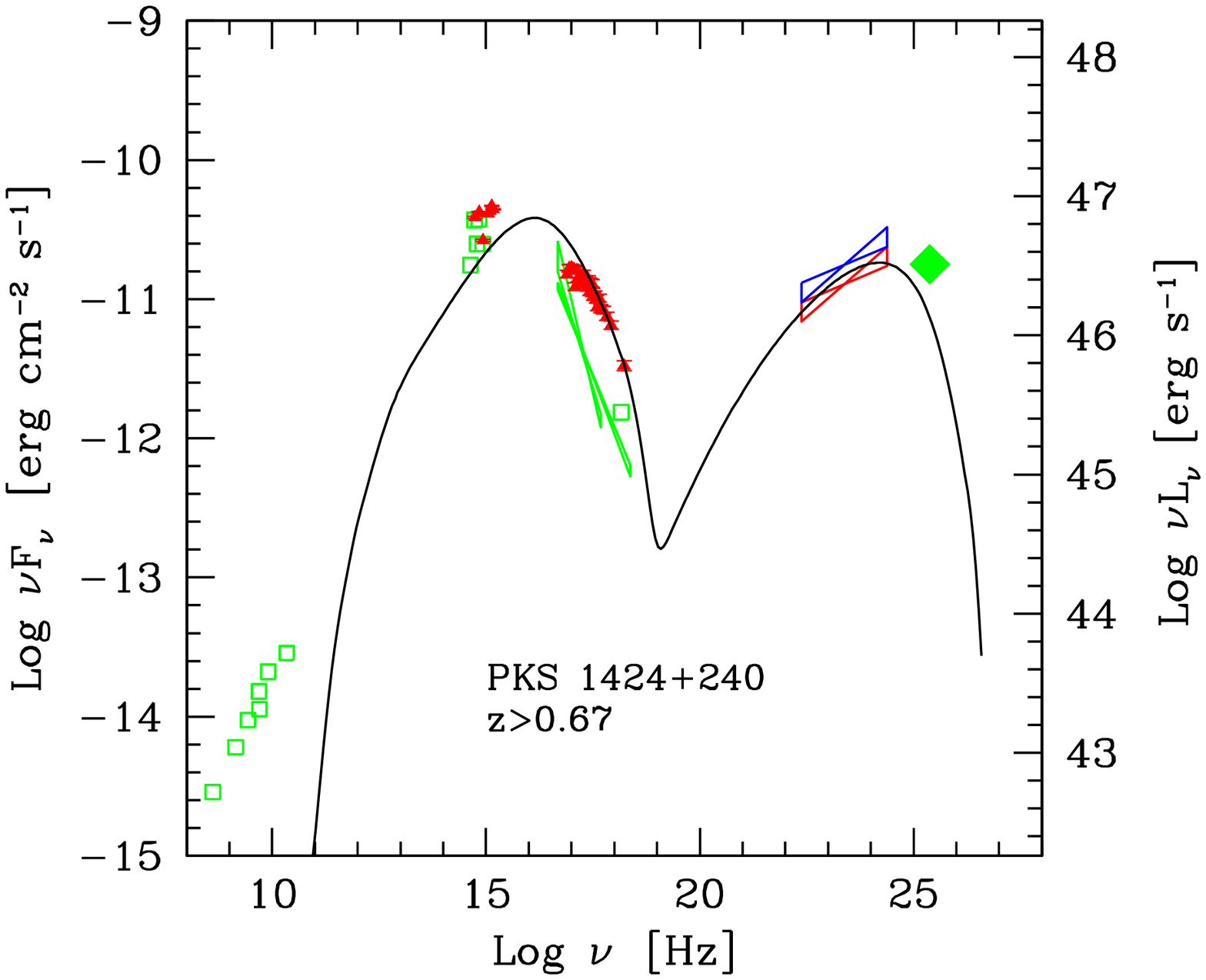,width=7cm,height=7cm}
\vskip -1.5 cm
\hskip 7 cm
\psfig{figure=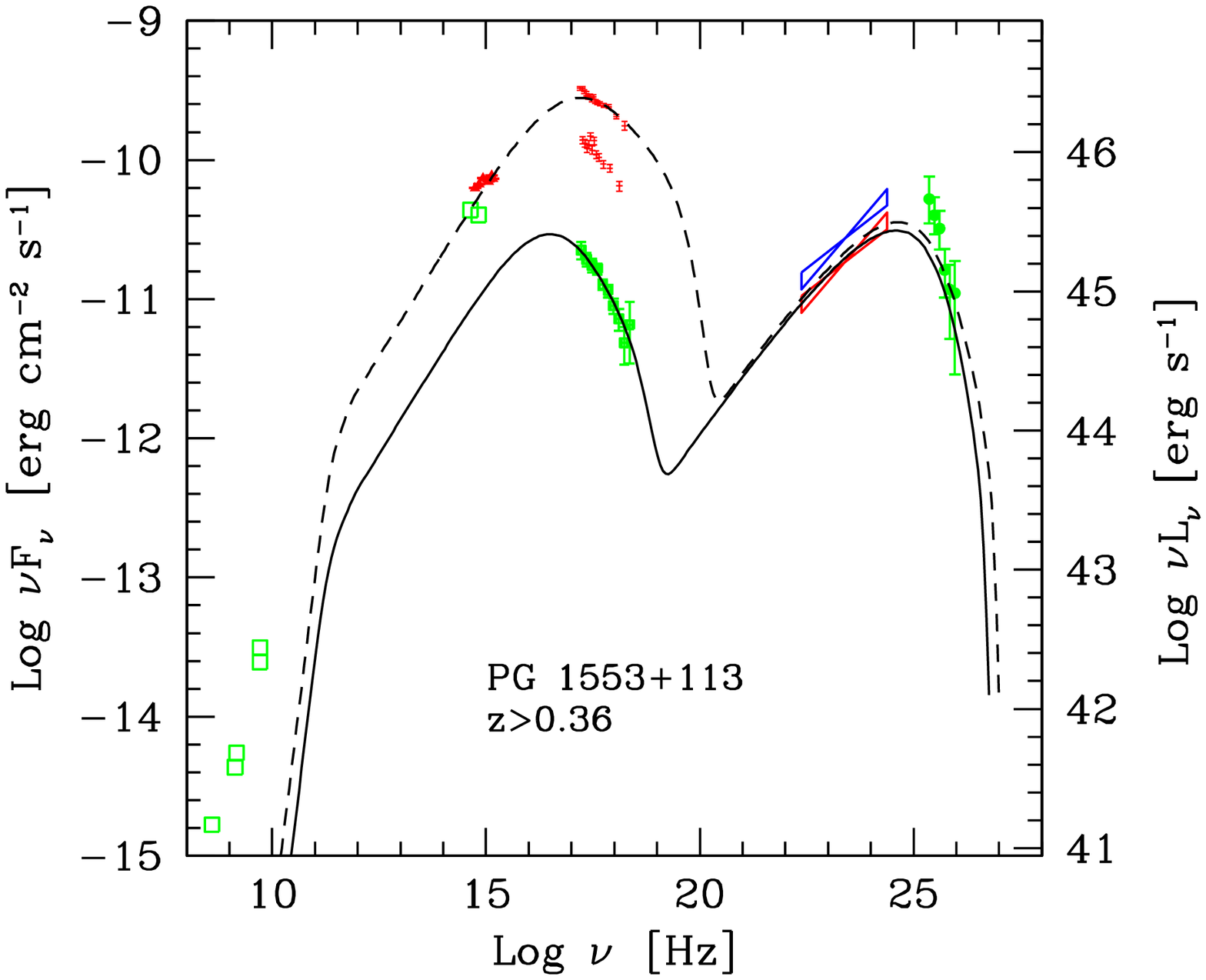,width=7cm,height=7cm}
\vskip -1.5 cm
\hskip 7 cm
\psfig{figure=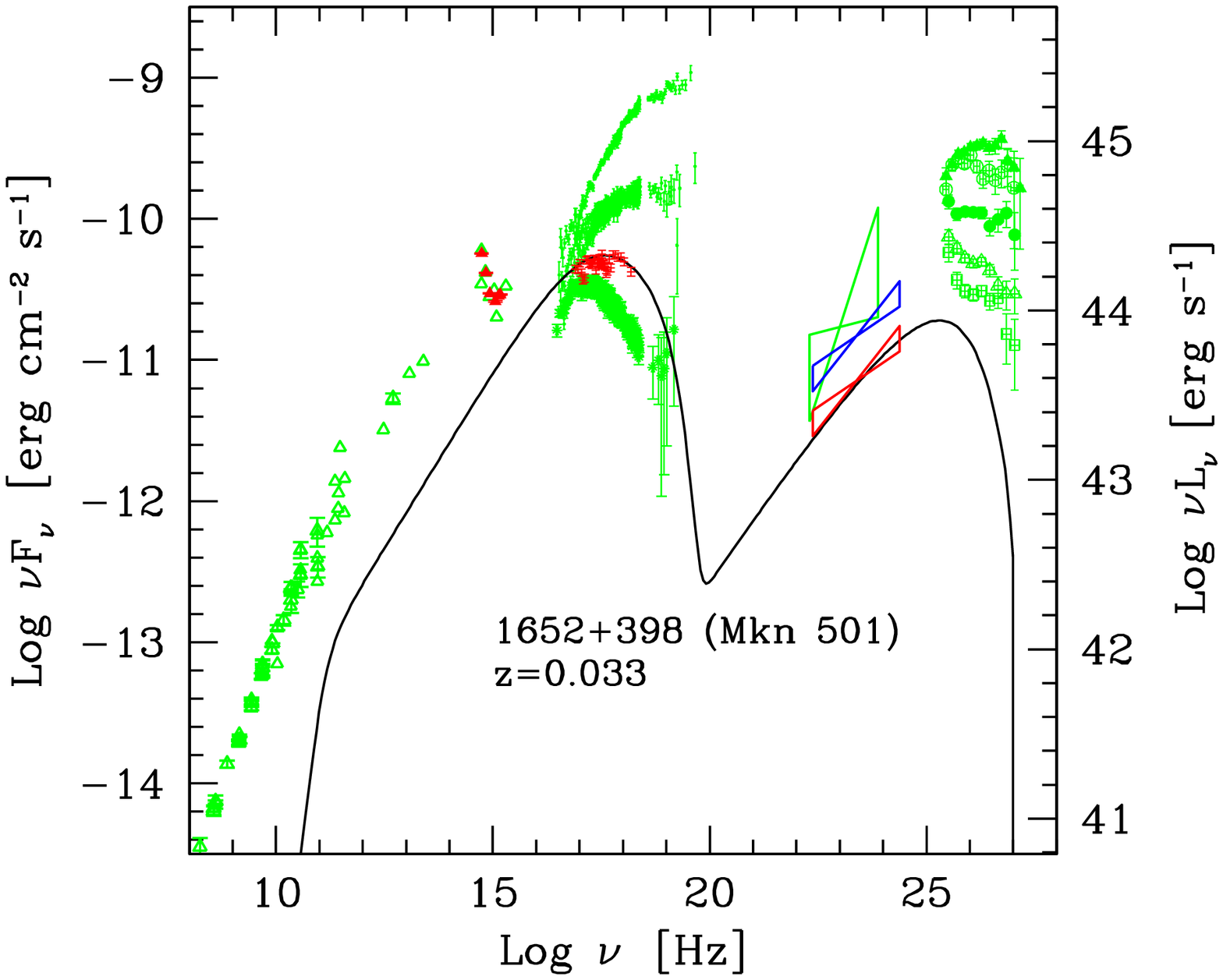,width=7cm,height=7cm}
\vskip -0.5cm
\caption{Spectral Energy Distribution of the BL Lacs detected by LAT and TeV telescopes. Historical data are from NED. Optical-UV and X-ray data from {\it Swift} are from this work. Bow ties report the power law model fitting the LAT data (Abdo et al. 2009). See text for more details. TeV spectra have been deabsorbed with the "lowSFR" model of Kneiske et al. (2004). References for the TeV data: 0219+428: Acciari et al. (2009a);  0716+71: Anderhub et al. (2009); 1011+496: Albert et al. (2007b); 1101+384: Zweerink et al. (1997), Maraschi et al. (1999), Albert et al. (2007c); 1219+285: Acciari et al. (2008); 1424+240: the datapoint is the observed (not deabsorbed) flux reported by Ong et al. (2009); 1553+113: Albert et al. (2007d); 1652+398: Albert et al. (2007a). The solid line is the result of the one-zone leptonic model.}
\label{sed}
\end{figure*}

Models are shown in Figs. \ref{sed},\ref{sedno} and \ref{sedcand}  and the parameters used to reproduce the data are reported in Table \ref{tableparam}. We assume a redshift of $z=0.3$ for all the sources with unknown distance. For the sources with more than one model we indicated the parameters for the low ({\it l}) and high ({\it h}) state.

The parameters of this simple one-zone model can be completely constrained when the value of peak frequency and flux of both emission components and an estimate of the minimum variability timescale are available (Tavecchio et al. 1998). In most cases the synchrotron peak can be constrained by the optical and X-ray data, while the IC peak is less determined, since the LAT spectra generally cover only the raising (or in some cases the declining) part of the bump. In the fit procedure we decide to locate the IC peak just above (or below) the LAT band. With this choice (see also below) we minimize the required Doppler factor.

Inspection of Table \ref{tableparam} shows that the parameters are quite similar in most of the sources, with some notable exceptions. In Fig. \ref{fitparam}  we summarize the results showing the values of the magnetic field and the electron break Lorentz factor $\gamma _{\rm b}$.  Open circles (blue triangles) report the values corresponding to TeV BL Lacs detected (not detected) by LAT, while green open squares indicates LAT BL Lacs not detected in the TeV band.
Most of the LAT sources are localized in the region corresponding to magnetic field between 0.1 and 1 G and $\gamma _{\rm b}$ is between $10^4$ and $10^5$. BL Lac itself requires a slightly larger magnetic field and a lower value of the break Lorentz factor. The reason is that its peaks (both synchrotron and IC) are located at lower energies (note that it is characterized by one of the steepest LAT spectrum, Fig.\ref{histo}, locating the IC peak well below 100 MeV).

Sources not detected by LAT seem to clearly separate, being characterized by low magnetic field and large electron Lorentz factor. Inspection of the SEDs reveals that  these sources could be divided in two groups: (i) sources rather similar to those detected by LAT, whose parameters fall quite close to the region populated by the LAT BL Lacs; (ii) peculiar sources: in this group there are particular sources (1ES 0229+200, 1ES 1101-232, RGB J0152+017, 1ES 0347-121, H1426+428 and, even if not so extreme, H 2356-309), characterized by a very hard TeV continuum (de-absorbed). The interpretation of these spectra is still a debated issue (e.g. Katarzynski et al. 2005, Stecker et al. 2007, Aharonian et al. 2008b, Boettcher et al. 2008). To reproduce the SED of these sources one needs to use rather extreme parameters, in particular a rather large energy for the emitting electrons. A possible model invokes the emission from relativistic electrons distributed as a power law in energy with a very high value of the minimum energy. In these conditions it is possible to get a rather hard IC continuum, $F(\nu)\propto \nu ^{1/3}$ (see Katarzynski et al. 2005 and Tavecchio et al. 2009 for a detailed discussion of the cases of 1ES 1101-232 and 1ES 0229+200, respectively). 

\begin{figure}
\vskip -0.3cm
\psfig{figure=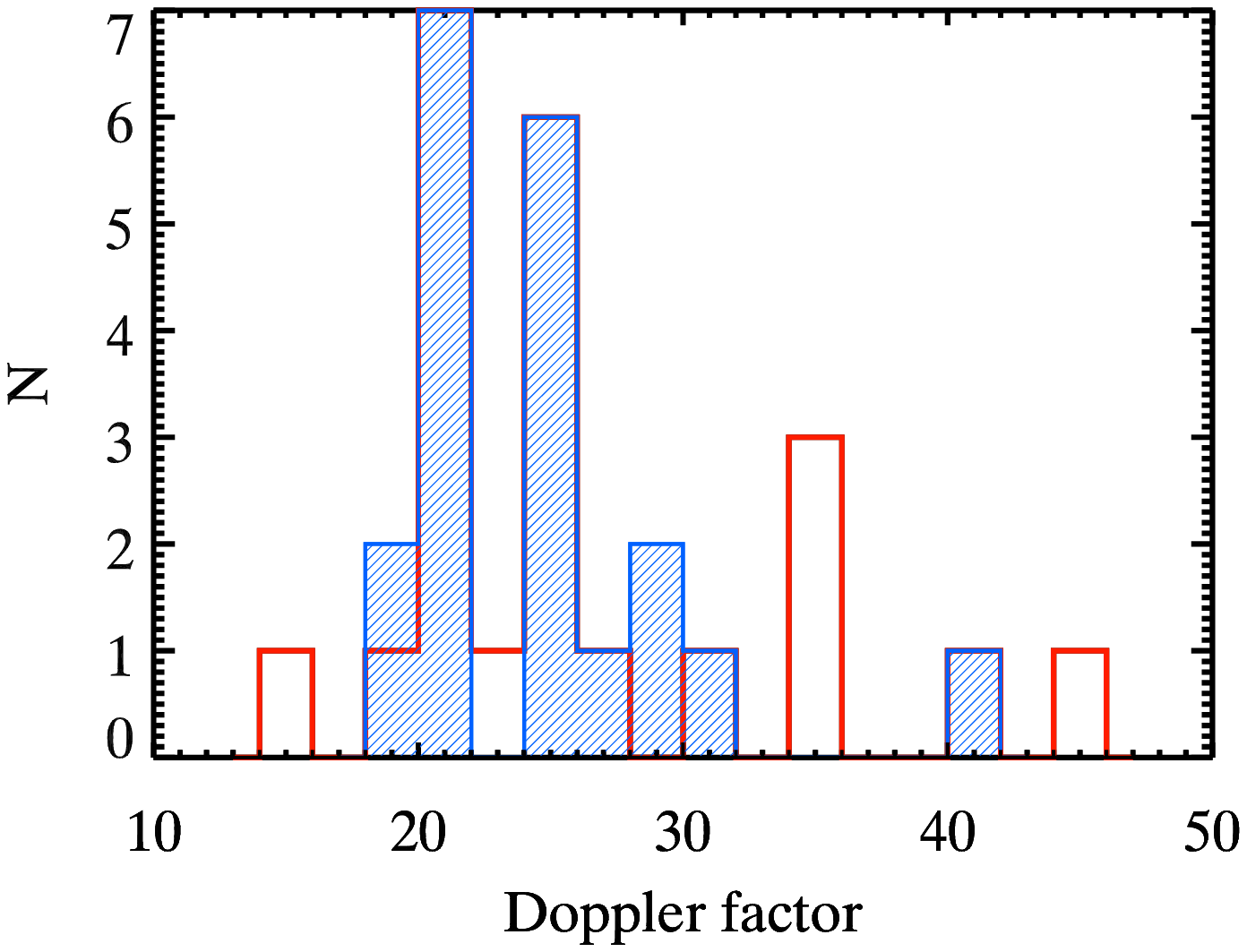,width=8cm,height=7.2cm}
\vskip -0.5cm
\psfig{figure=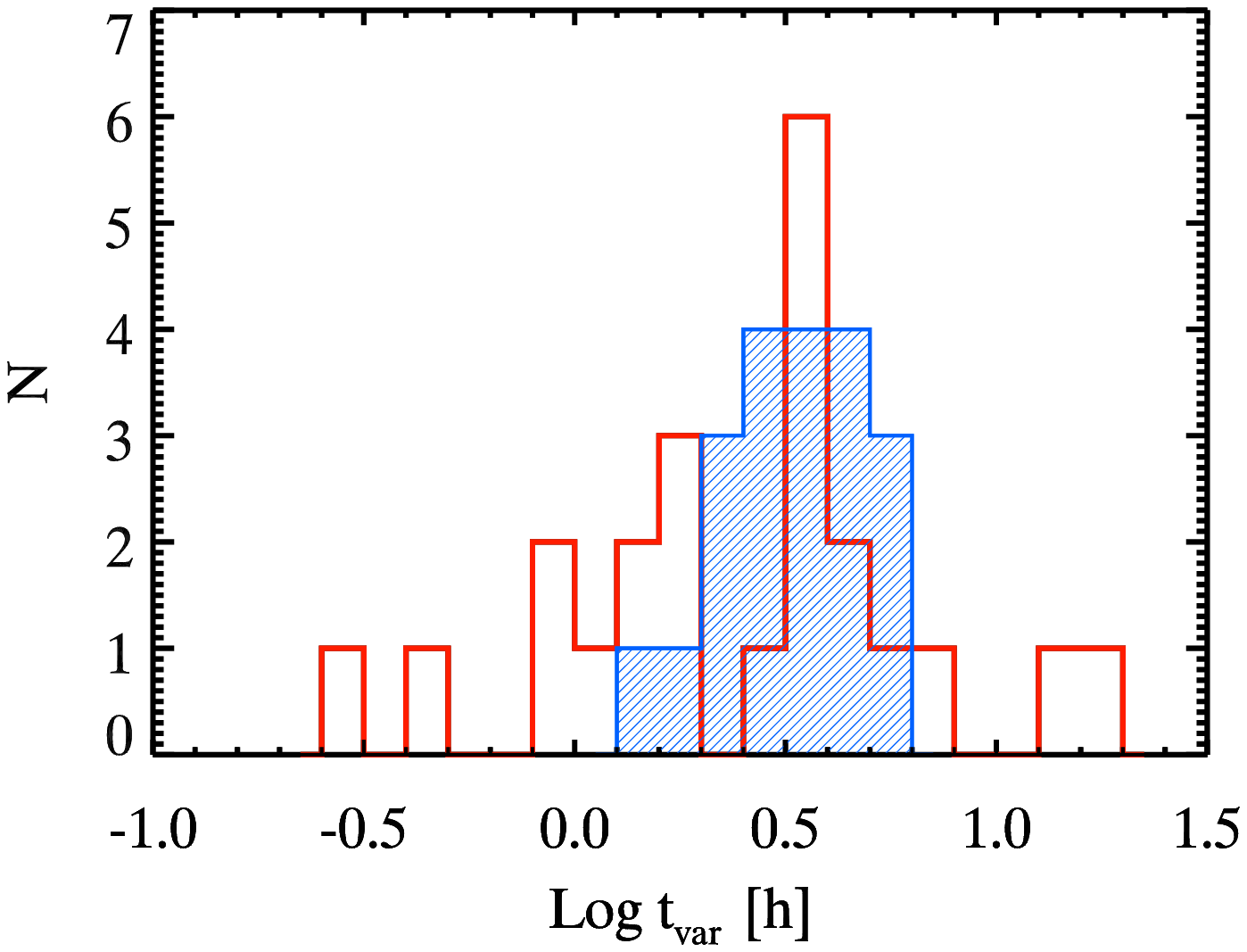,width=8cm,height=7.2cm}
\vskip -0.15 cm
\caption{Distributions of the Doppler factors (upper panel) and minimum variability timescales (lower panel) derived through the modelling of the SED. The shaded area indicates the distribution of sources not detected in the TeV band.}
\label{deltatvar}
\end{figure}

Based on these differences, we can predict that sources belonging to the group (i) will probably be detected in deeper observations of LAT. Instead, sources of type (ii) are intrinsically different: the presence of the very hard IC continuum, peaking above 1 TeV, makes very difficult for LAT to detect them, even with prolonged exposures. In this sense, LAT is probably able to probe only a portion of the TeV BL Lac population, i.e. the sources with the peak located not far from 100 GeV. All sources with a peak at higher energies will probably escape the detection of LAT. This fact has important consequences for the use of LAT data in the search for new TeV candidates.

In Fig. \ref{deltatvar} (upper panel) we show the distribution of the Doppler factors used in the models (the shaded histogram is for non-TeV sources). Most sources have $\delta$ between 20 and 30. Few extreme sources require $\delta >40$. TeV sources requiring these extreme values are mainly those with a very hard TeV spectrum. Large $\delta$ in the case of non-TeV sources are required for those BL Lacs with hard LAT spectra and synchrotron peak in the optical band (see below).

In the bottom panel of Fig. \ref{deltatvar} we show the minimum variability timescale (as measured in the observer frame), $t_{\rm var}=(1+z)R/c\delta$ derived from the modeling. The bulk of the sources have $t_{\rm var}$ around 1 and 10 hours. These are typical timescales predicted with the standard one-zone SSC model. Faster variations (such as those observed in PKS 2155-304 and Mkn 501, Aharonian et al. 2007, Albert et al. 2007a) require modifications of this scheme (e.g. Ghisellini \& Tavecchio 2008b, Ghisellini et al. 2009b, Giannios et al. 2009, Neronov et al. 2008).

\subsection{Comments on specific sources}

Inspection of SEDs reveals few cases (e.g. 1717+177, 2322+396) in which the model does not provide a good fit of the LAT spectrum. All these sources are characterized by a synchrotron peak located close to the optical band and a hard LAT spectrum, implying a IC peak above 10 GeV. In these conditions, the SSC model requires a rather large Doppler factor, $\delta>50$, and a low magnetic field (e.g. Tavecchio \& Ghisellini 2008 for a discussion). In the case of 2322+396 even with $\delta=40$ (and $B=0.007$ G; corresponding to the lowest open square in Fig. \ref{fitparam} ) the fit is not completely satisfactory.
A possibility to solve this problem is to assume that, as already discussed, the bow tie is somewhat misleading, since the power law is not a good model for the high-energy bins of the actual LAT spectrum. Therefore, contrary to the impression given by the power law fit, the peak would be located at few GeV, alleviating the requirements on the model parameters.

Notable is the case of 1514-241 (AP Lib). In this case it is rather difficult to reconcile the extremely steep optical-UV continuum, the hard X-ray spectrum and the (non simultaneous) LAT spectrum. The model strongly underpredicts the X-ray flux. The first obvious solution to this problem is to assume that the non simultaneous X-ray spectrum is not representative of the real state of the X-ray emission during the LAT observations. Another possibility to reconcile the X-ray and the LAT spectra is to assume that the emission originates in a structured jet (Ghisellini et al. 2005), in which the radiative interplay between a fast inner core (the spine) and a slower outer layer would result in a complex spectrum, with more than one component contributing in the X-ray-$\gamma$-ray band. The detailed discussion of this point is beyond the scope of this paper.

Another interesting source is S5 0716+71. Although the SSC model reproduces well the SED when only the LAT data are considered, the fit of the TeV emission observed in April 2008 by MAGIC although feasible (Anderhub et al. 2009), is difficult, requiring a rather broad high-energy component, extending above 300 GeV. Also in this case a viable solution invokes the emission from a structured jet (Tavecchio \& Ghisellini 2009, Anderhub et al. 2009).

\section{Prospects for the detection in the TeV band}

The search for new TeV BL Lacs is a rather difficult task for Cherenkov telescopes. Their limited field of view does not allow an all-sky survey. The method usually applied is then to select good candidates based on indirect estimators of the level of the TeV emission, such as the X-ray luminosity (e.g. Stecker et al. 1996). With the launch of {\it Fermi} and the availability of an all-sky survey at $\gamma$-rays up to hundreds of GeV we have now the unprecedented possibility to choose the best TeV candidates selecting those sources showing an intense emission up to 10-50 GeV (the maximum energy probed by the LAT with moderate exposures). This selection based on the LAT flux and spectrum is the most effective way to select TeV candidates, since it uses information on the level and spectrum of the emission {\it just below} the threshold of the current Cherenkov telescopes. Fig. \ref{histo} could thus already provide some interesting clues on some of the best TeV candidates. Apart BL Lac and S5 0716+71 (characterized by relatively soft LAT spectra), the other known TeV BL Lacs appear to be bright and relatively hard in the LAT band. Within this group, 1215+303, 1717+177, 2322+396 and Ap Lib, should be considered as optimal candidates. An outstanding source is MS 1050.7+4946. Its exceptionally hard spectrum makes it an optimal candidate but the flux is rather low, probably limiting the possibility to detect it.

\begin{figure}
\vskip -1 cm
\psfig{figure=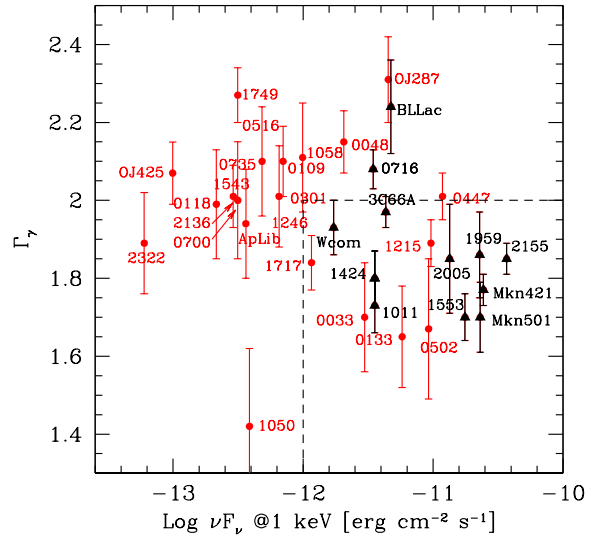,width=8.5cm,height=8.5cm}   
\vskip -0.3cm
\caption{Photon index of the LAT spectrum versus the X-ray flux at 1 keV (in $\nu F_{\nu}$) for the LAT BL Lacs. Black triangles indicate the BL Lac detected in the TeV band). Besides the two IBL BL Lac and S50716+71 all the TeV BL  Lacs, characterized by a hard spectrum ($\Gamma<2$) and a bright X-ray continuum, are confined in the area defined by the dashed line.
}
\label{plane2}
\end{figure}

Fig. \ref{plane2} shows the LAT photon index as a function of the X-ray flux. Here the TeV sources (again, excluding BL Lac and S50716+71) are located in the region of hard GeV spectra and large X-ray fluxes. This is simply due to the fact  that the majority of the known TeV sources (mostly HBLs) are bright X-ray sources and have the IC peak above 100 GeV, thus showing a hard GeV spectrum. Notably, some objects not yet detected in the TeV band lye in the same locus of the TeV BL Lacs, that we can approximately define by the rectangle plotted with dashed lines. Together with some of the sources already noted above (1215+303, 1717+177, 2322+396), there are other sources that should be considered as ideal TeV candidates: 0447--439, 0502+675, 0133+388, 0033-1938. Note also that 1424+240, very recently detected by VERITAS and MAGIC (Ong et al. 2009, Teshima et al. 2009), belongs to this group of sources. This reinforce the idea that the sources included in the area defined by the dashed lines should be considered the best candidates for the TeV detection. Note again the peculiar position of MS 1050.7+4946.

\begin{figure}
\vskip -1 cm
\psfig{figure=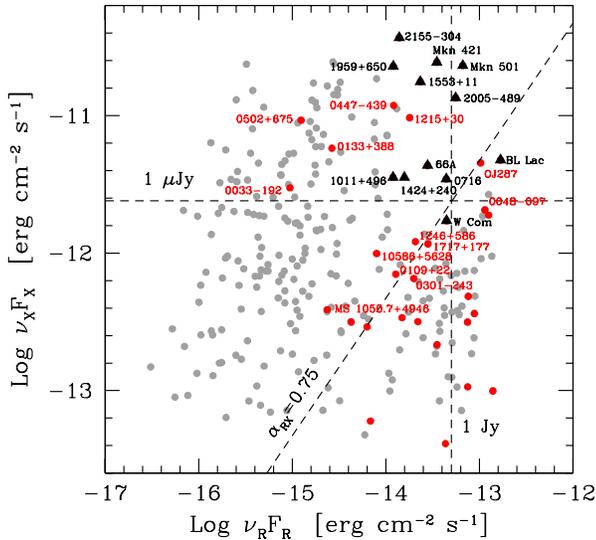,width=8.5cm,height=8.5cm}     
\vskip -0.3cm
\caption{X-ray versus radio flux for the BL Lacs in the A09 list. Filled triangles indicates known TeV BL Lacs. Gray points are from the original plot of Costamante \& Ghisellini (2002). Dashed lines indicate the X-ray flux level of 1$\mu$Jy (horizontal) and the radio flux of 1 Jy (vertical). The diagonal dashed line indicates the locus characterized by an effective slope between the radio and the X-ray band of $\alpha _{RX}=0.75$, conventionally separating HBLs ($\alpha _{RX}<0.75$) and LBLs ($\alpha _{RX}>0.75$).
}
\label{plane}
\end{figure}

Finally, it is also interesting to compare the positions of the LAT BL Lacs with the other blazars  in the radio-X-ray flux plot of CG02. They proposed to select the TeV candidates among the blazars with the largest X-ray and radio fluxes. The theoretical idea behind this choice is that the level of the synchrotron X-ray emission is a proxy for the number of high-energy electrons, while the radio flux is a measure of the energy density of the soft photons providing the seeds for the IC process. The brightest TeV BL Lacs should be those with the largest radio and X-ray fluxes.
This methods have been proved to be quite effective: indeed most (though not all) the TeV BL Lacs detected in the past years belong to the objects selected by CG02. In Fig. \ref{plane} we report a revision of the CG02 plot including all the LAT BL Lacs for which both radio and X-ray fluxes are available. TeV BL Lacs (shown by triangles) are concentrated in the left-top side of the diagram. Quite interestingly, almost all the LAT BL Lacs are characterized by a large radio flux. We interpret this fact as a confirmation of the idea behind the CG02 plot: the brightest $\gamma$-ray emitters are in fact those with the largest radio flux. The large span in the X-ray flux, instead, comes from the presence of both HBL and LBL in the A09 list, conventionally separated in this plane by the line at $\alpha _{RX}=0.75$ (Giommi \& Padovani 1994). Also in this diagram we can extract some particularly interesting candidates looking at the area occupied by the known TeV BL Lacs: again we find 1215+303, 0447-439, 0133+388. 
Two other outstanding sources are 0033-0192 and 0502+675. They are the sources with the lowest radio flux among the LAT BL Lacs, but they have rather hard GeV spectra and large X-ray flux (see Fig.\ref{plane2}), suggesting an important emission at TeV energies.

An important {\it caveat} in considering the results of these plots is that, when selecting the best candidates, one must also consider the effect of the absorption by the EBL, especially important above 50 GeV and for sources at redshift $z>0.1$. It is possible that sources appearing good candidates accordingly to the plots discussed above are actually penalized by strong absorption due to a relatively large distance.

In order to provide a criterion to select sources for TeV telescopes including also the effect of absorption, 
we calculated the expected fluxes of all the sources in the band covered by Cherenkov telescopes by using two independent way to estimate the intrinsic flux of the source: (i) in the first case we simply assume that the spectrum measured by LAT extends unbroken up to TeV energies; (ii) as a second estimate we use the high-energy spectrum predicted through the application of the emission model to the SEDs (in Fig. \ref{sedcand} we show both the intrinsic and the absorbed spectrum). Of course the first choice is rather generous, since we know that the actual spectrum of TeV blazars generally steepens above 100 GeV. On the other hand, in the second case the predicted fluxes are rather small, since the model IC component generally decreases quite fast above the LAT band due to the decreasing scattering efficiency. However, as discussed above, we cannot exclude the contribution from the emission of other components at TeV energies, thus resulting in larger fluxes than those expected with our simple one-zone modelling. Considering both estimates should thus give a reasonable idea of the real TeV fluxes from these sources. Also here absorption is calculated using the "lowSFR" model of Kneiske et al. (2004).

The results are reported in Table \ref{tevcand}. We report the fluxes above 50 GeV and 300 GeV predicted with the extrapolation of the LAT spectrum ($L$) and from the model ($M$). We recall that for all the sources with unknown redshift we fix it to $z=0.3$. With the largest threshold only few sources could be accessible with current instruments, while lowering the energy at 50 GeV all the sources should be characterized by fluxes larger than $10^{-12}$ ph cm$^{-2}$ s$^{-1}$. Not surprisingly, the favorite sources are those already noted above, in particular 0447-439, 0502+675, 1215+303 and 1717+177. MS 1050.7+4946 is the preferred source accordingly to both ways to estimate the intrinsic flux. Less bright, but still interesting are 0033-1938, 0133+388, 2322+396.  In particular the first source, though showing a rather intense intrinsic TeV emission, is located at large redshift ($z=0.61$), determining a relatively low observed flux.

\begin{table}
\centering
\begin{tabular}{lcccc}
\hline
\hline
source        & F$^L_{>50}$ & F$^L_{>300}$& F$^M_{>50}$& F$^M_{>300}$\\
\hline
0033-1938       & 8.3& 4.7& 1.7 & 0.15\\      
0048--097       & 6.0& 12.3& 0.5& 0.01\\     
0109+22         & 5.8& 12.9& 0.3& 0.05\\     
0118--272        &3.6 & 1.8&0.3 & 0.02\\     
0133+388        & 1.8& 79.5&2.5 & 0.93\\     
0301--243        & 5.3& 17.1& 0.55& 0.22\\     
0447--439        & 2.0& 172.7& 5.9& 7.7\\     
0502+675         &26.7 & 58.8 & 15.4& 13.8\\     
0712+5033       &4.4 & 10.6& 0.77& --\\     
0735+178          & 3.2& 3.2& 0.33 & 0.01\\     
0814+425          &7.0 & 3.8& 1.8& 0.08\\     
0851+202          &2.0 & 3.1& 0.14& --\\     
1050.7+4946     &2.2 & 364.6& 8.6 & 32.3\\
10586+5628      & 3.4& 18.9& 1.1& 2.1\\     
1215+303          & 3.1& 268.4& 4.9& 6.3\\     
1514--241          & 1.0& 96.7& 0.33 & --\\     
15429+6129      & 3.5 & 9.1& 0.4& 0.1\\     
1717+177          & 3.0& 261.4& 4.5& 12.5\\     
1749+096          & 5.2& 7.7& 0.76& 0.1\\     
2136-428           & 1.1& 27.1& 1.2& 0.13\\     
2322+396          & 7.5& 22.7& 0.75& 0.06\\     
\hline
\hline
\end{tabular}
\vskip 0.4 true cm
\caption{Predicted flux above 50 and 300 GeV extrapolating the LAT spectrum ($L$) and from the SED model ($M$) for the LAT BL Lac not yet detected in the TeV band. Fluxes above 50 GeV are in units of $10^{-11}$ ph cm$^{-2}$ s$^{-1}$, fluxes above 300 GeV in units of $10^{-13}$ ph cm$^{-2}$ s$^{-1}$. For sources with unknown redshift we use $z=0.3$. 
}
\label{tevcand}
\end{table}

\section{Summary and conclusions}

We have studied all the BL Lac objects detected by the LAT instruments onboard {\it Fermi} in its first three months of operations. Among them there are 12 sources detected also in the TeV band. To these we also added in our study the remaining 12 TeV BL Lacs not detected by LAT.
For all these sources we have built the SED using also optical-UV and X-ray data from {\it Swift}. The SEDs have been modeled with a simple one-zone leptonic model in order to derive the main physical parameters of the emission region. We found that the sources have rather similar Doppler factors, distributed around $\delta=20-30$. Some exceptional source require larger values, up to $\delta=40$.
Typical magnetic fields are in the range $B=0.1-1$ G, while the Lorentz factor of the electrons emitting most of the power (at the peaks of the SED) are between $\gamma_{\rm b}=10^3$ and $10^5$. 

Interestingly, while a fraction of TeV BL Lacs not yet detected by LAT share the same values of the parameters with the LAT detected objects, several sources are characterized by rather extreme values (very small $B$ and large electron Lorentz factor), suggesting perhaps a real difference in the physical conditions of the emission region in these sources. In fact, different possibilities to produce hard spectra have been  discussed (see Tavecchio et al. 2009 for a discussion). In particular, Katarzy{\'n}ski et al. (2005) showed that if the electrons (assumed to follow a power law distribution with energy) are characterized by a relatively large value of the minimum energy, a very hard spectrum can be achieved ($F_{\nu}\propto \nu^{1/3}$).  The same hard spectrum would be visible in the synchrotron component below the soft X-rays. Stecker et al. (2007), instead, pointed out that, under specific conditions, diffusive shock acceleration produces power law electron distribution much harder than the canonical $\gamma ^{-2}$. In both cases there would be the possibility to have SSC spectra with slopes harder then the canonical value $\Gamma =1.5$, the limiting slope commonly assumed in deriving upper limits to the EBL (e.g. Aharonian et al. 2006).

The full-sky survey at $\gamma$-rays performed by LAT offers a unique opportunity for the study of the high-energy emission of BL Lacs. For sources already discovered in the TeV band, we can now study in detail the high-energy component in the GeV band, and how it connects with the emission at higher energies. The comparison between the variability in the two bands (GeV and TeV) is expected to give important clues on the emission mechanisms and dynamics of the relativistic electrons in the jet (e.g. Aharonian et al. 2009). The possibility to have a continuous monitoring of the sources can provide good alerts for the Cherenkov telescopes, allowing a better understanding of the global behaviour of the high energy emission of BL Lacs including their duty cycles.

Particularly interesting, as we have shown in this paper, is the prospect to use the LAT data to select good candidates to be observed by Cherenkov telescopes. The first three months of observations already provided several good sources worth to be considered. Due to the strong variability of these objects at high-energies, is likely that many other interesting BL Lacs, even with bright emission, would be revealed in the future. However, the LAT selection is probably biased against the sources with extremely hard TeV spectra, for which our model predict a rather low flux in the MeV--GeV band, well below the capabilities of LAT. 

\section*{Acknowledgments}
FT thanks M. Persic and A. Stamerra for useful discussions. We thank the referee, F. Stecker, for his encouraging report.
This work was partly financially supported by a 2007 COFIN-MIUR grant and by ASI under contract ASI I/088/06/.
This research has made use of the NASA/IPAC Extragalactic Database (NED) which is operated by the Jet Propulsion Laboratory, Caltec, under contract with the NASA. We acknowledge the use of public data from the Swift data archive. This research has made use of data obtained from the High Energy Astrophysics Science Archive Research Center (HEASARC), provided by NASAÕs GSFC.


\vskip 28truecm

\section{Appendix}

\begin{table*}
\centering
\vspace*{3truecm}
\begin{tabular}{lcccccc}
\hline
\hline
source     &OBS date    &$\Gamma  $& $E_b$ & $\Gamma_2$  & $\chi2{\rm /dof}$ & $F_{\rm 0.3-10}$  \\
        \quad \, [1]       &       [2]    &     [3] &    [4]    &    [5]    &     [6]    &   [7] \\
\hline
00311-1938 & 21/11/2008 & $2.38\pm0.1$  & -- & -- & 38.8/32 & 0.72 \\ 
0048-097   & 04/06/2008 & $2.38\pm0.24$ & -- & -- & 4.3/4   & 0.58 \\ 
0109+22    & 31/05/2006 & $2.16\pm0.1$  & -- & -- & 52/48   & 0.24 \\ 
0219+428   & 2/10/2008  & $2.7\pm0.15$ & -- & -- & 14.8/20& 5.8\\
           & 3/10/2008  & $2.75\pm0.08$& --& --& 41/30& 11.1\\ 
0301-243   &  & & & & &  \\ 
0447-439   & 19/12/2008 & $2.2\pm0.2$  & $0.9\pm0.1$ & $2.7\pm0.2 $& 83/98 & 2.84 \\ 
0502+675   & 04/01/2009 & $1.7\pm0.1$ & $1\pm-0.14$ & $2.37\pm0.05$ & 263/230 & 29.5 \\ 
0712+5033  & 21/1/2009 & $1.6\pm0.5$& -- & -- & --$^*$ & 0.06\\ 
0716+714   & 28/11/2008 & $2.56\pm0.11$ & -- & -- & 17/20 & 1.35 \\ 
0735+178   & 18/05/2007 & $1.55\pm0.45$ & -- & -- & --$^*$ & 0.14 \\ 
0814+425   & 04/01/2008 & $1.64\pm0.62$ & -- & -- & 2.4/2 & 0.85 \\ 
0851+202   & 11/11/2008 & $1.67\pm0.11$ & -- & -- & 17/19 & 1.17 \\ 
1011+496   & 8/5/2008  & $2.04\pm0.40$& $0.74\pm0.20$& $2.62\pm0.15$& 25.1/33&  25.5\\ 
1050.7+4946 & 02/03/2009 & $2.3\pm0.4$ & -- & -- & 8.6/4 & 0.08\\ 
1101+384    & 3/12/2008  & $1.7\pm0.5$ & $0.8\pm0.22$ & $2.3\pm0.2$ & 23/140 & 96.2\\
1219+285    & 7/6/2008& $2.43\pm0.05$& -- & -- & 66.48/69& 10.2\\ 
1424+240    & 11-19/6/2009 & $2.2\pm0.08$& $1.24\pm0.21$& $3.0\pm0.1$& 413/491& 3.1\\ 
1514-241 	 & 08/04/2007   & $1.7\pm0.3$ & -- & -- & 3/12 &  0.65 \\
15429+6129 & 18-20/1/2009 & $2.5\pm0.4$& -- & -- & 0.3/2& 0.05\\
1553+11  & 6/10/2005& $2.24\pm0.03$& -- & -- & 203/221& 72.8\\ 
                & 20/4/2005& $2.38\pm0.06$&-- & --& 84/73& 27.0\\ 
1652+398 & 12/5/2008  & $1.68\pm0.15$& $0.9\pm 0.14$& $2.1\pm0.06$ & 129/126 &16.4\\ 
1717+177 & 08/01/2009 & $1.72\pm0.37$ & -- & -- &  --$^*$ & 0.16\\
1749+096 & 26/02/2007 & $2.13\pm0.21$& -- & -- & 23/50 & 17.6\\
1959+650 & 27/10/2008& $2.07\pm0.05$ & -- & --   & 46/48     & 86.0 \\ 
2005$-$489  & 8/10/2007& $2.69\pm0.14$&-- & -- & 13/11& 5.7\\ 
2136-428 &  19/12/2008 & $2.3\pm0.4$& -- & -- & 2.6/2& 0.05\\
2155$-$304   & 17/8/2008& $2.63\pm0.04$& -- & -- & 100/91 & 85.9\\ 
                               & 17/10/2008& $2.39\pm0.08$&  -- & --& 42/33& 37.9\\ 
 2200+420  & 29/8/2008 & $1.97\pm 0.1$&    --   & --    & 34.6/41& 6.4\\                         
2322+396 &  2//3/2009& $3.0^{+1.5}_{-1.1}$& -- & -- & --$^{*}$& 0.05\\
\hline
0152+017  & 2/12/2007& $2.23\pm0.12$& -- & -- & 10/14& 5.2\\ 
                  & 13/1/2008& $2.33\pm0.18$& --& --& 8/11& 0.48\\ 
0229+200  & 8/8/2008& $1.8\pm0.1$& -- & -- &23.6/21 & 10.3\\ 
0347-121  & 3/10/2006& $1.95\pm0.05$& -- & -- & 80.4/70 & 4.4\\ 
0548$-$322   & 28/11/2006& $1.7\pm0.1$& $1.8\pm0.7$& $2.1\pm0.3$& 78/71& 21.0\\ 
0710+591   & 26/2/2009& $1.80\pm0.05$& -- & -- &  70.52& 47.0\\ 
0806+524  & 13/3/2008& $2.5\pm0.1$& -- & -- & 16/16& 9.26 \\ 
1133+704 & 10/5/2008& $1.7\pm0.2$  & $1.2\pm0.2$        &$2.4\pm0.2$   & 40/36  & 17.8\\ 
			& 30/10/2008& $2.44\pm0.1$  & --                      &--                     & 12.5/14  & 8.4\\
			& 24/11/2008& $1.35\pm0.5$  & $1.1\pm0.3$     &$2.0\pm0.1$   & 91/92  & 52.3\\
1426+428    & 9/6/2008 & $2.51\pm 0.23$& --& --& 212/219 & 5.1\\ 
\hline
\hline
\end{tabular}
\vskip 0.4 true cm
\caption{Results of the X--ray analysis. [1]: name. [2]: observation date (dd/mm/yyyy). [3]: photon index. [4]: break energy (broken power law model). [5]: high energy photon index (broken power law model). [5]: value of the $\chi^2$ and degrees of freedom. [6]: unabsorbed 0.3--10 keV flux in units of 10$^{-11}$  erg cm$^{-2}$ s$^{-1}$. $^*$ poorly determined spectrum, the C-Statistics was used.}
\label{xrt}
\end{table*}

\newpage

\begin{table*}
\centering
\begin{tabular}{lccccccc}
\hline
\hline
source        &OBS date    &  V & B & U & UVW1 & UVM2 & UVW2  \\
                   &dd/mm/yyyy     & & & & & &     \\
\hline
00311-1938 &  21/11/2008    & $16.38\pm0.03$ & $16.70\pm0.02$ & $15.80\pm0.02$ & $15.70\pm0.02$ & $15.67\pm0.02$ & $15.74\pm0.01$ \\ 
0048-097   &  04/06/2008   & -- & --- & -- & -- & -- & $15.16\pm0.01$ \\ 
0109+22    & 31/05/2006     & $14.79\pm0.01$ & $15.23\pm0.01$ & $14.38\pm0.01$ & $14.46\pm0.01$ & $14.45\pm0.01$ & $14.56\pm0.01$ \\ 
0219+428   & 2/10/2008    & $14.46\pm0.07$  & $14.73\pm0.01$ & $14.73\pm0.01$ & $13.97\pm0.01$ & -- &  $14.06\pm0.01$ \\ 
                  & 3/10/2008   & $14.39\pm0.01$  & $14.74\pm0.01$ & $13.89\pm0.01$ & $13.99\pm0.01$ & $14.00\pm0.01$ & $14.11\pm0.01$ \\ 
0301-243*   & 24/3/2009    & $15.46\pm 0.03$ & $15.78\pm 0.02$ & $14.91\pm 0.02$ & $14.88\pm 0.03$ & $14.77\pm 0.04$ & $14.89\pm 0.03$\\ 
0447-439   &19/12/2008     & $14.36\pm0.01$ & $14.62\pm0.01$ & $13.69\pm0.01$ & $13.56\pm0.01$ & $13.43\pm0.01$ &	$13.51\pm0.01$ \\ 
0502+675   & 04/01/2009     & $16.6\pm0.01$ & $17.01\pm0.01$ & $16.20\pm0.01$ & $16.35\pm0.01$ & $16.46\pm0.01$ & $16.49\pm0.01$ \\ 
0712+5033  &  21//1/2009  0.218 & $17.04\pm0.04$&$17.47\pm0.03$& $16.77\pm0.03$& $16.96\pm0.04$& $17.09\pm0.05$& $17.16\pm0.04$\\ 
0716+714   & 28/11/2008    & -- & -- & -- & -- & -- & $17.71\pm0.03$ \\
0735+178   & 18/05/2007    & -- & -- & -- & -- & 16.00$\pm 0.01$ & --\\ 
0814+425   & 04/01/2008    & -- & -- & -- & -- & --    & $18.27\pm0.02$ \\ 
0851+202   &11/11/2008     & $14.85\pm0.02$ & $15.32\pm0.02$ & $14.61\pm0.02$ & $14.75\pm0.02$ & $14.79\pm0.02$ & $15.01\pm0.02$ \\ 
1011+496   & 8/5/2008    & $15.28\pm0.03$ & $15.53\pm0.02$ &	$14.58\pm0.01$ & $14.42\pm0.01$ & $14.23\pm0.02$ & $14.30\pm0.01$ \\ 
1050.7+4946 & 02/03/2009   & -- & -- & 21.74 & $17.53\pm0.05$  & -- & -- \\ 
1101+384    & 3/12/2008  & --&-- & --& --& --& -- \\ 
1219+285    &  7/6/2008   & $14.68\pm0.01$ & $15.05\pm0.01$ & $14.20\pm0.01$ & $14.17\pm0.01$ & $14.01\pm0.01$ & $14.15\pm0.01$ \\ 
1424+240    & 11-19/6/2009   & $14.49\pm0.02$ &	$14.81\pm0.01$ & $14.49\pm0.02$ & $13.97\pm0.01$  & $13.9\pm0.01$ & $14.03\pm0.01$ \\ 
1514-241    & 08/04/2007    & -- & $16.02\pm0.03$ & $15.67\pm0.02$ & $16.15\pm0.01$ & -- & $16.39\pm0.09$ \\
15429+6129  & 18-20/1/2009  & $16.40\pm0.04$&$16.73\pm0.03$ &$15.87\pm0.03$&$15.88\pm0.04$ & $15.83\pm0.04$& $15.95\pm0.03$\\
1553+11     & 6/10/2005   & $13.94\pm0.01$ & $14.28\pm0.01$ & $13.34\pm0.01$ & $13.34\pm0.01$ & $13.29\pm0.01$ & $13.42\pm0.01$ \\
            & 20/4/2005   & $14.26\pm0.01$ & $14.68\pm0.01$ & $13.69\pm0.01$ & $13.74\pm0.01$ & $13.72\pm0.01$ & $14.01\pm0.01$ \\  
1652+398    & 12/5/2008   & $13.94\pm0.01$ & $14.66\pm0.01$ & $14.16\pm0.01$ & $14.16\pm0.01$ & $14.06\pm0.01$ & $14.12\pm0.01$ \\ 
1717+177    & 08/01/2009   & $17.71\pm0.09$ & $18.01\pm0.05$ & $17.27\pm0.04$ & $17.36\pm0.04$ & $17.37\pm0.05$ & $17.59\pm0.03$ \\
1749+096    & 26/02/2007   & $14.19\pm0.01$ & $14.76\pm0.01$ & $14.11\pm0.01$ & $14.60\pm0.01$ & $14.98\pm0.02$ & $15.06\pm0.01$ \\
1959+650    & 27/10/2008   & $15.09\pm0.03$ & $15.54\pm0.02$ & $14.84\pm0.02$ & $15.04\pm0.03$ & $15.12\pm0.03$ & $15.14\pm0.02$ \\ 
2005$-$489  & 8/10/2007   & $13.32\pm0.01$ & $13.71\pm0.01$ & $12.75\pm0.01$ & $12.66\pm0.01$ & $12.54\pm0.01$ & $12.67\pm0.01$ \\ 
2136-428 &19/12/2008   & $15.42\pm0.03$&$15.77\pm0.02$ &$14.90\pm0.02$& $14.92\pm0.03$ &$14.86\pm0.04$ & $14.98\pm0.03$\\
2155$-$304  & 17/8/2008  & $12.87\pm0.01$ & $13.18\pm0.01$ & $12.20\pm	0.01$ & $12.07\pm0.01$ & $11.91\pm0.01$ & $12.01\pm0.01$  \\ 
	       & 17/10/2008  & $13.44\pm0.01$ & $13.75\pm0.01$ & $12.79\pm0.01$ & $12.69\pm0.01$ & $12.56\pm0.01$ & $12.67\pm0.01$  \\ 
2200+420    & 29/8/2008  & $14.87\pm0.01$ & $15.69\pm0.01$ & $15.37\pm0.01$ & $16.04\pm0.02$ & $16.72\pm0.03$ & $16.73\pm0.02$ \\                
2322+396 & 2/3/2009 &$18.7\pm0.3$ &$18.9\pm0.1$ & $18.5\pm0.1$& $18.8\pm0.1$& $19.1\pm0.1$& $19.6\pm0.1$\\
\hline
0152+017  & 13/1/2008  & $15.59\pm0.03$ & $16.38\pm0.03$ & $15.84\pm0.03$ & $15.94\pm0.03$ & $15.86\pm0.03$ & $15.94\pm0.02$ \\ 
0229+200  & 8/8/2008  & $17.04\pm0.07$& $18.11\pm0.07$& $17.92\pm0.08$& $18.21\pm0.08$& $18.45\pm0.1$& $18.38\pm0.06$ \\ 
0347-121  & 3/10/2006  & $17.18\pm 0.07$ & $17.65\pm 0.05$ & $16.86\pm 0.04$ & $16.54\pm 0.04$  & -- & -- \\ 
0548$-$322     & 28/11/2006 & $16.3\pm0.03$ & $16.96\pm0.02$ & $16.61\pm	0.03$ & $16.72\pm0.03$ & $16.68\pm0.04$ & --  \\ 
0710+591    & 26/02/2009& $16.56\pm0.05$ & $17.21\pm0.04$ & $16.48\pm0.04$ & $16.41\pm0.04$ & $16.34\pm0.05$ & $16.40\pm0.03$  \\ 
0806+524    & 13/03/2008& $15.78\pm0.04$ & $16.17\pm0.02$ & $15.29\pm0.02$ & $15.27\pm0.02$ & $15.29\pm0.02$ &  $15.32\pm	0.02$ \\ 
1133+704   & 30/10/2008 & -- & -- & -- & -- &  $15.41\pm	0.03$ & $15.42\pm0.02$ \\ 
1426+428      & 09/06/2008 & $16.39\pm0.04$ & $16.99\pm0.03$ & $16.21\pm0.01$ & $15.99\pm0.02$ & $15.89\pm0.03$ & $15.86\pm0.02$  \\ 
\hline
\hline
\end{tabular}
\vskip 0.4 true cm
\caption{Results of the UVOT analysis. (*) Average of two observations performed the same day.}
\label{uvot}
\end{table*}

\begin{table*}
\centering
\begin{tabular}{lccccccccc}
\hline
\hline
source        & $\gamma _{\rm min}$ & $\gamma _{\rm b}$& $\gamma _{\rm max}$& $n_1$&$n_2$ &$B$ &$K$ &$R$ & $\delta $ \\
      \quad [1]         & [2]  & [3] & [4] & [5] & [6] & [7] & [8]  & [9] & [10]  \\
\hline
0219+428    {\it l}          &$1 $&$ 1.4\times 10^{4} $&$ 3\times 10^{5} $&$ 2 $&$ 4.8 $&$ 1.1 $&$ 3\times 10^{4}$&$ 6 $& 25   \\      
$\quad \quad \quad \quad $ {\it h}     &$1 $&$ 2.2\times 10^{4} $&$ 3\times 10^{5} $&$  2$&$ 4.8 $&$ 0.8$&$  5\times 10^{4} $&$ 6 $& 25 \\    
0716+71      &$ 1 $&$ 6\times 10^{3} $&$  8\times 10^{4} $&$  2$&$ 4.5 $&$ 0.2 $&$ 2.2\times 10^{4} $&$  8.3 $& 47 \\           
1011+496  &$ 30 $&$ 8\times 10^{4} $&$ 2\times 10^{5}$&$ 1.9 $&$ 4.3 $&$ 0.33 $&$ 1.6\times 10^{3} $&$  8.5 $& 25  \\       
1101+384           &$ 1 $&$ 10^{ 5}   $&$ 4\times 10^{ 5} $&$ 2 $&$ 4 $&$ 0.24 $&$ 8.5\times 10^{3} $&$ 4 $&   20 \\     
1219+285      &$10^{2} $&$ 1.5\times 10^{4}$&$ 10^{6} $&$  2 $&$ 4.2 $&$ 0.3 $&$ 4\times 10^{4} $&$ 3 $& 25  \\         
1424+240  &$ 3\times 10^{2}$&$ 2\times 10^{4}$&$2\times 10^{5}$&$  2 $&$ 4.4 $&$ 0.55 $&$ 1.3\times 10^{4} $&$ 7.6 $& 35 \\           
1553+113    {\it l} &$ 1 $&$4\times 10^{4} $&$ 3.5\times 10^{5} $&$ 2 $&$ 4.2 $&$ 0.4 $&$ 2\times 10^{4} $&$ 7.5 $& 20 \\     
 $\quad \quad \quad \quad $ {\it h}  &$ 1 $&$3\times 10^{4} $&$  5\times 10^{5} $&$  2 $&$ 3.8 $&$ 2 $&$ 3.35\times 10^{3} $&$  9  $& 23 \\        
1652+398            &$ 1 $&$ 1.2\times 10^{ 5}$&$ 6\times 10^{5}$&$ 2$&$ 3.9$&$ 0.313$&$ 4.5\times 10^{4} $&$1 $&20  \\  
1959+650 &$1  $&$ 5.7\times 10^{4}$&$ 6\times 10^{5}$&$ 1.9$&$ 3.4$&$ 0.4 $&$ 7\times 10^{ 2}$&$ 7.3       $& 18   \\  
2005-489  &$ 10^{2} $ & $ 1.3\times 10^{4}$ & $ 10^{7}$ & $  2 $ &$ 4.8 $& $ 0.7 $ & $ 1.6\times 10^{3} $ &$ 8 $ & 22  \\                
2155--304   {\it l}&$ 5\times 10^{1} $&$ 1.5\times 10^{4}$&$10^{6}$&$ 2$&$ 4.5$&$ 0.28$&$ 1.65\times 10^{4}$&$ 17.5 $&15\\   
$\quad \quad \quad \quad $ {\it h}  &$ 5\times 10^{1} $&$ 2.5\times 10^{4}$&$ 10^{5} $&$ 2$&$ 4.5 $&$ 0.9 $&$1.5\times 10^{3}$&$ 14.5  $& 20 \\    
2200+420              &$10$&$ 9\times 10^{2}   $&$ 2\times 10^{ 5}$&$ 2 $&$ 3.9 $&$ 1.4 $&$  8\times 10^{5}$&$ 2 $&15           \\   
\hline
0152+017     &$ 10 $&$ 3\times 10^{5} $&$ 10^{6} $&$ 2 $&$ 3.5 $&$ 0.03 $&$ 10^{4} $&$ 4 $& 25 \\                            
0229+200  & $7.5\times 10^5 $&$5\times 10^6 $  & $4\times 10^7$& 2.3 & 3.3 & $5\times 10^{-4}$& $9.5\times 10^5$& $35$ & 40 \\ 
0347--121   &$ 3\times 10^{3} $&$ 2\times 10^{6} $&$ 4\times 10^{6}$&$ 2.2 $&$ 3.5 $&$ 0.0025 $&$ 2.2\times 10^{3}$&$  60 $& 44\\ 
0548--322$^{*}$ &$1.5\times 10^4 $&$2.8\times 10^5 $&$1.5\times 10^7 $&$2 $&$4.1 $&$0.1 $&$10^4 $&$3.6 $& 25\\ 
0710+591$^{*}$  &$2.5\times 10^4 $&$5\times 10^5 $&$10^7 $&$2 $&$4.1 $&$0.13 $&$2.5\times 10^4 $&$4 $&25\\ 
0806+524  &$ 4\times 10^{3} $&$ 5\times 10^{4}$&$ 2\times 10^{6} $&$ 2 $&$ 4.5 $&$ 0.39 $&$ 8\times 10^{3} $ & 5.7 & 20\\        
1101--232   &$ 4\times 10^{4} $&$ 3.5\times 10^{5} $&$  2\times 10^{6}$&$  1.8 $&$ 4.3 $&$ 0.02 $&$ 1.9\times 10^{2} $&13 & 41 \\       
1133+704          &$ 10^{4}$&$ 5\times 10^{4}$&$ 2\times 10^{6}$&$ 2$&$4$&$ 0.15$&$ 1.2\times 10^{4}$&$ 4$& 20  \\ 
1218+304  &$10^{4} $&$ 1.3\times 10^{5} $&$ 10^{7} $&$ 2 $&$ 4.5 $&$ 0.2 $&$ 1.2\times 10^{5} $&$ 1 $& 35 \\                 
1426+428  {\it l}    &$10^{2} $&$ 6\times 10^{4} $&$ 10^{6} $&$ 2 $&$ 4 $&$ 0.18 $&$ 2.8\times 10^{3} $&$ 5.9 $& 25   \\             
$\quad \quad \quad \quad $ {\it h} &$7\times 10^{3} $&$ 1\times 10^{4} $&$ 6\times 10^{6} $&$ 2 $&$ 3.1 $&$ 0.033 $&$ 5\times 10^{5} $&$ 3 $& 35   \\             
2344+514 &$ 1 $&$  10^{4}    $&$ 7\times 10^{ 5}$&$ 2$&$3.2 $&$ 0.1   $&$ 3\times 10^{4}    $&$ 3.5$& 25   \\   
2356-309       &$10^{3} $&$ 2\times 10^{5} $&$ 3\times 10^{7}$&$ 2$&$ 3.8$&$ 0.12 $&$ 6\times 10^{3} $ &3 & 35 \\                 
\hline
\hline
\end{tabular}
\vskip 0.4 true cm
\caption{Input model parameters for all the TeV BL Lacs with (upper panel) and without (lower panel) LAT detection. [1]: source.  [2], [3] and [4]: minimum, break and maximum electron Lorentz factor.  [5] and [6]: slope of the electron energy distribution below and above $\gamma _b$. [7]: magnetic field [G]. [8]: normalization of the electron distribution in units of cm$^{-3}$. [9]: radius of the emission zone in units of $10^{15}$ cm. [10]: Doppler factor. $*$: no high energy data yet available.}
\label{tableparam}
\end{table*}

\begin{table*}
\centering
\begin{tabular}{lccccccccc}
\hline
\hline
source        & $\gamma _{\rm min}$ & $\gamma _{\rm b}$& $\gamma _{\rm max}$& $n_1$&$n_2$ &$B$ &$K$ &$R$ & $\delta $ \\
       \quad [1]         & [2]  & [3] & [4] & [5] & [6] & [7] & [8]  & [9] & [10]  \\
\hline
0033-1938             &$100 $&$ 2\times 10^{4} $&$ 10^{6} $&$ 2 $&$ 4 $&$ 0.4 $&$ 2\times 10^{4}$&$ 6.5 $& 26   \\      
0048--097             &$100 $&$ 5\times 10^{3} $&$ 2\times 10^{5} $&$ 2 $&$ 4.2 $&$ 0.46 $&$ 9\times 10^{4}$&$ 7 $& 18   \\      
0109+22             &$30 $&$ 7\times 10^{3} $&$ 7\times 10^{4} $&$ 2 $&$ 4.5 $&$ 0.2 $&$ 3.5\times 10^{4}$&$ 7.6 $& 30   \\      
0118--272             &$10 $&$ 10^{4} $&$ 1.5\times 10^{5} $&$ 2 $&$ 4.6 $&$ 0.35 $&$ 2\times 10^{4}$&$ 10 $& 25   \\      
0133+388             &$1 $&$ 2\times 10^{4} $&$ 10^{6} $&$ 2 $&$ 4.2 $&$ 0.8 $&$ 5.5\times 10^{3}$&$ 8 $& 20   \\      
0301--243             &$10 $&$  10^{4} $&$ 2.5\times 10^{5} $&$ 2 $&$ 4.6 $&$ 0.33 $&$ 1.4\times 10^{4}$&$ 10 $& 2   \\      
0447--439             &$100 $&$ 1.5\times 10^{4} $&$  10^{6} $&$ 2 $&$ 4.4 $&$ 0.4 $&$ 2\times 10^{4}$&$ 5.1 $& 20   \\      
0502+675             &$5\times 10^3 $&$ 7\times 10^{4} $&$ 10^{6} $&$ 2 $&$ 4.1 $&$ 0.75 $&$ 3.5\times 10^{3}$&$ 10 $& 20   \\      
0712+5033             &$60 $&$ 10^{4} $&$ 3\times 10^{4} $&$ 2 $&$ 4.5 $&$ 0.17 $&$ 1.3\times 10^{5}$&$  $& 25   \\      
0735+178             &$30 $&$ 10^{4} $&$ 8\times 10^{4} $&$ 2 $&$ 4.6 $&$ 0.13 $&$ 7\times 10^{4}$&$ 8 $& 25   \\      
0814+425             &$70 $&$ 2\times 10^{4} $&$ 1.5\times 10^{5} $&$ 2 $&$ 4.6 $&$ 0.042 $&$ 3.3\times 10^{5}$&$ 6 $& 25   \\      
0851+202             &$250 $&$ 2\times 10^{3} $&$ 5\times 10^{4} $&$ 2 $&$ 4.2 $&$ 0.5 $&$ 4\times 10^{5}$&$ 7 $& 18   \\      
1050.7+4946             &$7\times 10^3$&$ 10^{5} $&$ 5\times 10^{6} $&$ 2 $&$ 4.6 $&$ 0.025 $&$  10^{5}$&$ 3.8 $& 20   \\      
10586+5628            &$30 $&$ 8\times 10^{3} $&$ 10^{6} $&$ 2 $&$ 4 $&$ 0.1 $&$ 7\times 10^{4}$&$ 5 $& 25   \\      
1215+303             &$10^3 $&$ 2\times 10^{4} $&$ 10^{7} $&$ 2 $&$ 4.4 $&$ 0.1 $&$ 9\times 10^{4}$&$ 4.5 $& 20   \\      
1514--241             &$1 $&$ 2\times 10^{4} $&$ 5\times 10^{4} $&$ 2 $&$ 4.9 $&$ 0.012 $&$  10^{4}$&$ 10 $& 40   \\      
15429+6129             &$40 $&$ 1.4\times 10^{4} $&$ 3\times 10^{5} $&$ 2 $&$ 4.8 $&$ 0.085 $&$ 2.3\times 10^{4}$&$ 9 $& 28   \\      
1717+177             &$500 $&$ 8\times 10^{3} $&$ 6\times 10^{5} $&$ 2 $&$ 3.6 $&$ 0.05 $&$ 6.3\times 10^{5}$&$ 3 $& 21   \\      
1749+096             &$200 $&$ 2\times 10^{3} $&$ 1.05\times 10^{5} $&$ 2 $&$ 4 $&$ 0.75 $&$ 1.6\times 10^{5}$&$ 6 $& 25   \\      
2136-428             &$200 $&$ 1.4\times 10^{4} $&$ 9\times 10^{4} $&$ 2 $&$ 4.8 $&$ 0.12 $&$ 3\times 10^{4}$&$ 9 $& 28   \\      
2322+396             &$50 $&$ 3\times 10^{4} $&$ 9\times 10^{4} $&$ 2 $&$ 4.9 $&$ 0.007 $& $8\times 10^{4}$&$ 9 $& 40   \\      
\hline
\hline
\end{tabular}
\vskip 0.4 true cm
\caption{Input model parameters for the LAT BL Lacs not yet detected in the TeV band. See Table \ref{tableparam} for definitions.}
\label{tableparamnotev}
\end{table*}

\newpage


\setcounter{figure}{5}
\begin{figure*}
\vskip -1 cm
\psfig{figure=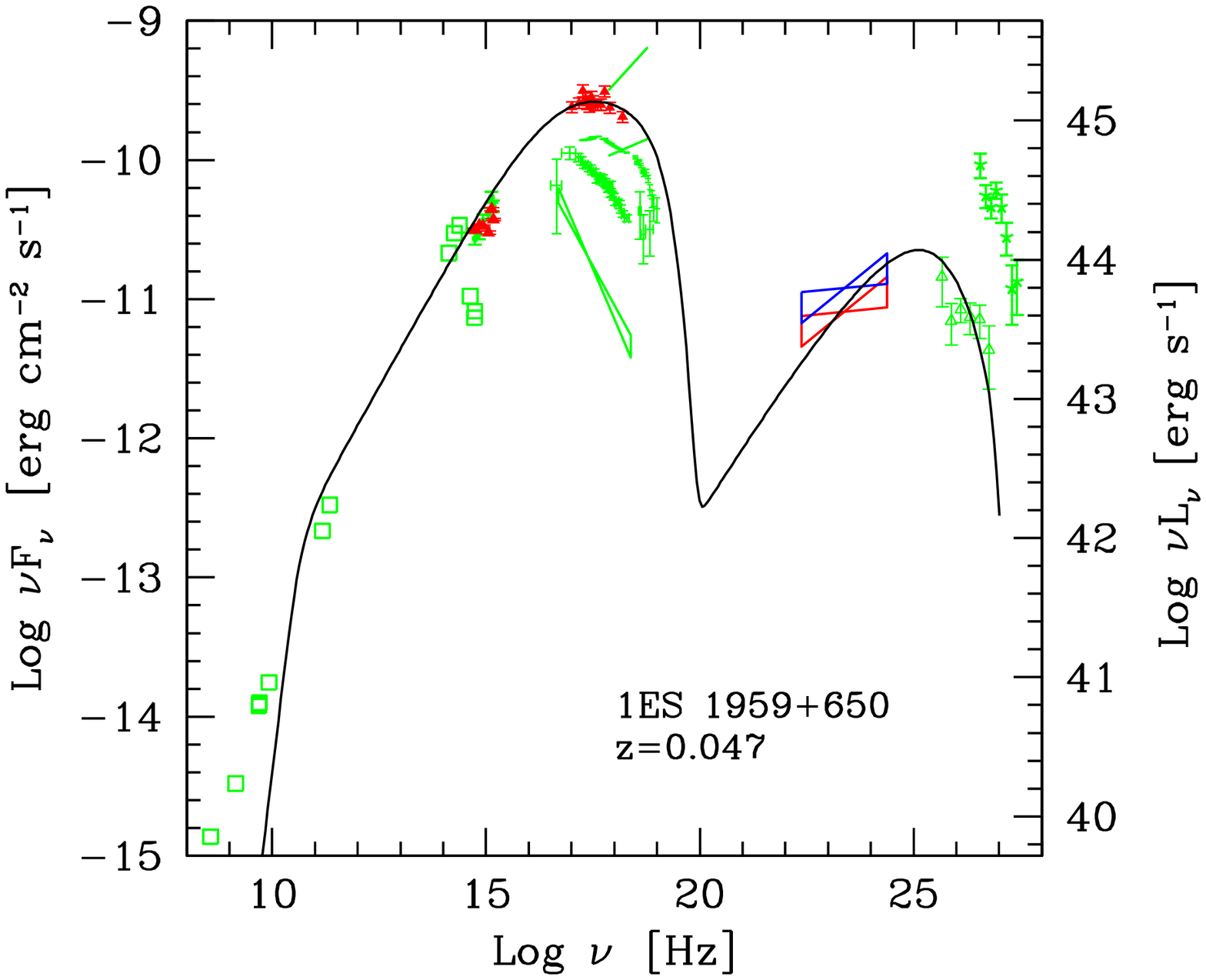,width=7cm,height=7cm}
\vskip -1.5 cm
\psfig{figure=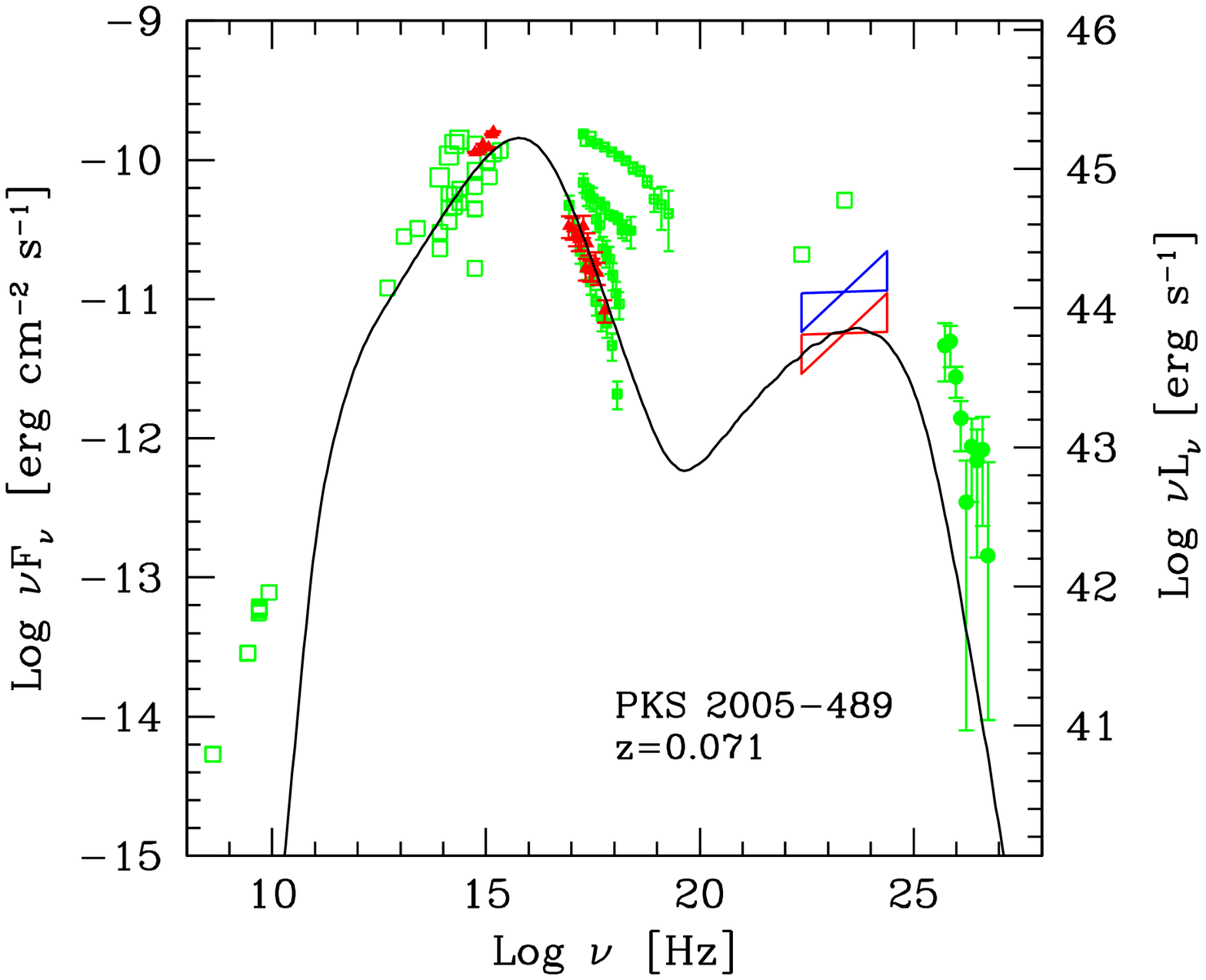,width=7cm,height=7cm}
\vskip -12.55 cm
\hskip 7 cm
\psfig{figure=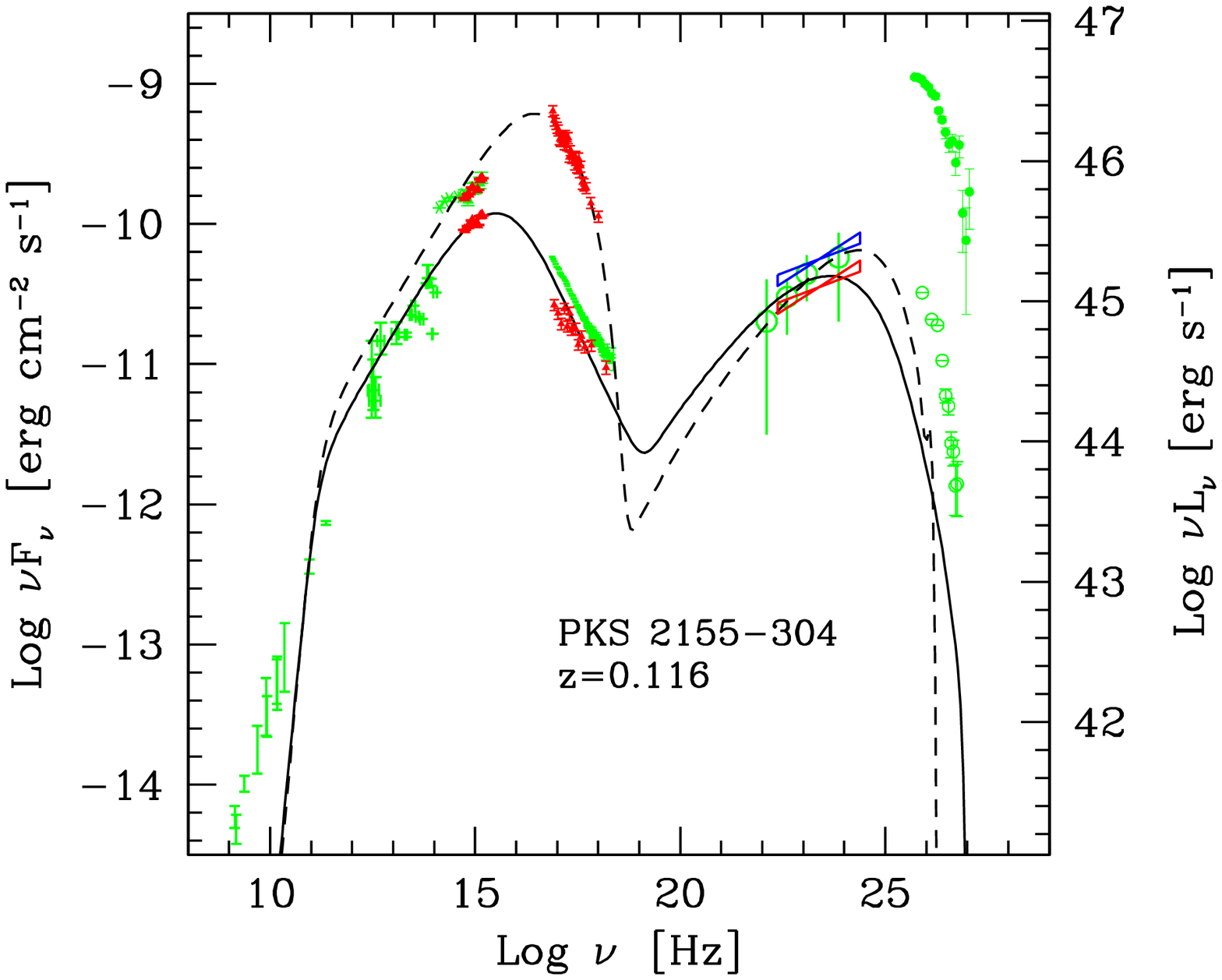,width=7cm,height=7cm}
\vskip -1.5cm
\hskip 7 cm
\psfig{figure=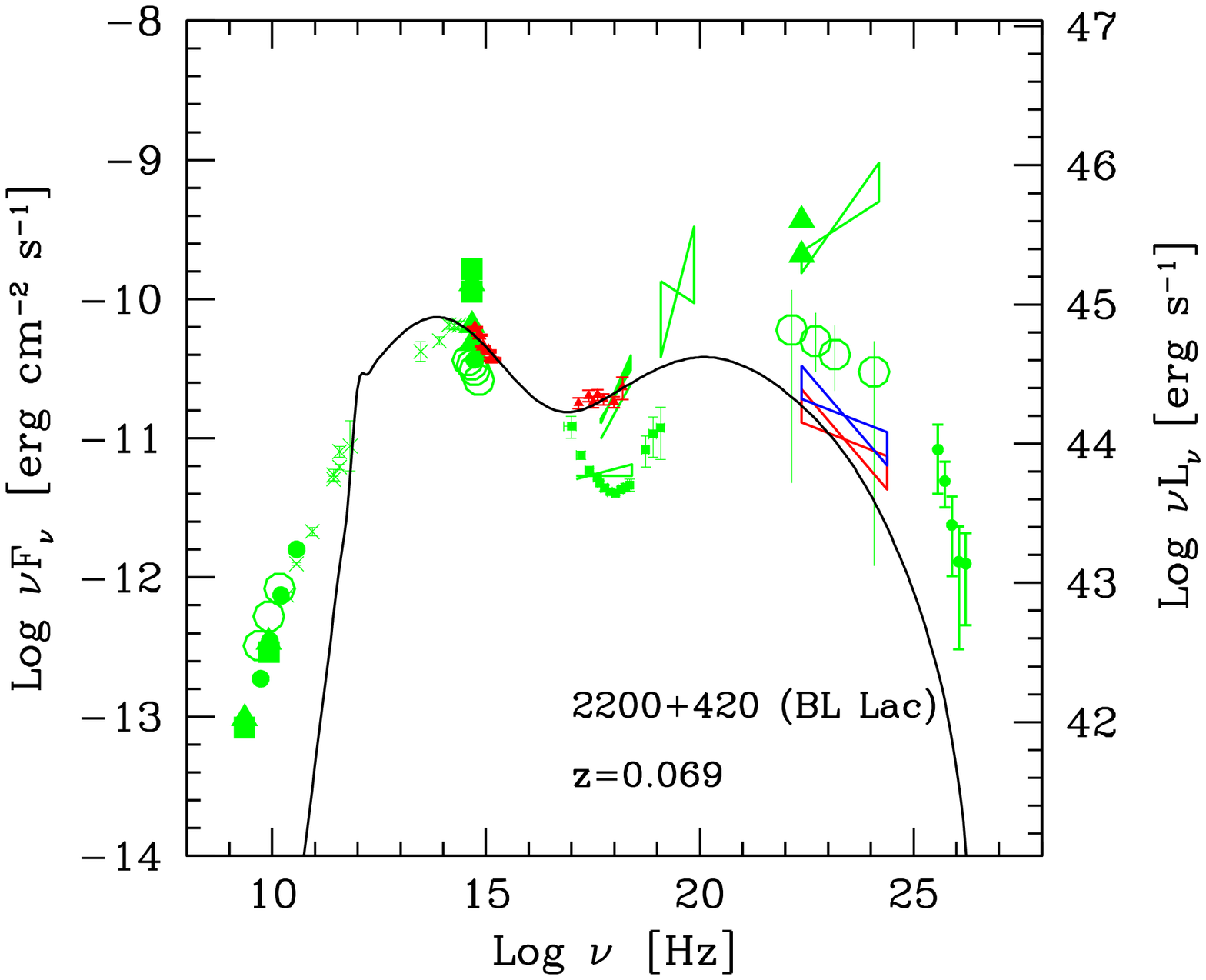,width=7cm,height=7cm}
\vskip -0.5cm
\caption{--continue-- 1959+650: Tagliaferri et al. (2008), Aharonian et al. (2003b); 2005-489: Aharonian et al. (2005a); 2155-304: Aharonian et al. (2005b), Aharonian et al. (2007a); BL Lac: Albert et al. (2007e).
}
\end{figure*}


\begin{figure*}
\vskip -1 cm
\psfig{figure=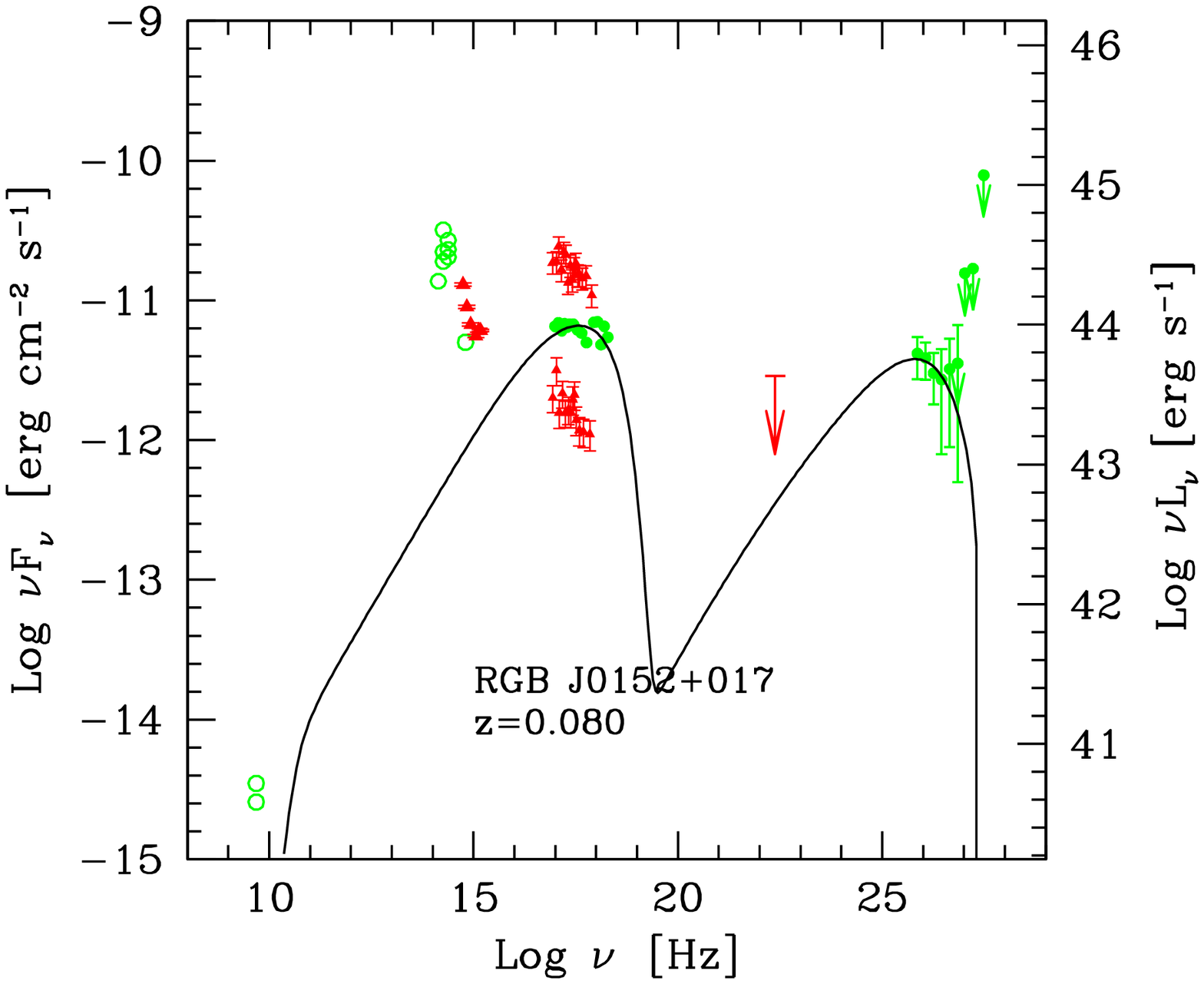,width=7cm,height=7cm}
\vskip -1.5 cm
\psfig{figure=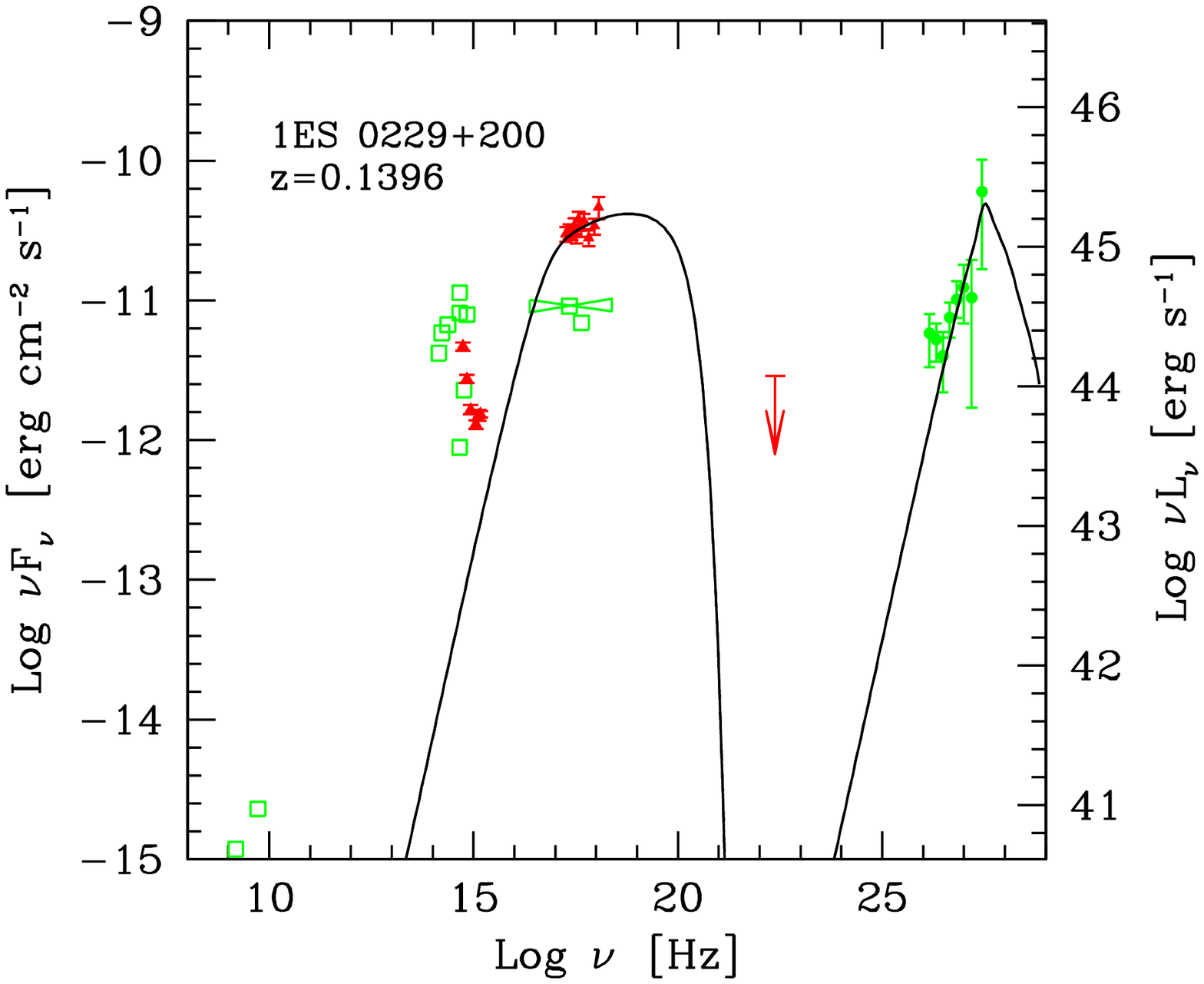,width=7cm,height=7cm}
\vskip -1.5 cm
\psfig{figure=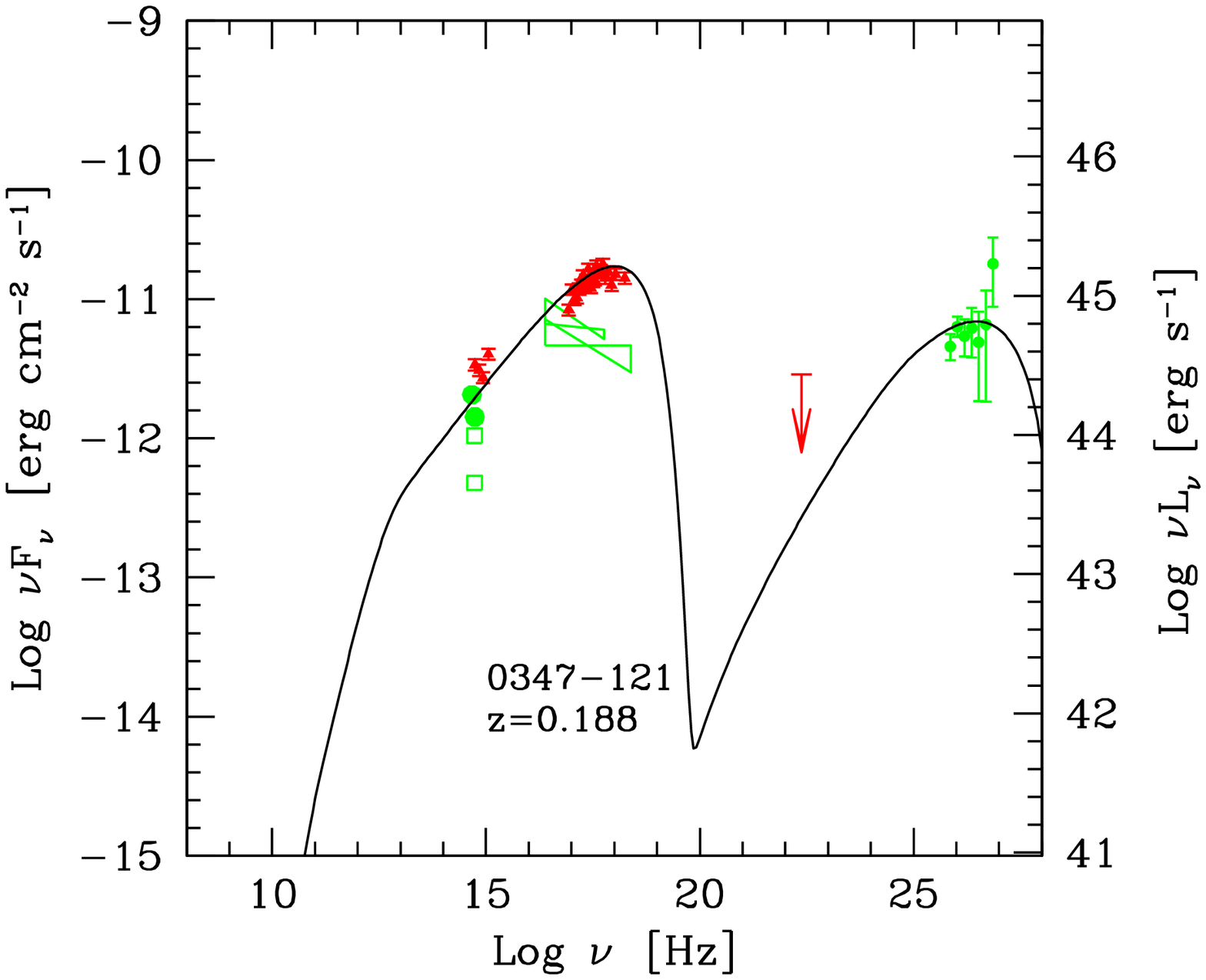,width=7cm,height=7cm}
\vskip -1.5 cm
\psfig{figure=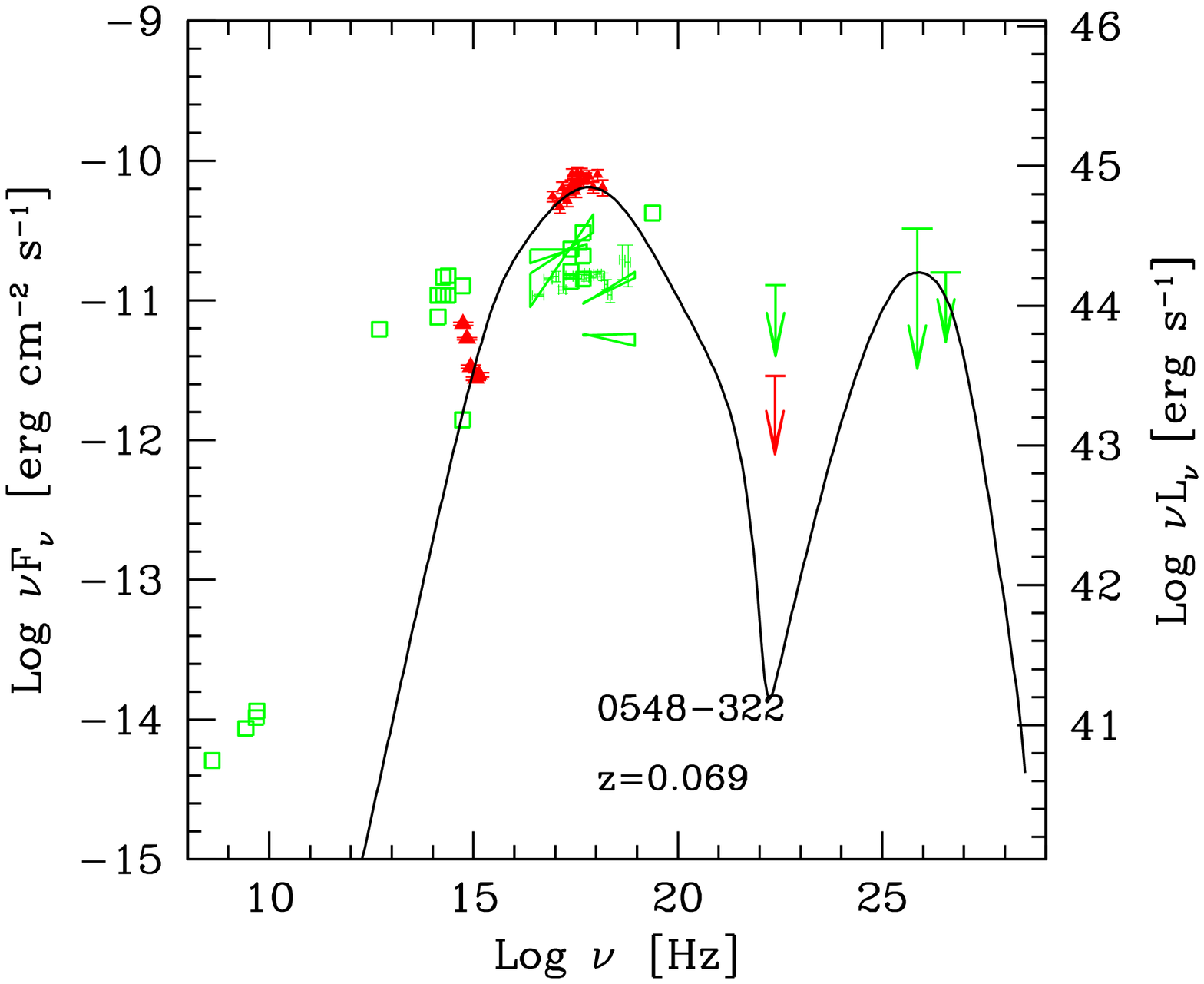,width=7cm,height=7cm}
\vskip -23.6 cm
\hskip 7 cm
\psfig{figure=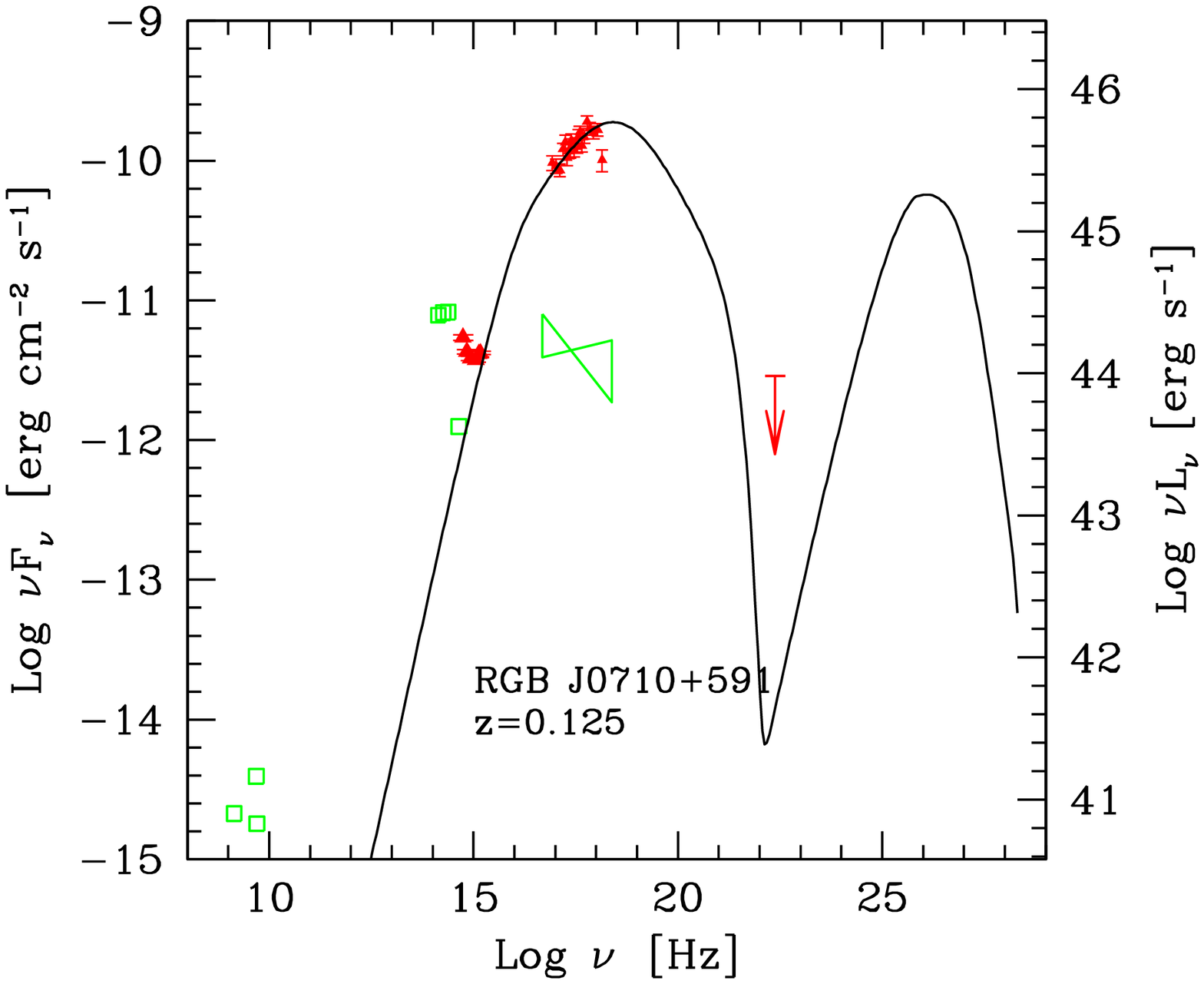,width=7cm,height=7cm}
\vskip -1.5 cm
\hskip 7 cm
\psfig{figure=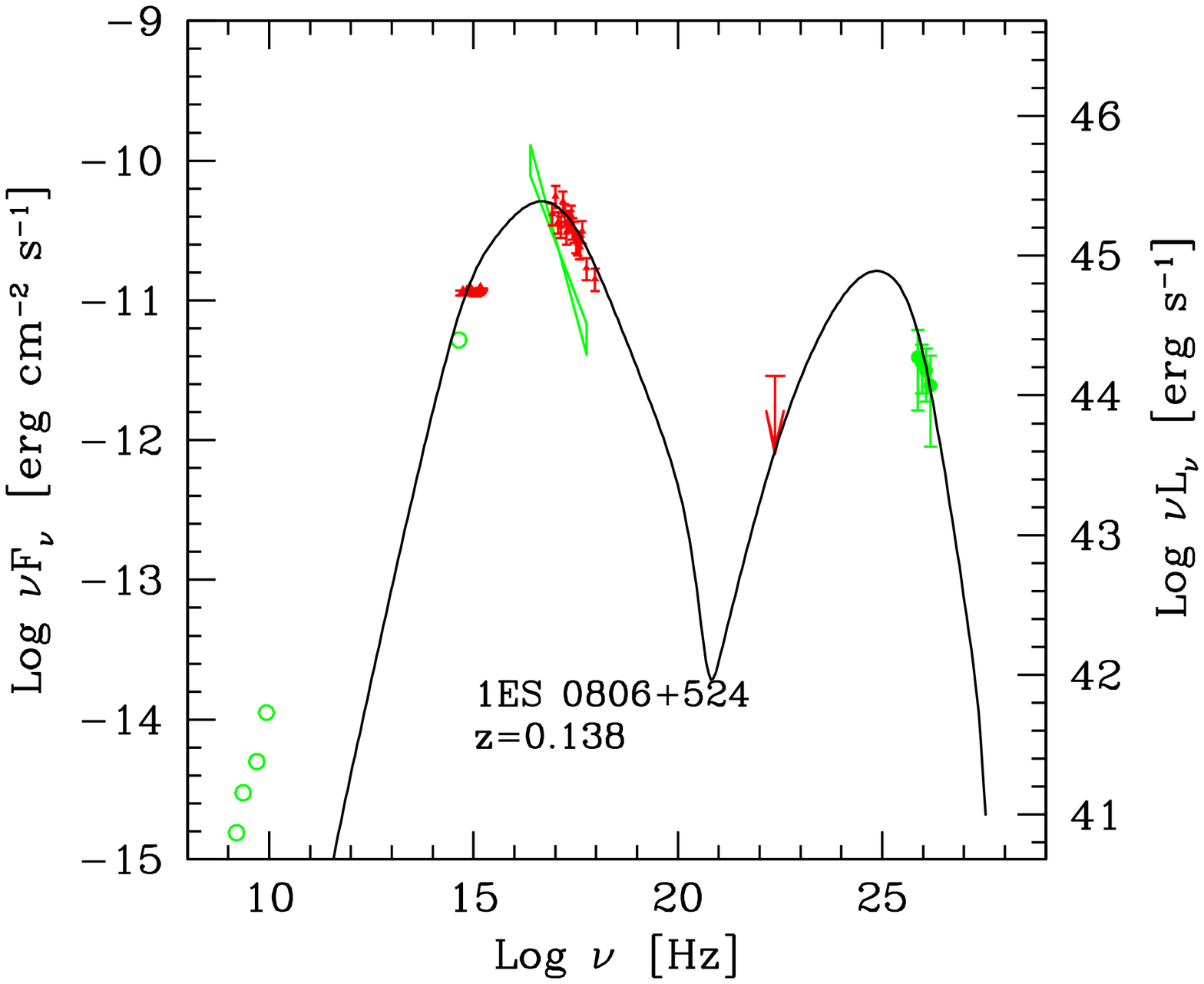,width=7cm,height=7cm}
\vskip -1.5 cm
\hskip 7 cm
\psfig{figure=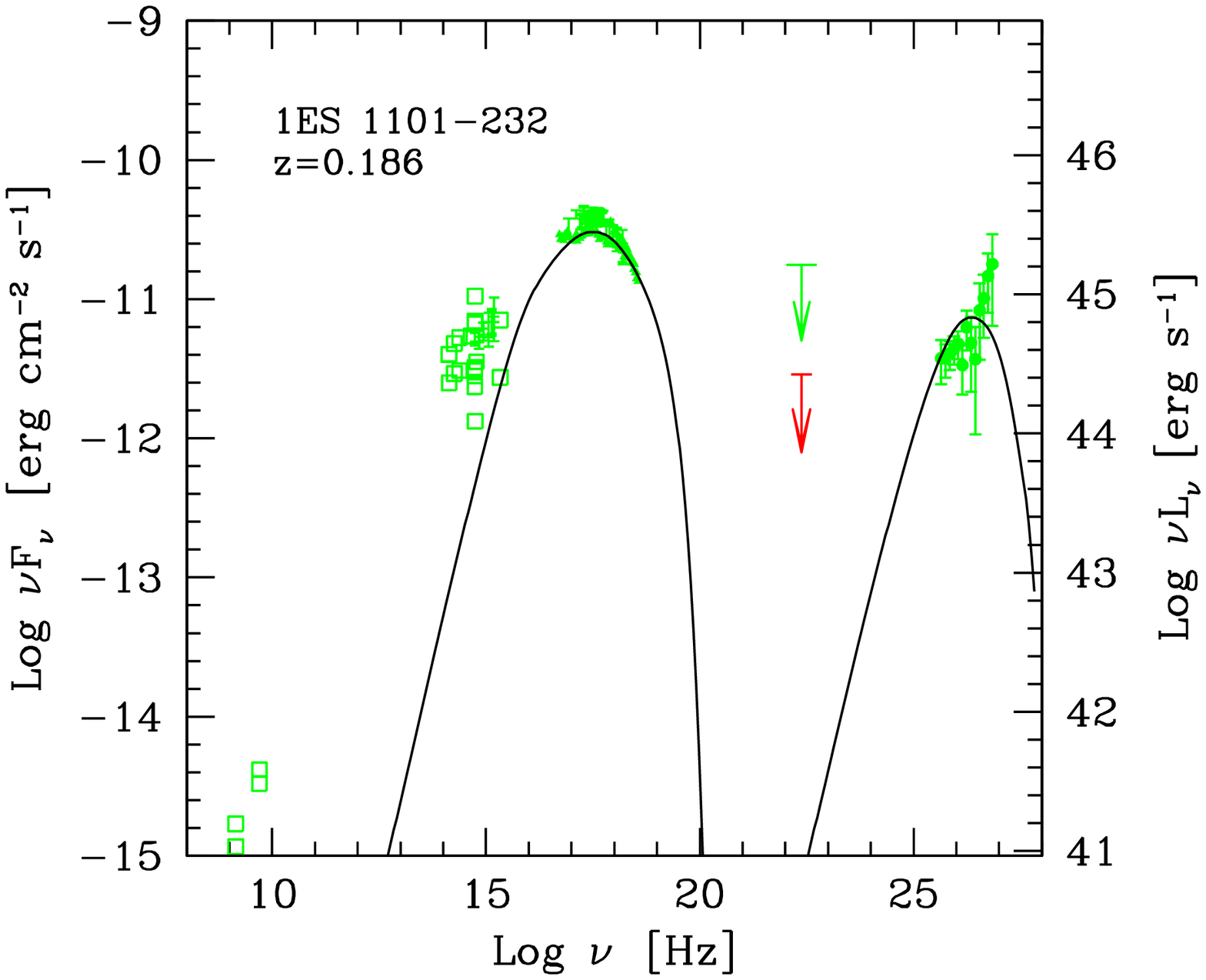,width=7cm,height=7cm}
\vskip -1.5 cm
\hskip 7 cm
\psfig{figure=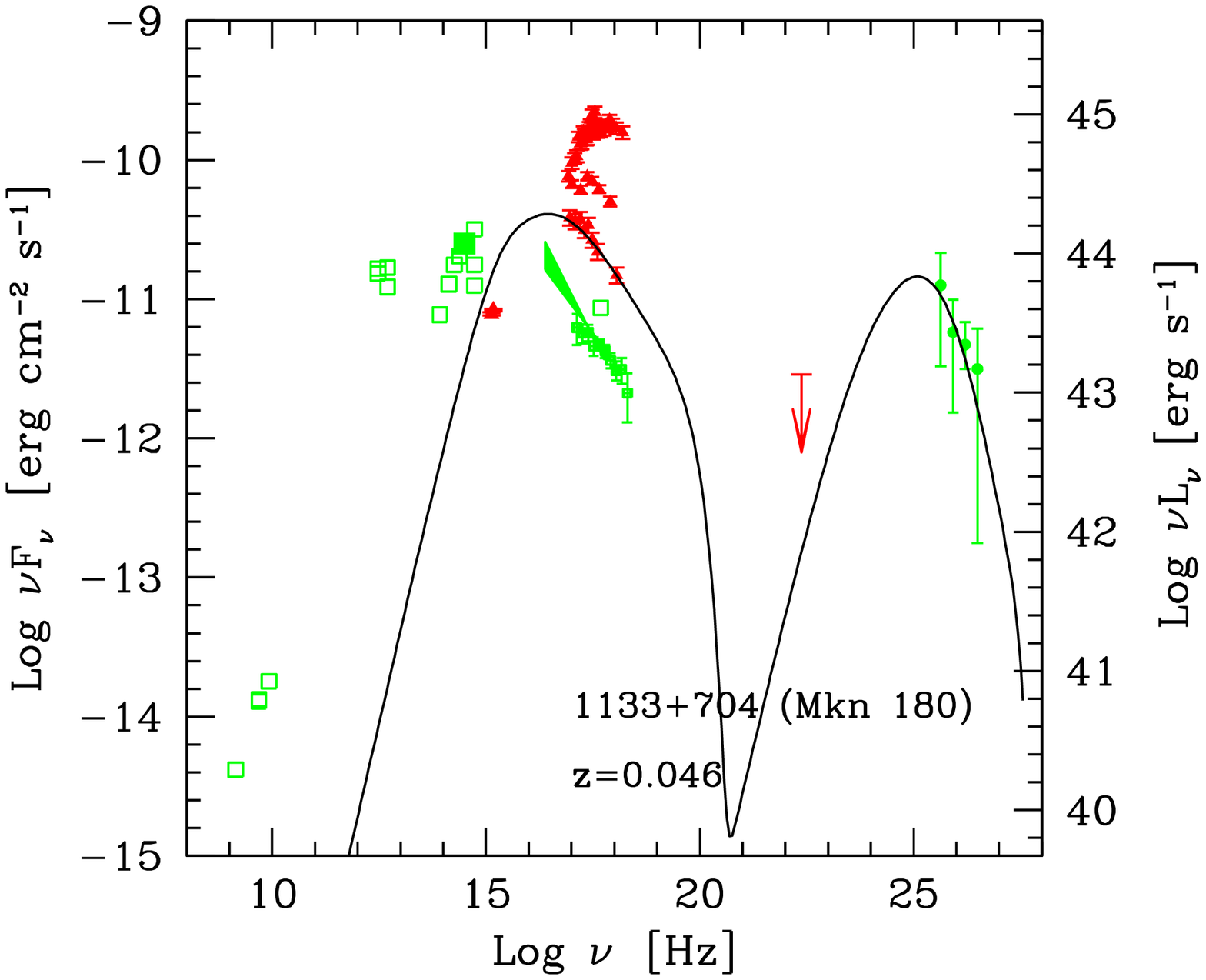,width=7cm,height=7cm}
\vskip -0.5cm
\caption{Spectral Energy Distribution of the BL Lacs detected by TeV telescopes but not by LAT. TeV spectra have been deabsorbed with the "lowSFR" model of Kneiske et al. (2004). References for the TeV data: 0152+017: Aharonian et al. (2008c); 0229+200: Aharonian et al. (2007b); 0347-121: Aharonian et al. (2007c); 0548-322 and 0710+591: no TeV spectrum published yet; 0806+524: Acciari et al. (2009b); 1101-232: Aharonian et al. (2006); 1133+704: Albert et al. (2006a). The solid line is the result of the one-zone leptonic model.
}
\label{sedno}
\end{figure*}

\setcounter{figure}{6}
\begin{figure*}
\vskip -1 cm
\psfig{figure=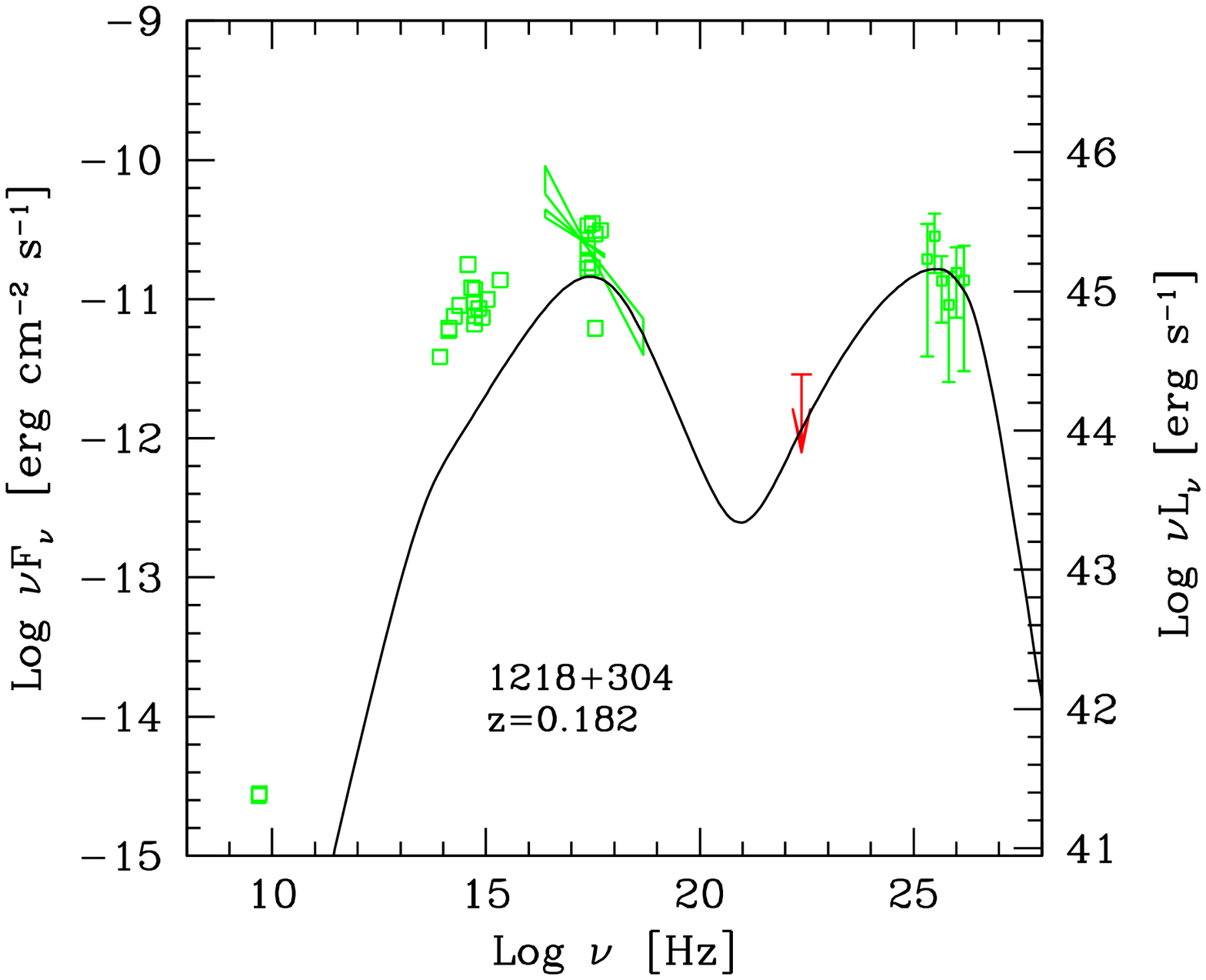,width=7cm,height=7cm}
\vskip -1.5 cm
\psfig{figure=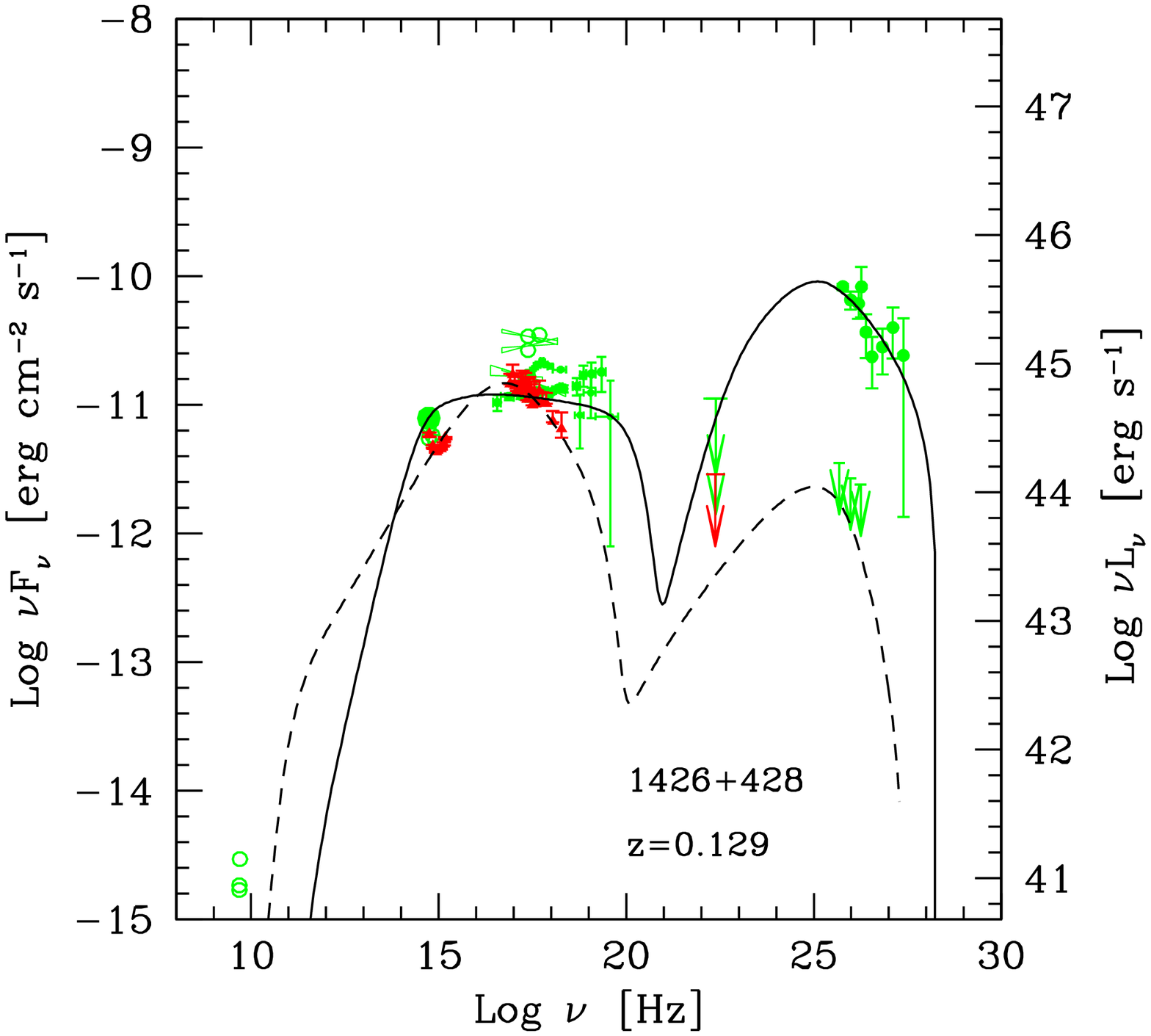,width=7cm,height=7cm}
\vskip -12.55 cm
\hskip 7 cm
\psfig{figure=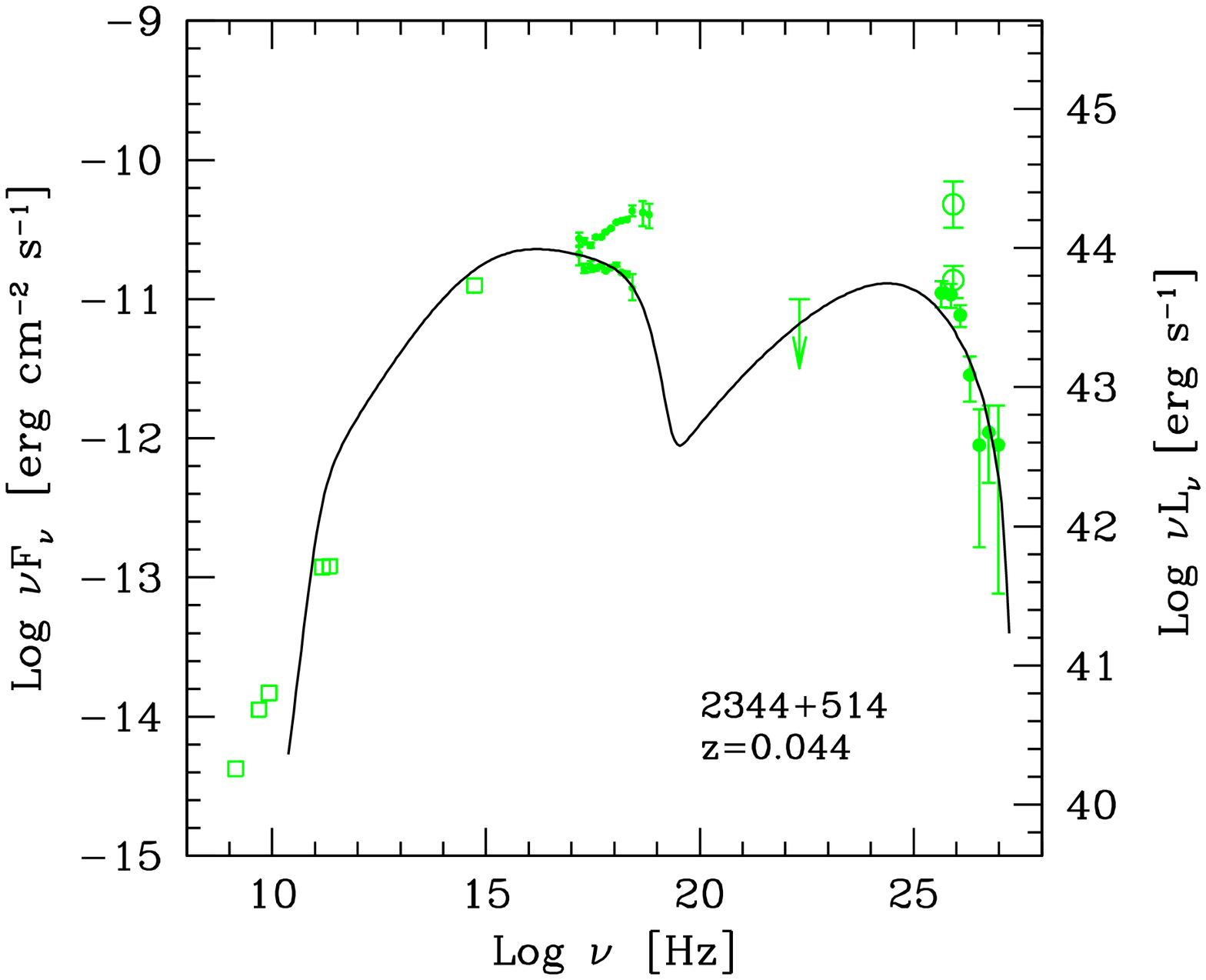,width=7cm,height=7cm}
\vskip -1.5 cm
\hskip 7cm
\psfig{figure=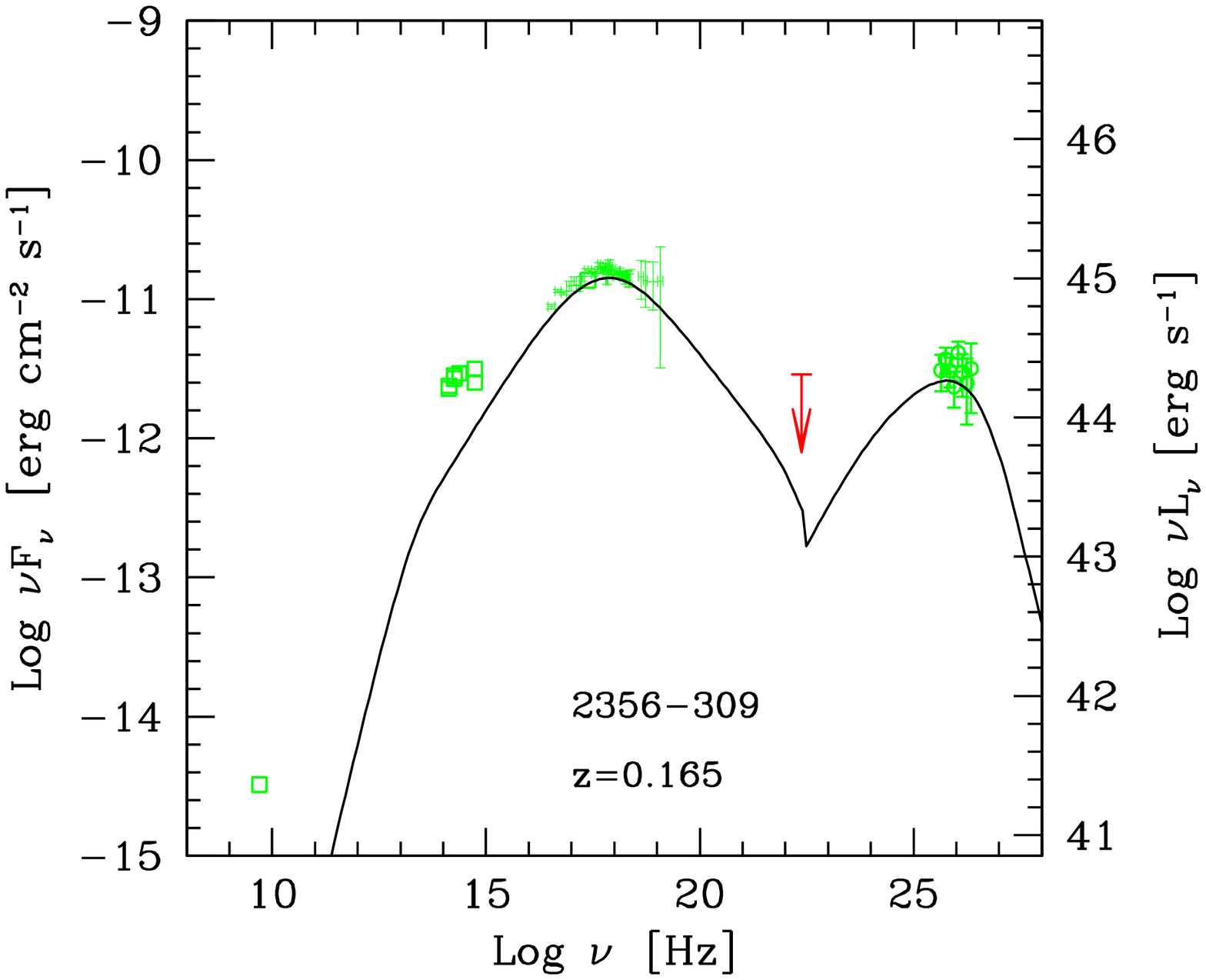,width=7cm,height=7cm}
\vskip -0.5cm
\caption{--continue-- 1218+304: Albert et al. (2006b); 1426+428: Aharonian et al. (2003), Leonardo et al. (2009); 2344+514: Albert et al. (2007f); 2356-309: Aharonian et al. (2006).
}
\end{figure*}


\begin{figure*}
\vskip -1 cm
\psfig{figure=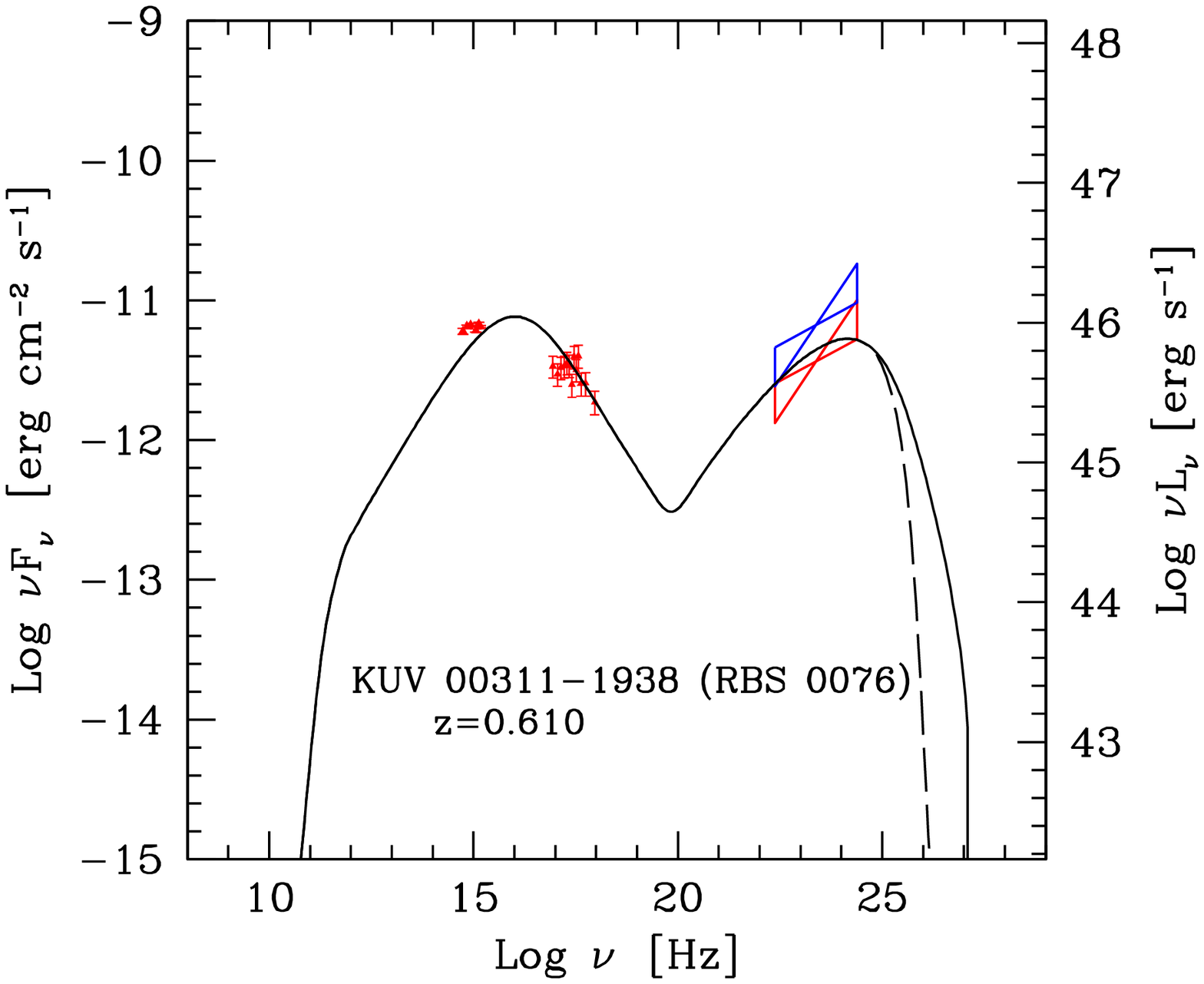,width=7cm,height=7cm}
\vskip -1.5 cm
\psfig{figure=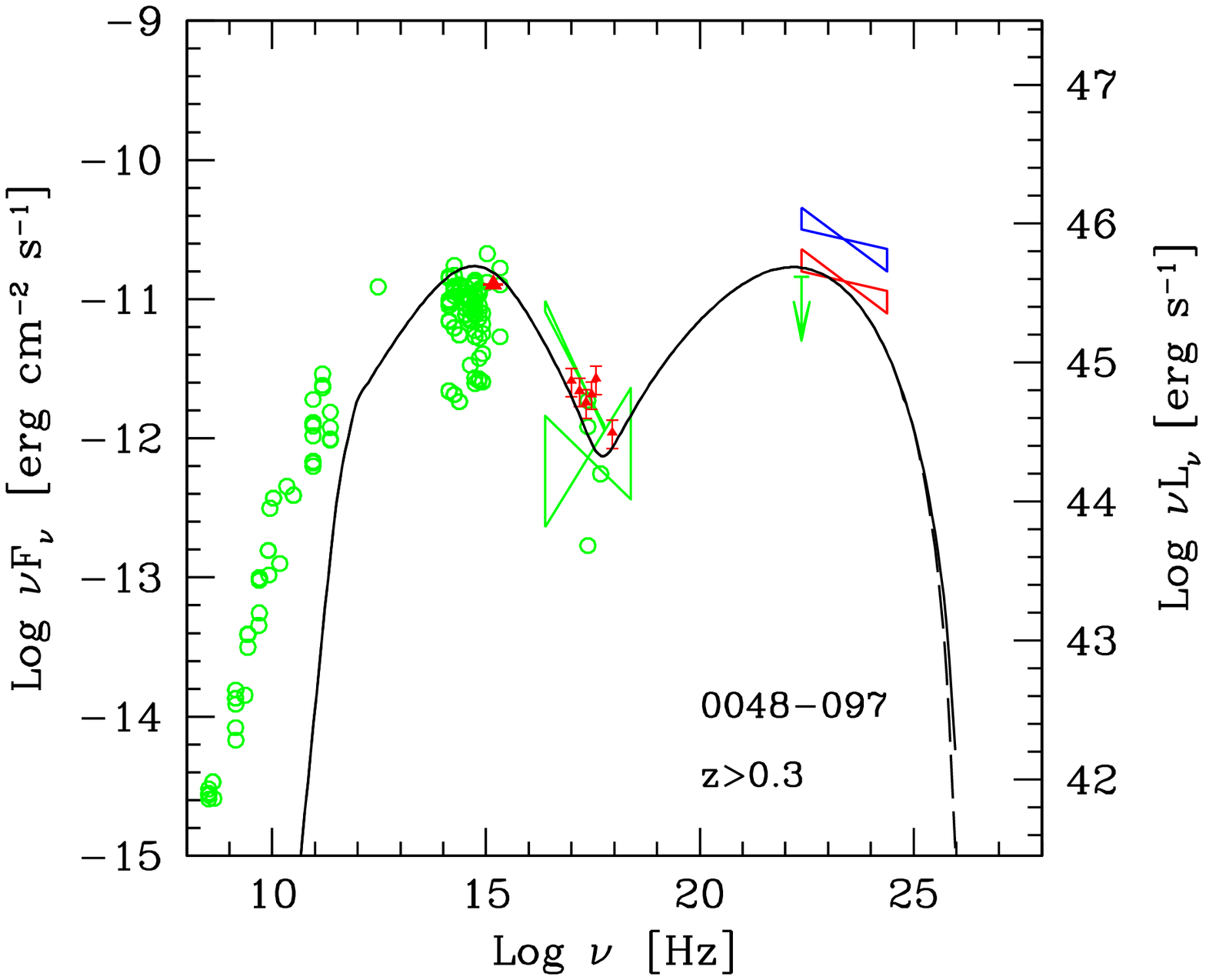,width=7cm,height=7cm}
\vskip -1.5 cm
\psfig{figure=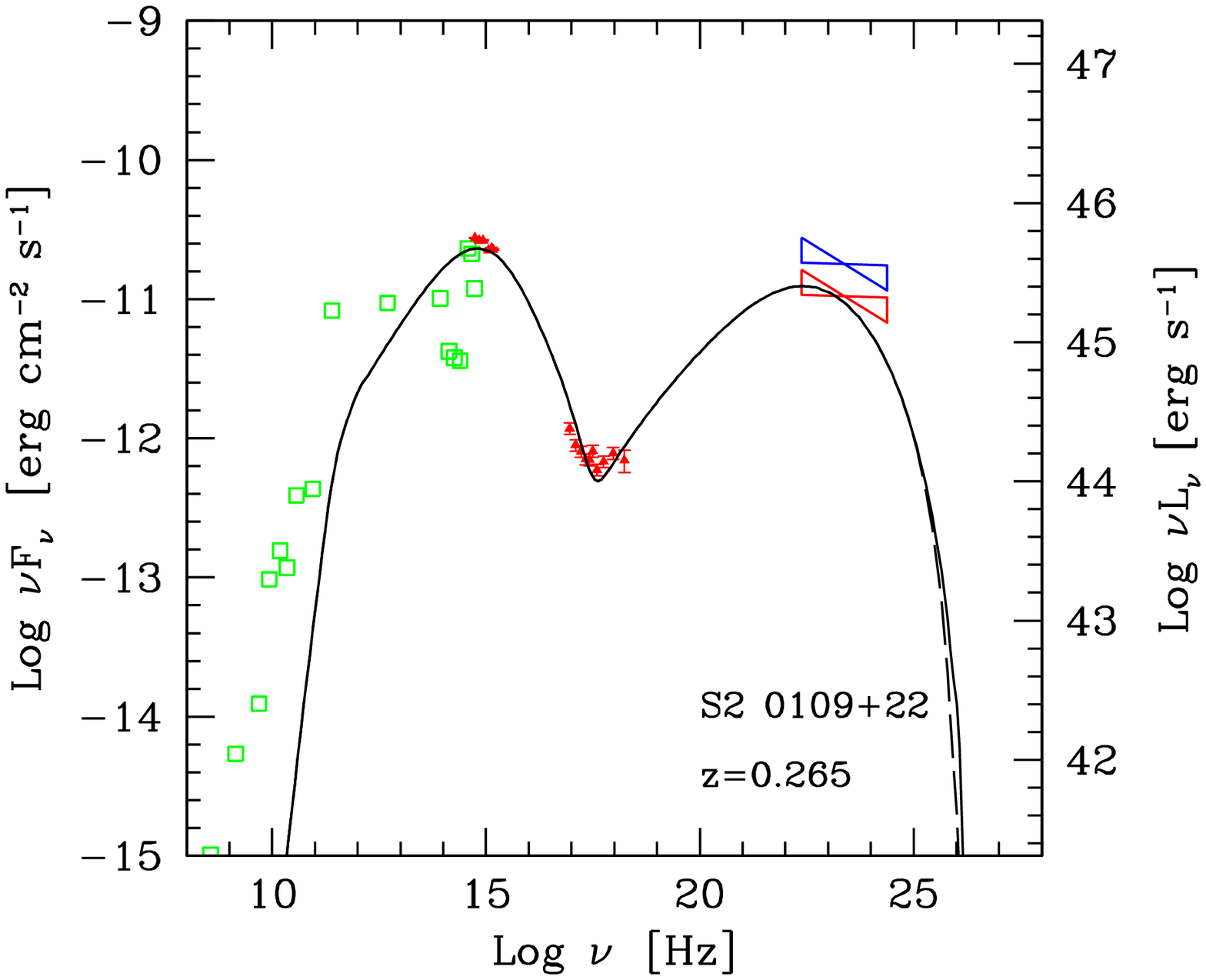,width=7cm,height=7cm}
\vskip -1.5 cm
\psfig{figure=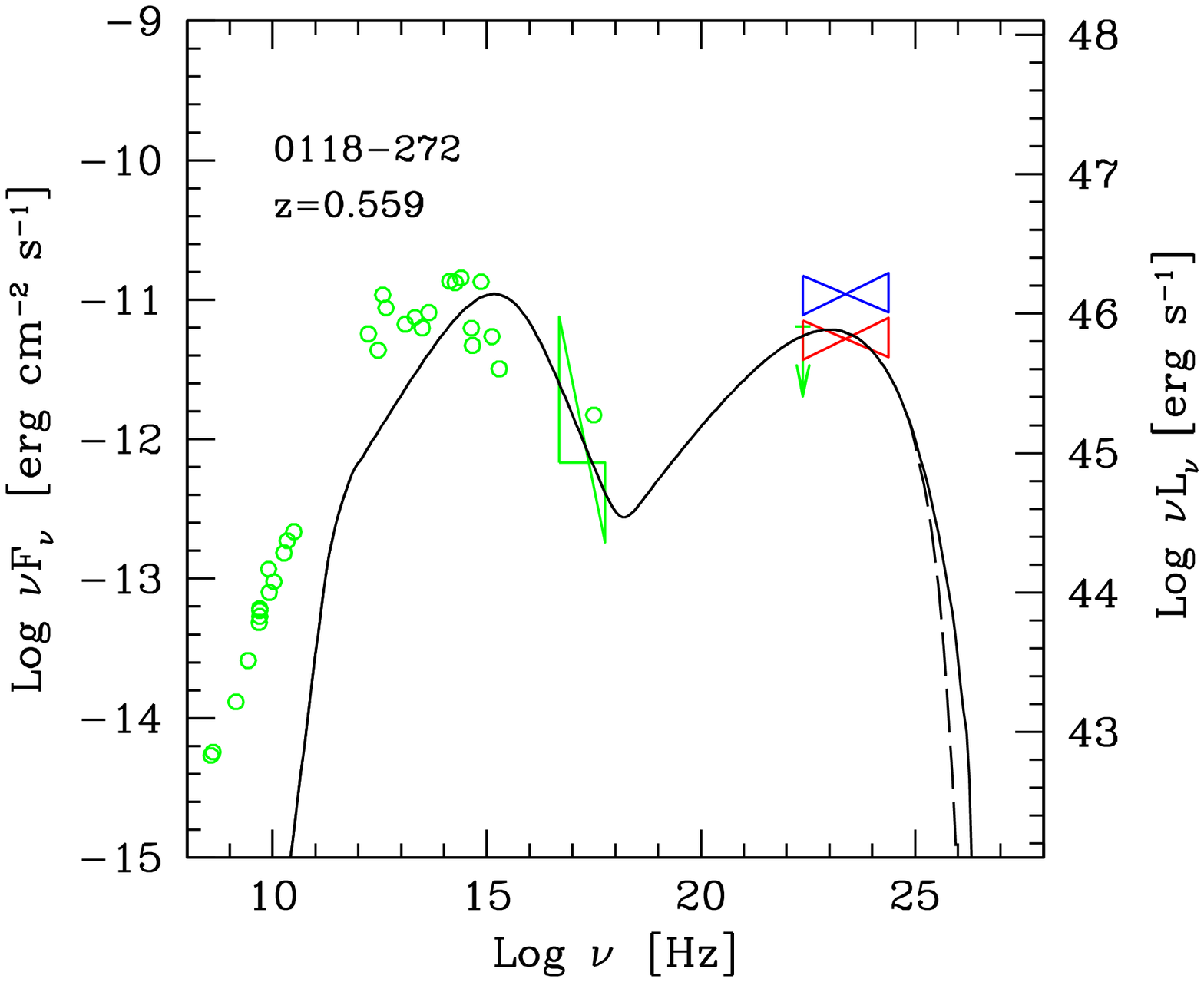,width=7cm,height=7cm}
\vskip -23.6 cm
\hskip 7 cm
\psfig{figure=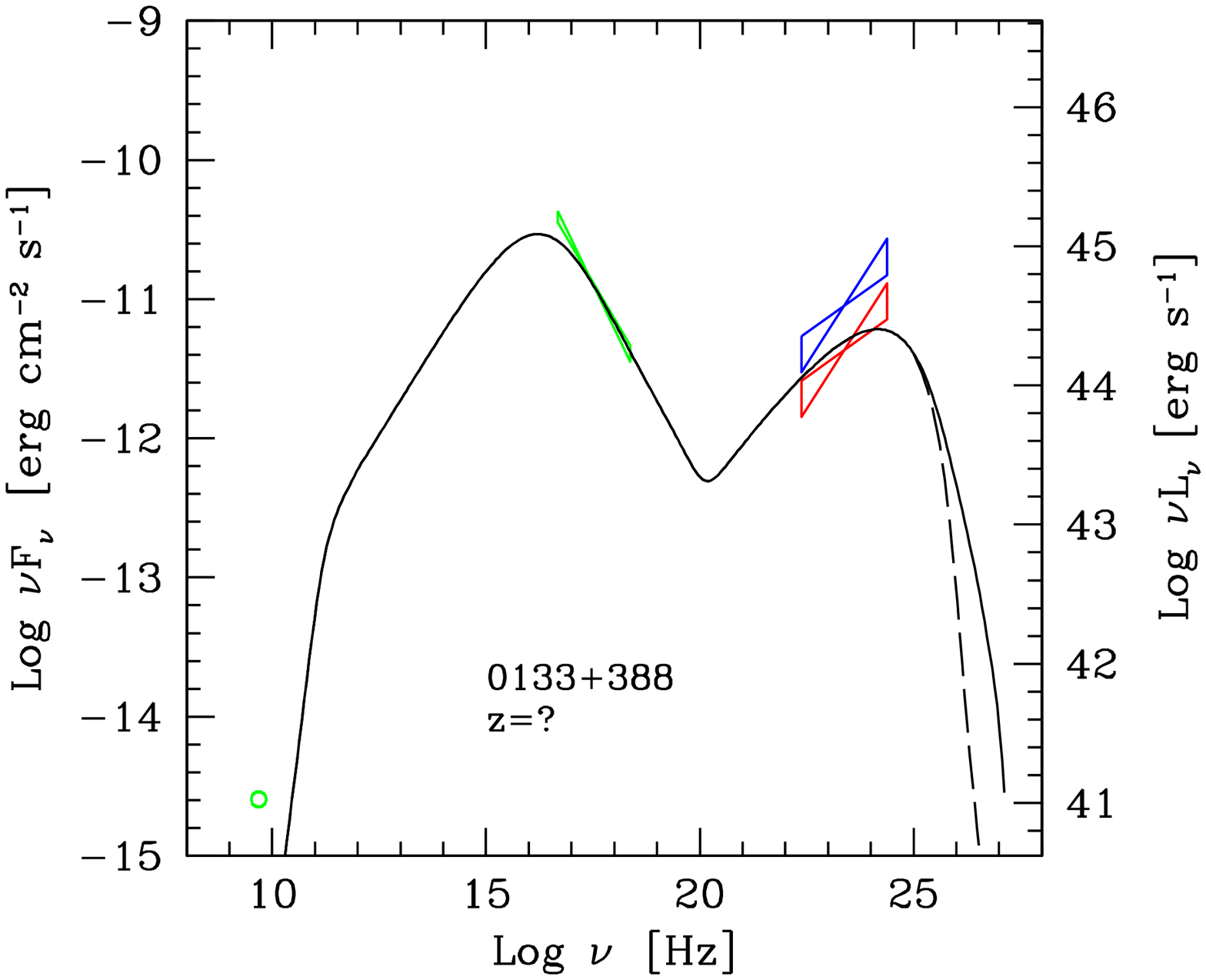,width=7cm,height=7cm}
\vskip -1.5 cm
\hskip 7 cm
\psfig{figure=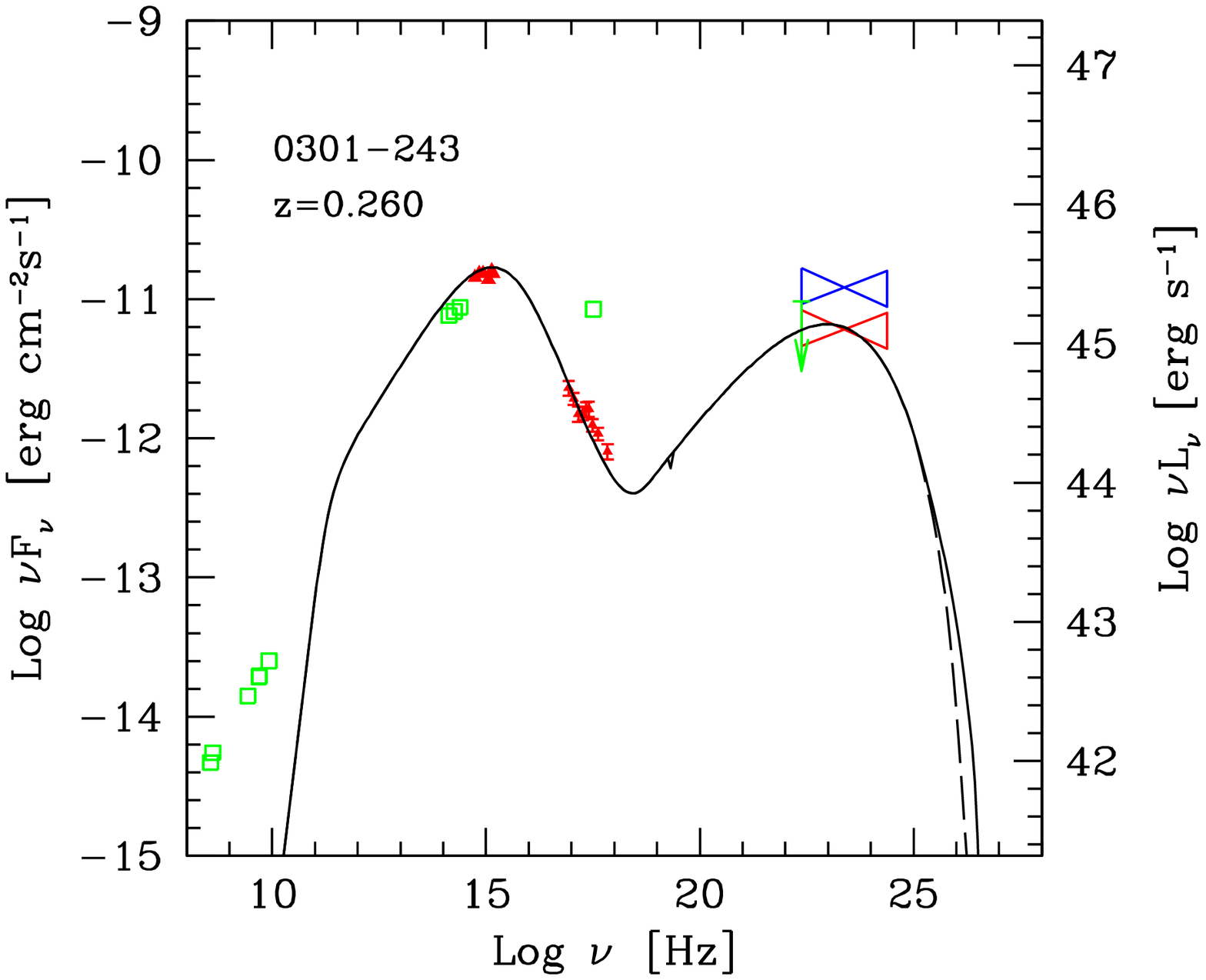,width=7cm,height=7cm}
\vskip -1.5 cm
\hskip 7 cm
\psfig{figure=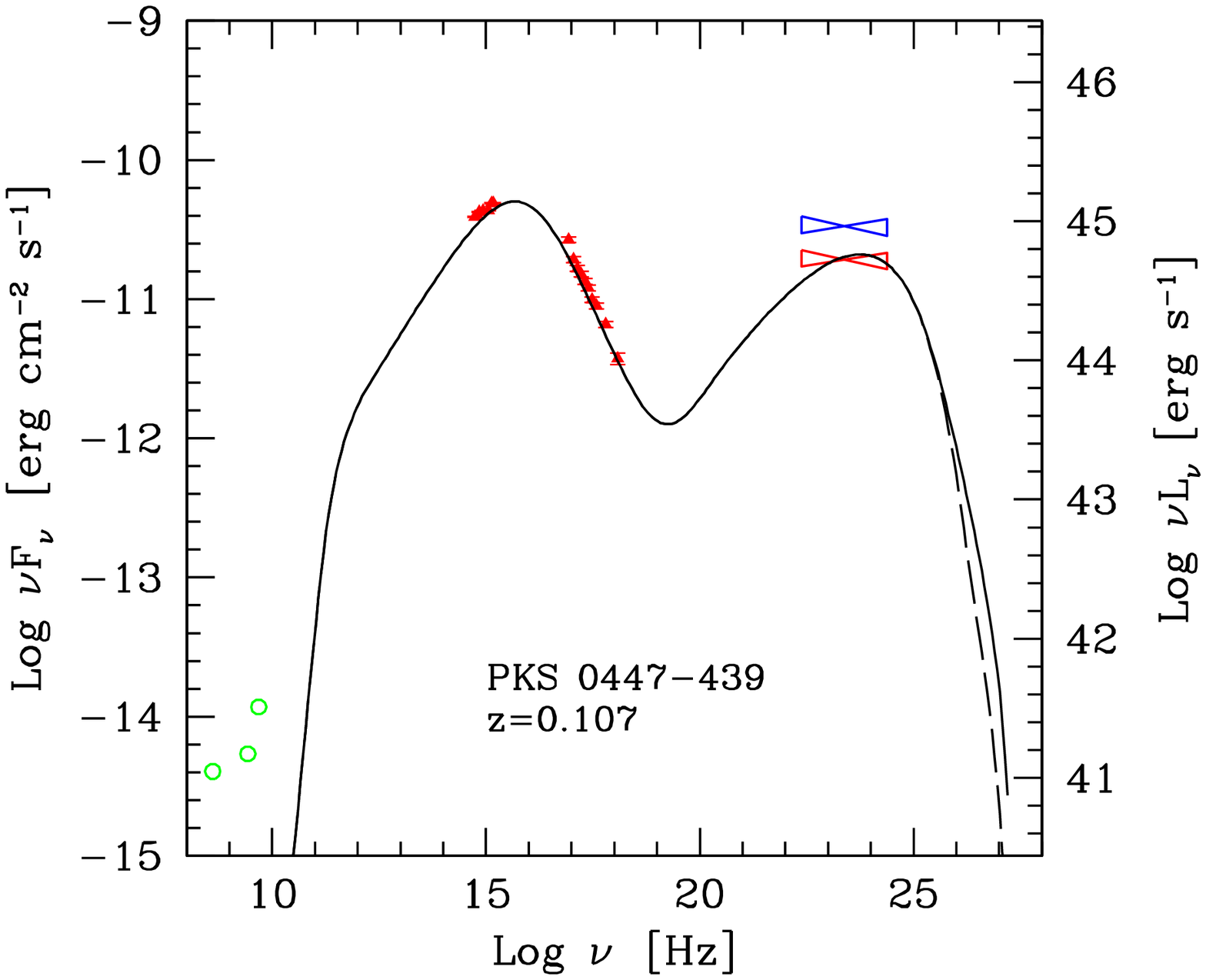,width=7cm,height=7cm}
\vskip -1.5 cm
\hskip 7 cm
\psfig{figure=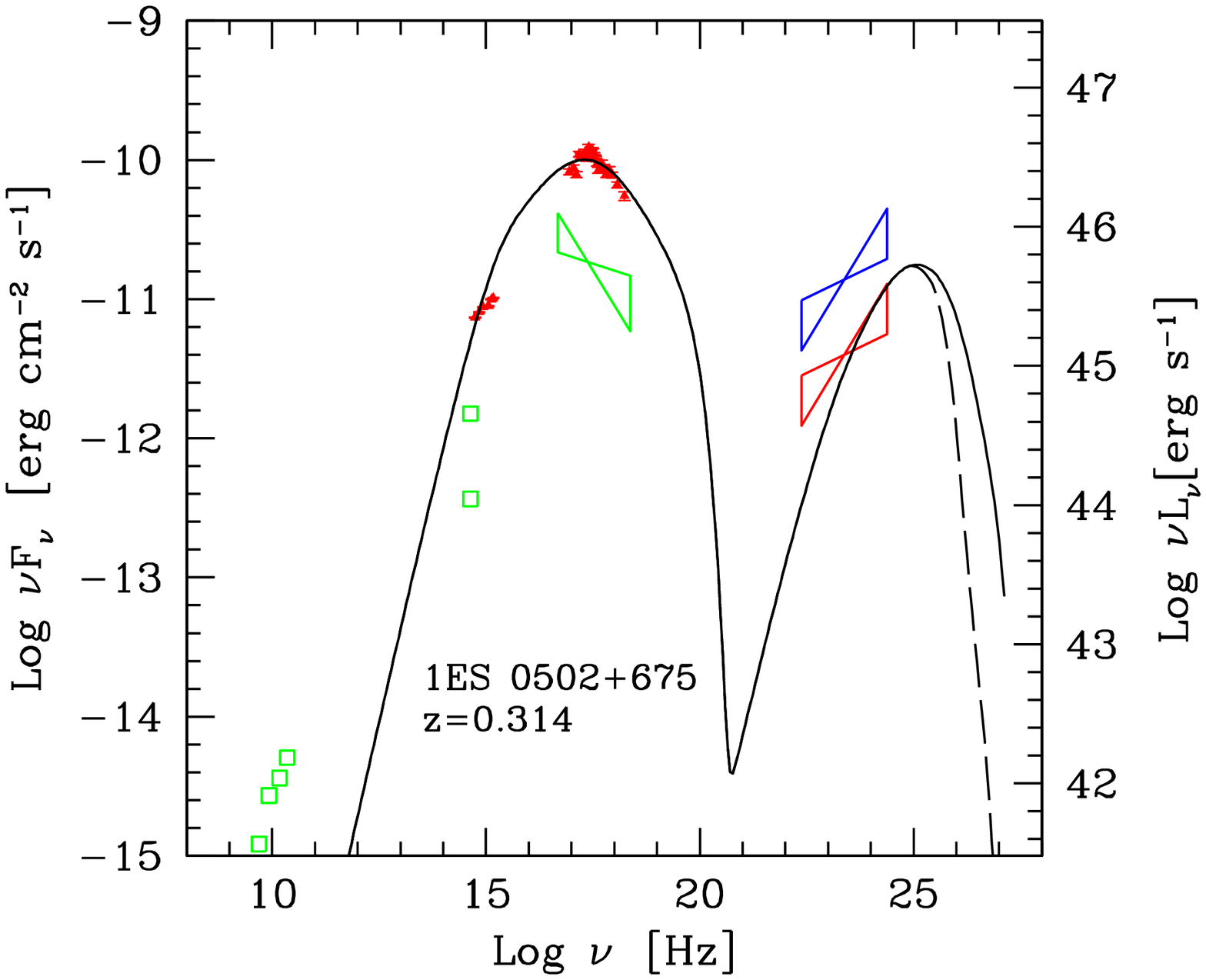,width=7cm,height=7cm}
\vskip -0.5cm
\caption{Spectral Energy Distribution of the BL Lacs detected by LAT but not by TeV telescopes. The solid line is the result of the one-zone leptonic model, the dashed line is the same model taking into account the absorption of $\gamma$-rays by the extragalactic background light (using the Kneiske et al. 2004 lowSFR model).
}
\label{sedcand}
\end{figure*}

\setcounter{figure}{7}
\begin{figure*}
\vskip -1 cm
\psfig{figure=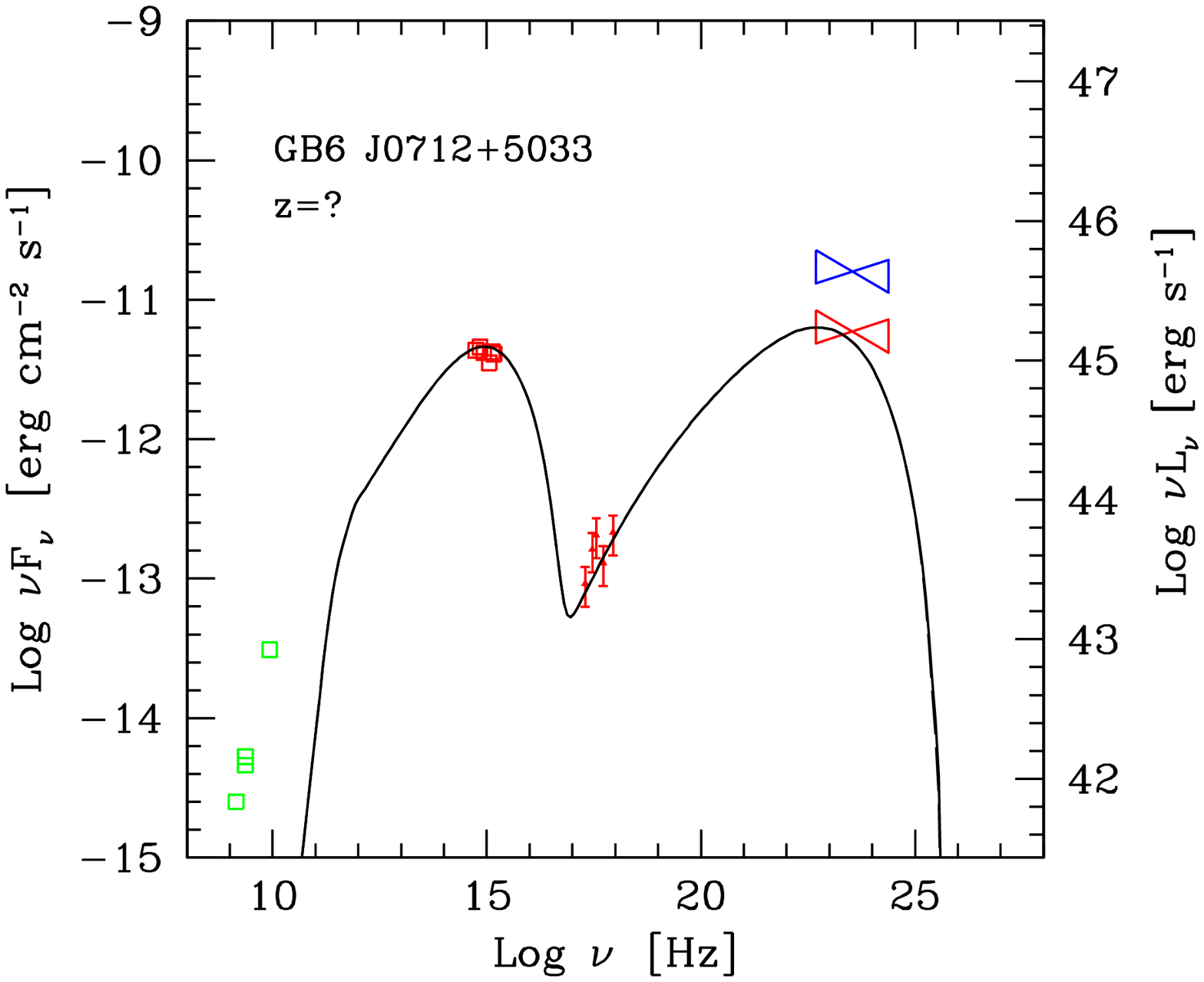,width=7cm,height=7cm}
\vskip -1.5 cm
\psfig{figure=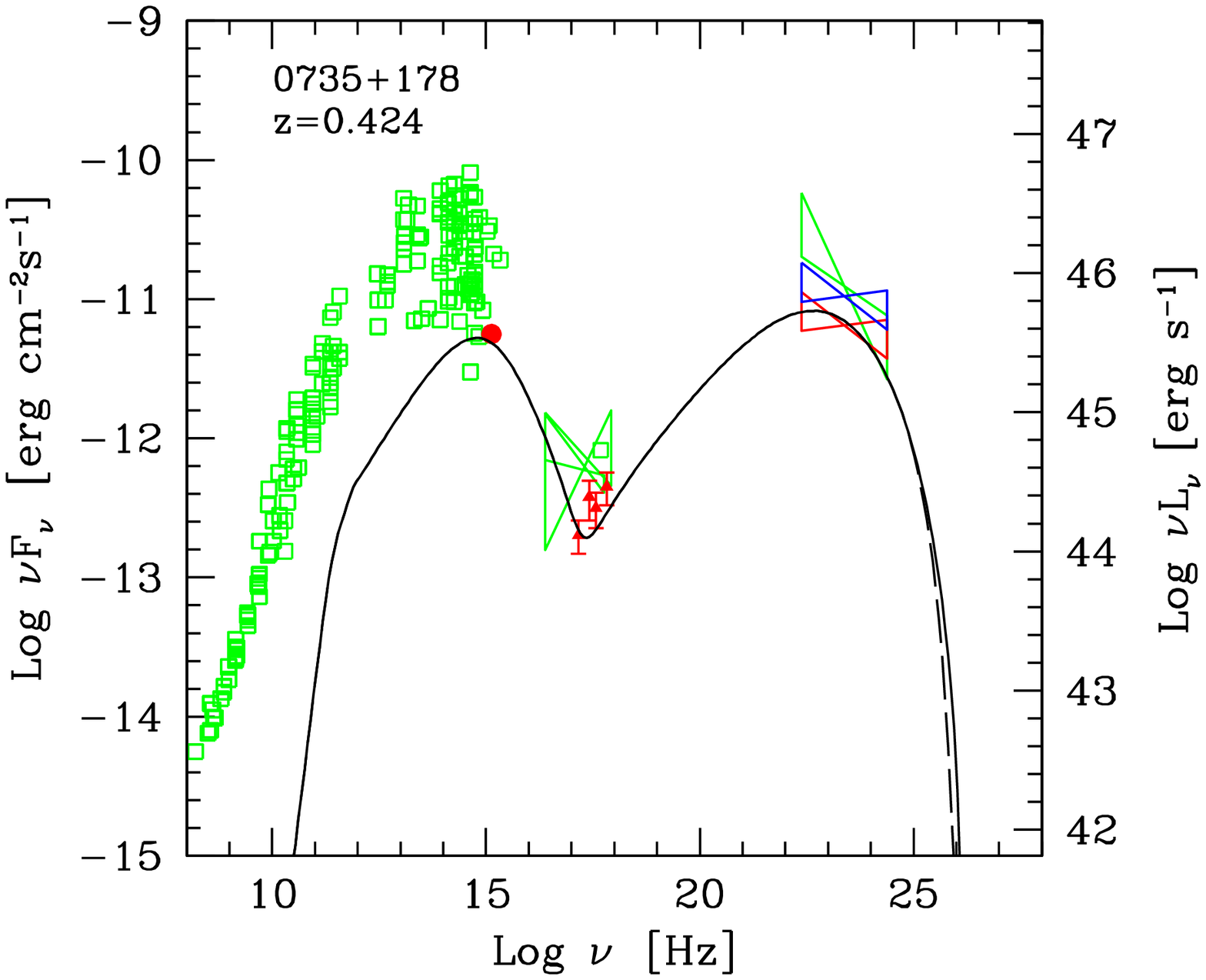,width=7cm,height=7cm}
\vskip -1.5 cm
\psfig{figure=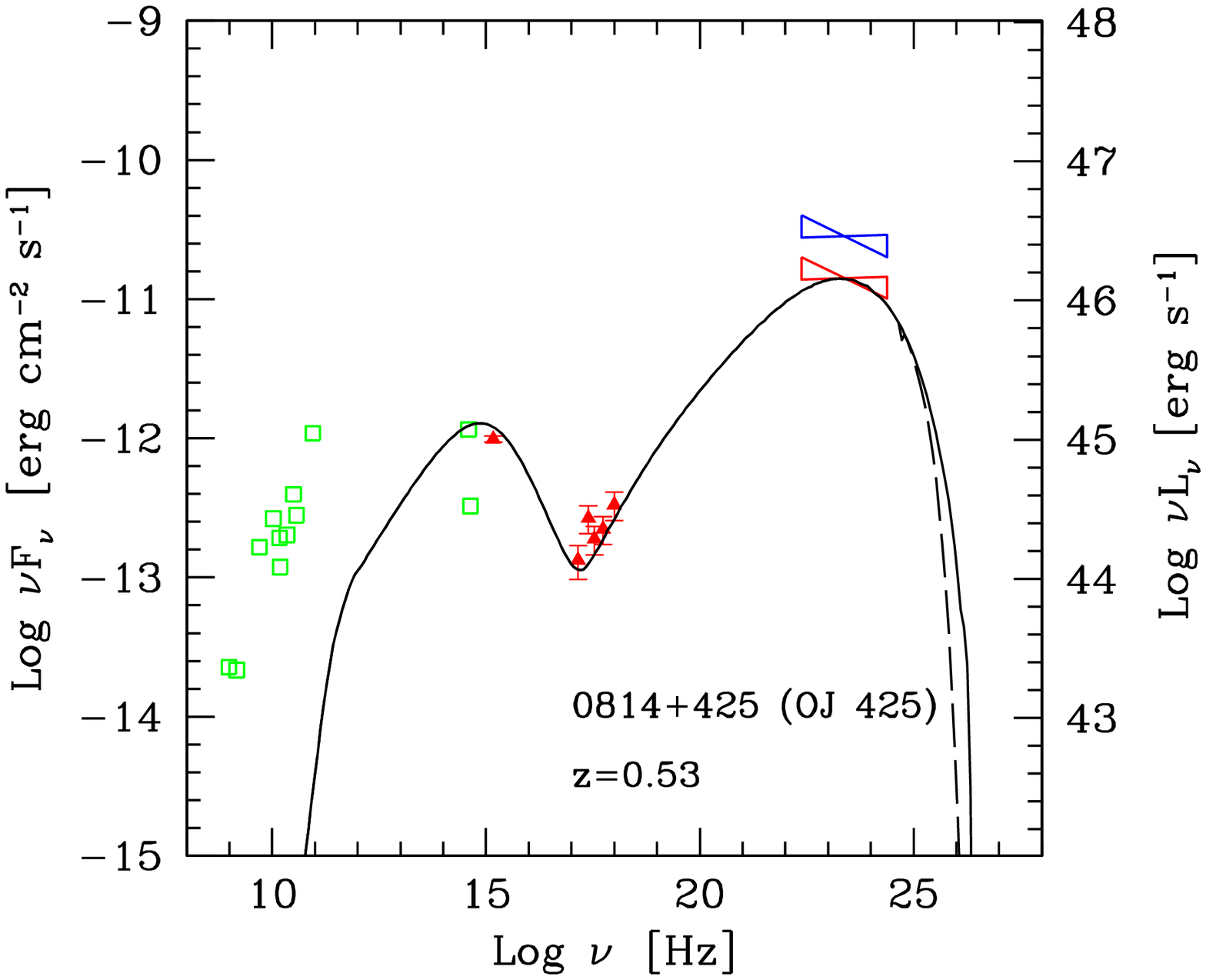,width=7cm,height=7cm}
\vskip -1.5 cm
\psfig{figure=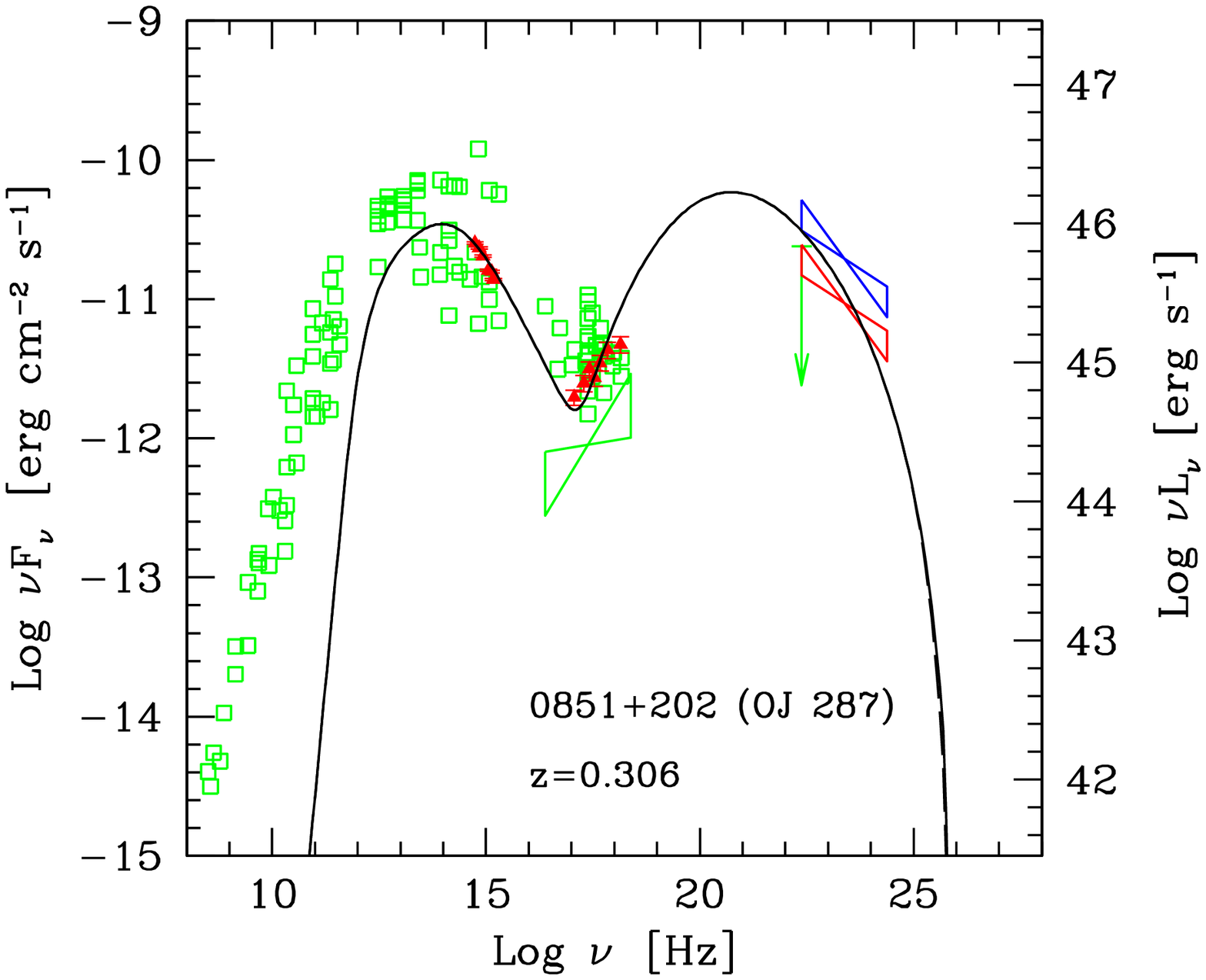,width=7cm,height=7cm}
\vskip -23.6 cm
\hskip 7 cm
\psfig{figure=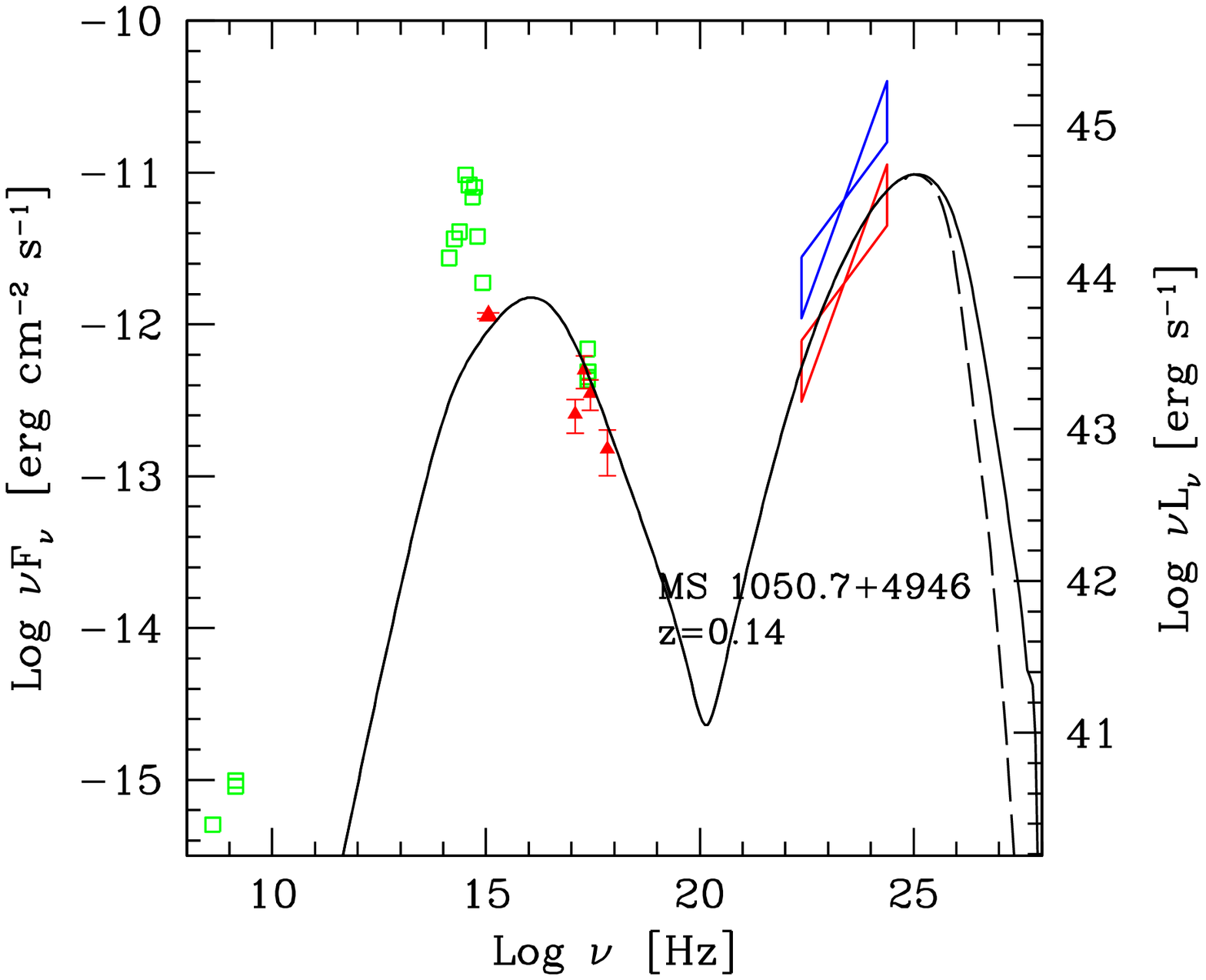,width=7cm,height=7cm}
\vskip -1.5 cm
\hskip 7 cm
\psfig{figure=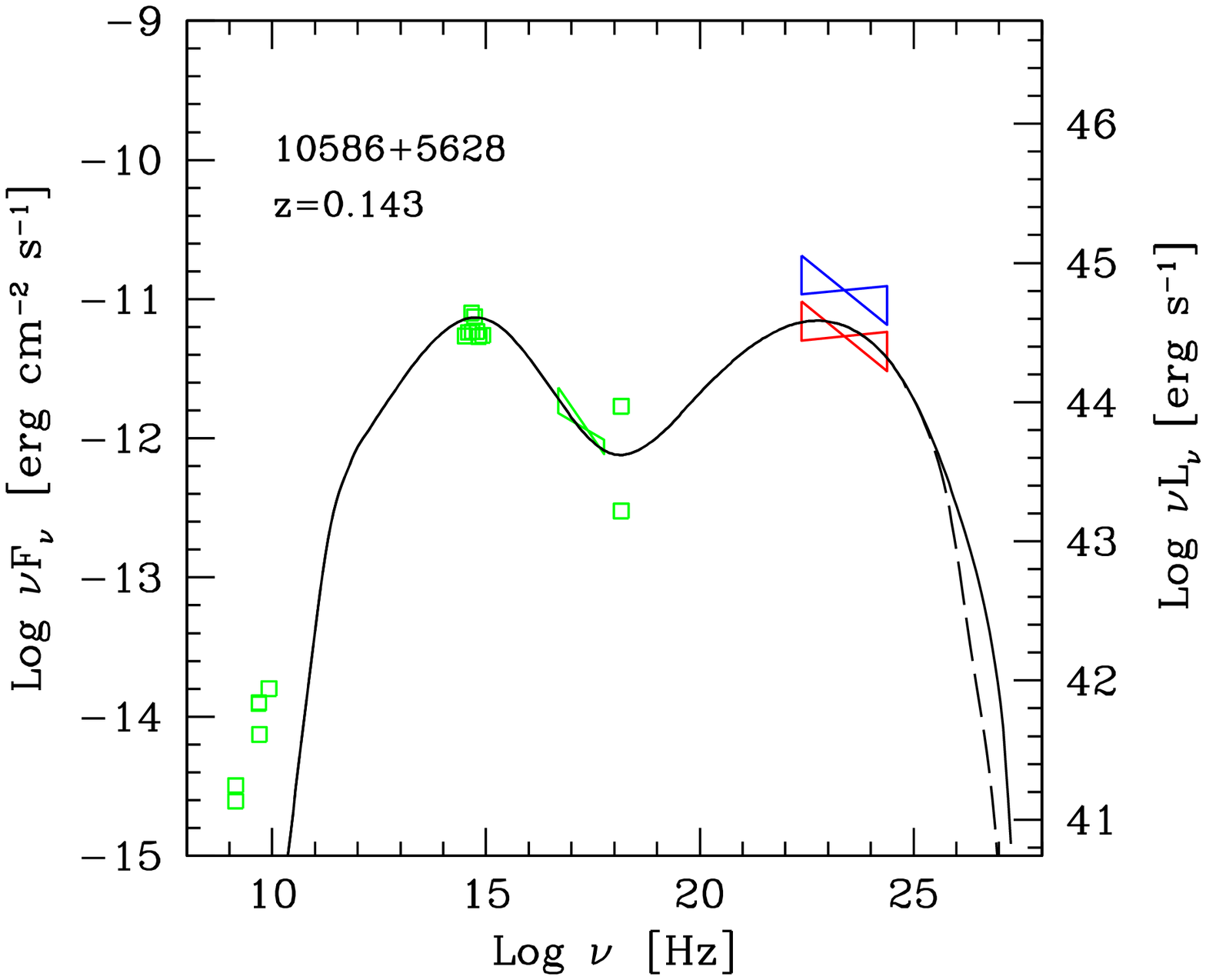,width=7cm,height=7cm}
\vskip -1.5 cm
\hskip 7 cm
\psfig{figure=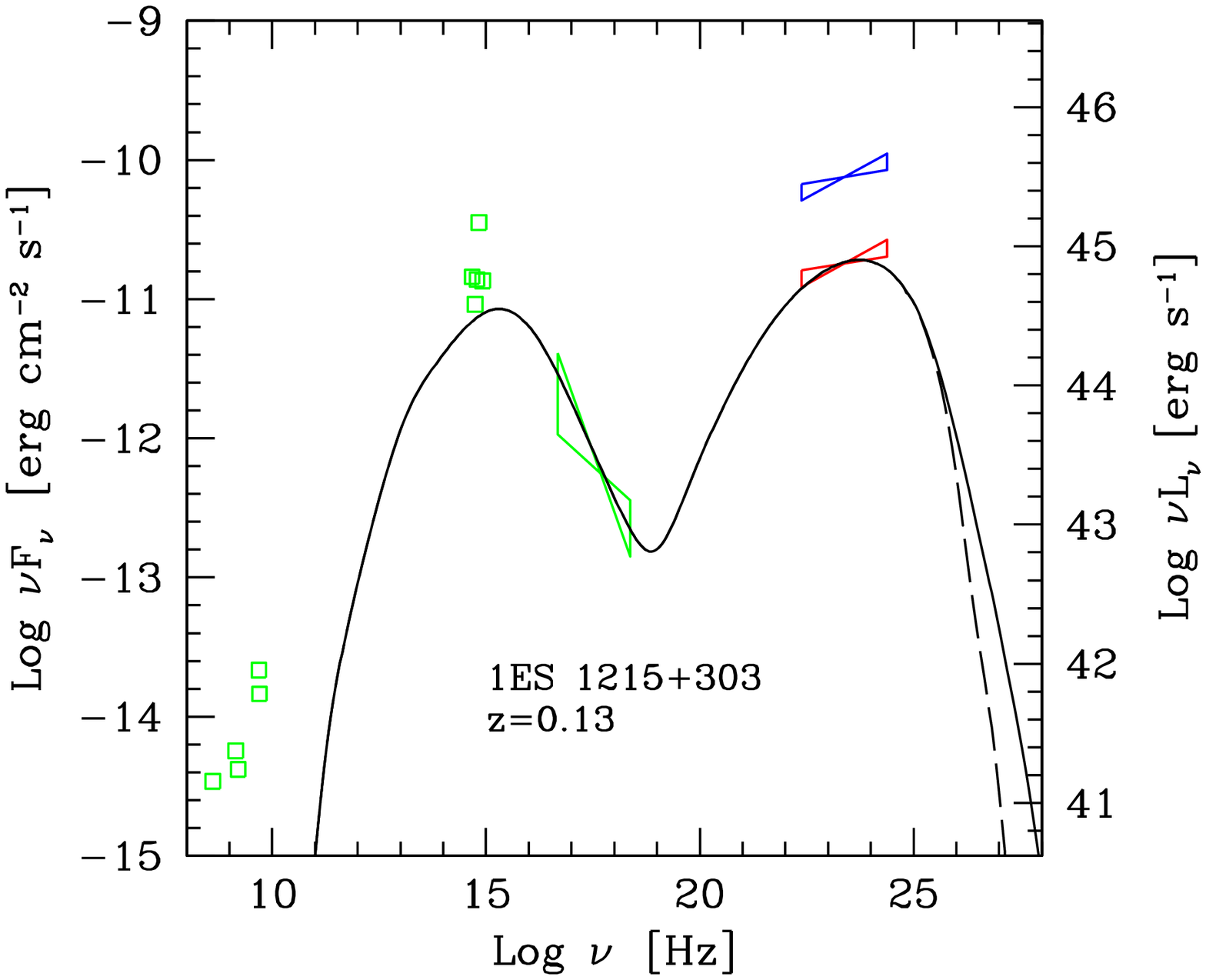,width=7cm,height=7cm}
\vskip -1.5 cm
\hskip 7 cm
\psfig{figure=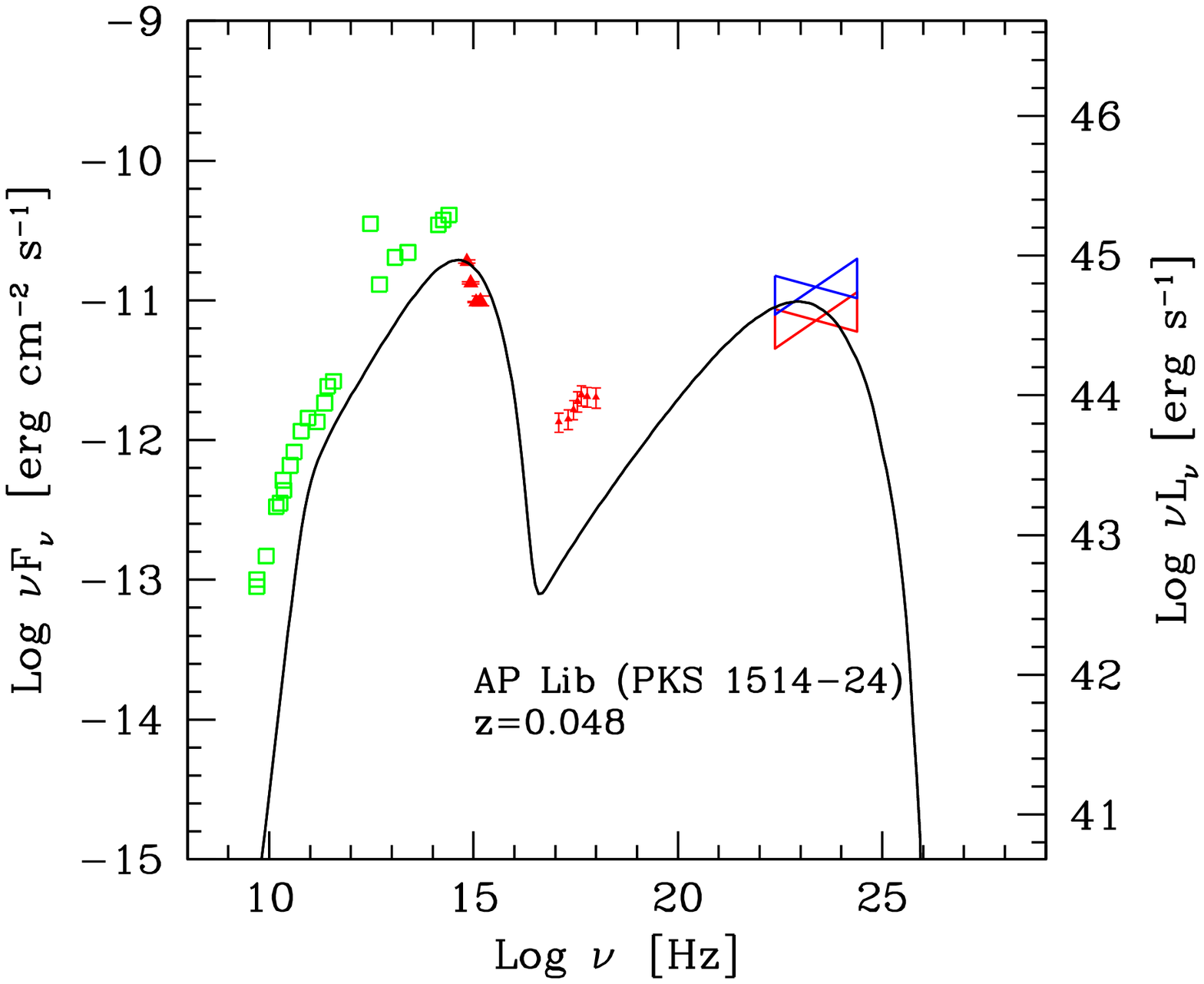,width=7cm,height=7cm}
\vskip -0.5cm
\caption{--continue--
}
\end{figure*}

\setcounter{figure}{7}
\begin{figure*}
\vskip -1 cm
\psfig{figure=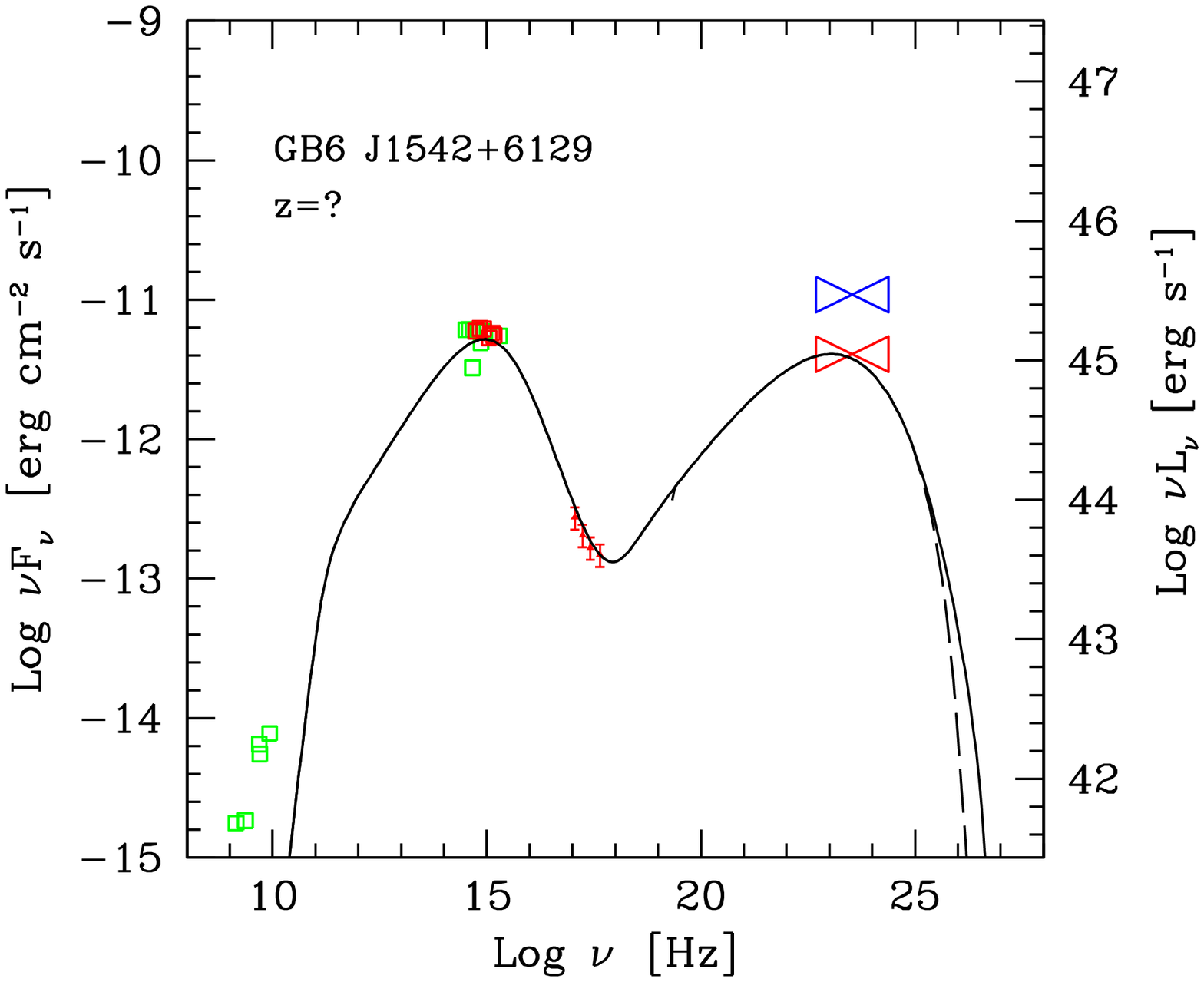,width=7cm,height=7cm}
\vskip -1.5 cm
\psfig{figure=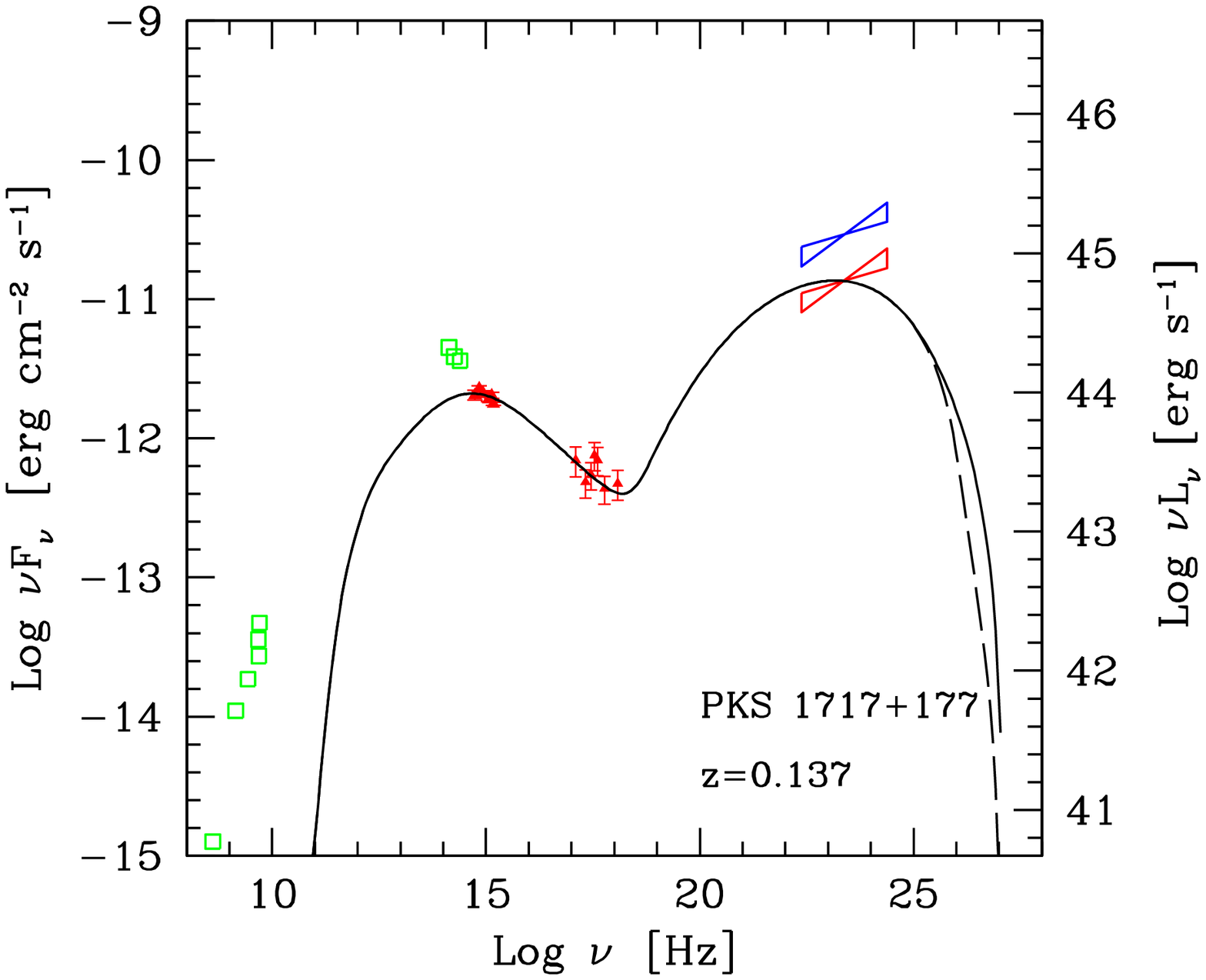,width=7cm,height=7cm}
\vskip -1.5 cm
\psfig{figure=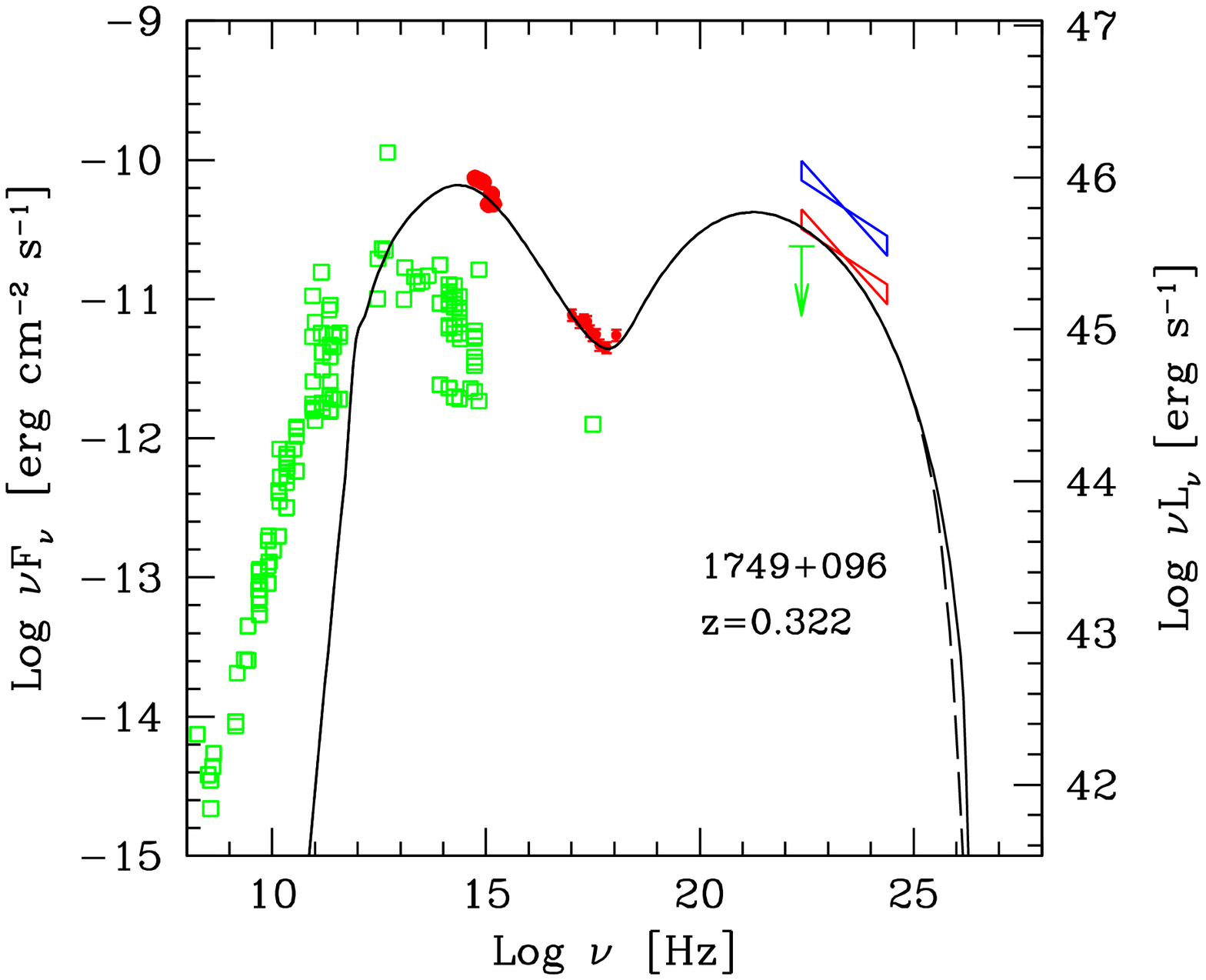,width=7cm,height=7cm}
\vskip -18.1cm
\hskip 7cm
\psfig{figure=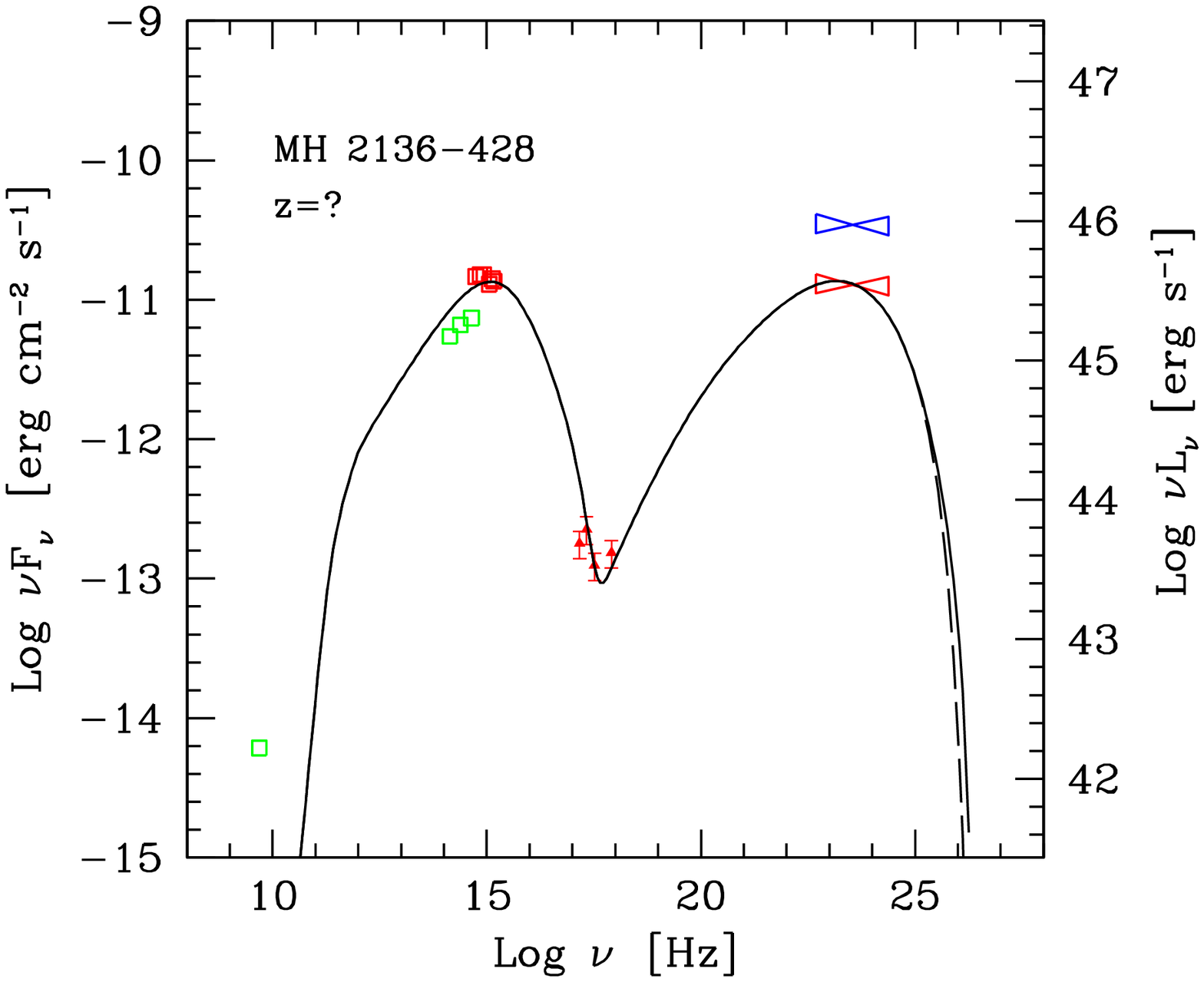,width=7cm,height=7cm}
\vskip -1.5cm
\hskip 7 cm
\psfig{figure=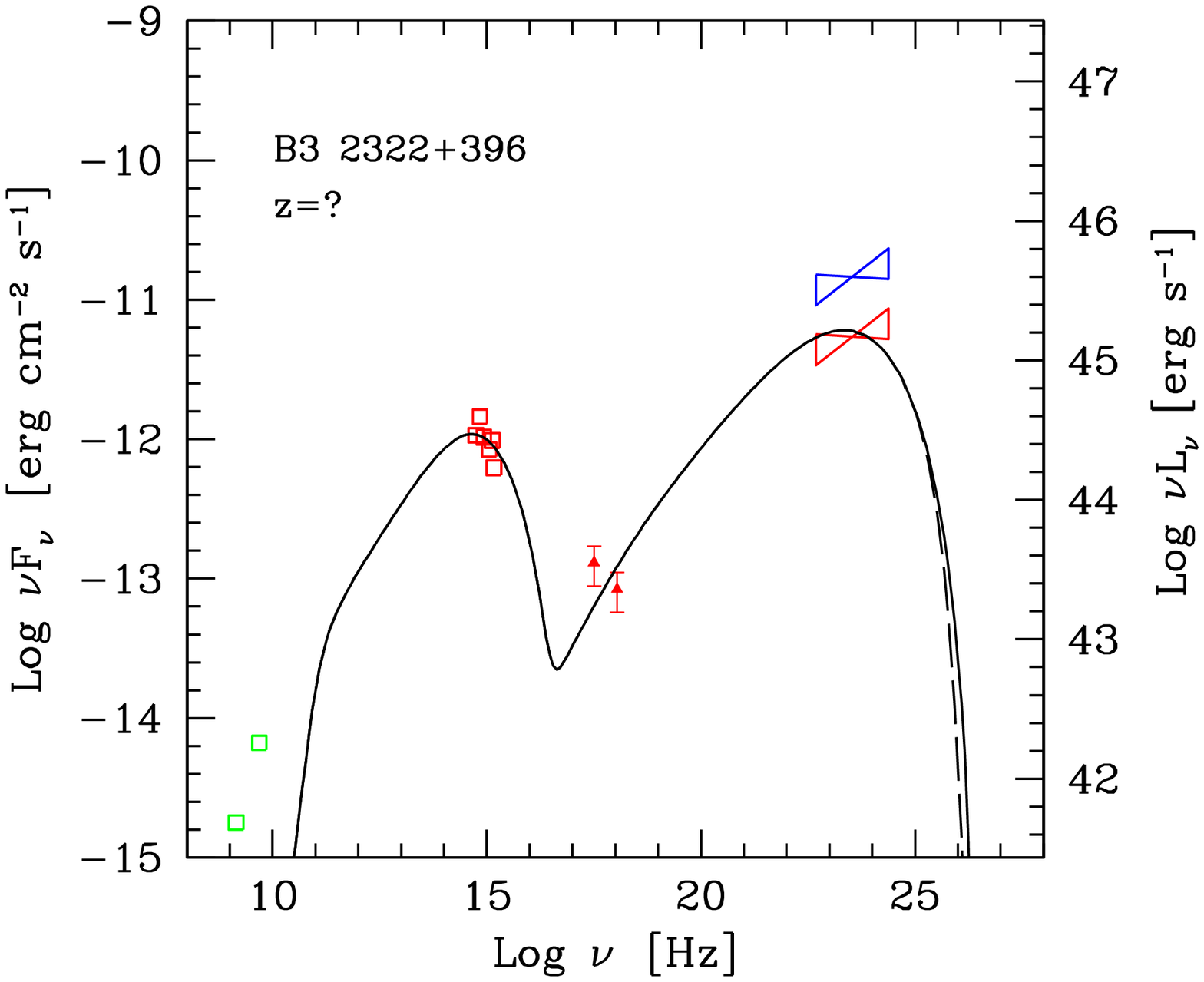,width=7cm,height=7cm}
\vskip 5 cm
\vskip -0.5cm
\caption{--continue--
}
\end{figure*}


\begin{thebibliography}{}

\bibitem[\protect\citeauthoryear{Abdo et al.}{2009}]{2009ApJS..183...46A}  Abdo A.~A., et al., 2009a, ApJ, 700, 597 (A09)

\bibitem[\protect\citeauthoryear{Abdo et al.}{2009}]{2009ApJ...699..976A} Abdo A.~A., et al., 2009b, ApJ, 699, 976 

\bibitem[\protect\citeauthoryear{Acciari et al.}{2008}]{2008ApJ...684L..73A} Acciari V.~A., et al., 2008, ApJ, 684, L73 

\bibitem[\protect\citeauthoryear{Acciari et al.}{2009}]{2009ApJ...693L.104A} Acciari V.~A., et al., 2009a, ApJ, 693, 
L104

\bibitem[\protect\citeauthoryear{Acciari et al.}{2009}]{2009ApJ...690L.126A} Acciari V., et al., 2009b, ApJ, 690, L126 

\bibitem[\protect\citeauthoryear{Aharonian et al.}{2003}]{2003A&A...403..523A} Aharonian F., et al., 2003a, A\&A, 403, 523 

\bibitem[\protect\citeauthoryear{Aharonian et al.}{2003}]{2003A&A...406L...9A} Aharonian F., et al., 2003b, A\&A, 406, L9 

\bibitem[\protect\citeauthoryear{Aharonian et al.}{2005}]{2005A&A...436L..17A} Aharonian F., et al., 2005a, A\&A, 436, L17 

\bibitem[\protect\citeauthoryear{Aharonian et al.}{2005}]{2005A&A...430..865A} Aharonian F., et al., 2005b, A\&A, 430, 865 


\bibitem[\protect\citeauthoryear{Aharonian et al.}{2006}]{2006Natur.440.1018A} Aharonian F., et al., 2006, Natur, 440, 
1018 

\bibitem[\protect\citeauthoryear{Aharonian et al.}{2007}]{2007ApJ...664L..71A} Aharonian F., et al., 2007a, ApJ, 664, L71 

\bibitem[\protect\citeauthoryear{Aharonian et al.}{2007}]{2007A&A...475L...9A} Aharonian F., et al., 2007b, A\&A, 475, L9 

\bibitem[\protect\citeauthoryear{Aharonian et al.}{2007}]{2007A&A...473L..25A} Aharonian F., et al., 2007c, A\&A, 473, L25 

\bibitem[Aharonian et al.(2008)]{2008RPPh...71i6901A} Aharonian, F.,Buckley, J., Kifune, T., \& Sinnis, G.\ 2008a, Reports on Progress in Physics, 71

\bibitem[\protect\citeauthoryear{Aharonian, Khangulyan, \& Costamante}{2008}]{2008MNRAS.387.1206A} Aharonian F.~A., Khangulyan D., Costamante L., 2008b, MNRAS, 387, 1206 

\bibitem[\protect\citeauthoryear{Aharonian et al.}{2008}]{2008A&A...481L.103A} Aharonian F., et al., 2008c, A\&A, 481, L103 


\bibitem[\protect\citeauthoryear{Aharonian et al.}{2009}]{2009ApJ...696L.150A} Aharonian F., et al., 2009, ApJ, 696, L150 

\bibitem[\protect\citeauthoryear{Albert et al.}{2006}]{2006ApJ...648L.105A} Albert J., et al., 2006a, ApJ, 648, L105 

\bibitem[\protect\citeauthoryear{Albert et al.}{2006}]{2006ApJ...642L.119A} Albert J., et al., 2006b, ApJ, 642, L119 

\bibitem[\protect\citeauthoryear{Albert et al.}{2007}]{2007ApJ...669..862A} Albert J., et al., 2007a, ApJ, 669, 862 

\bibitem[\protect\citeauthoryear{Albert et al.}{2007}]{2007ApJ...667L..21A} Albert J., et al., 2007b, ApJ, 667, L21 

\bibitem[\protect\citeauthoryear{Albert et al.}{2007}]{2007ApJ...663..125A} Albert J., et al., 2007c, ApJ, 663, 125 

\bibitem[\protect\citeauthoryear{Albert et al.}{2007}]{2007ApJ...654L.119A} Albert J., et al., 2007d, ApJ, 654, L119 

\bibitem[\protect\citeauthoryear{Albert et al.}{2007}]{2007ApJ...666L..17A} Albert J., et al., 2007e, ApJ, 666, L17 

\bibitem[\protect\citeauthoryear{Albert et al.}{2007}]{2007ApJ...662..892A} Albert J., et al., 2007f, ApJ, 662, 892 

\bibitem[\protect\citeauthoryear{Anderhub}{2009}]{2009arXiv0907.2386A} Anderhub H., 2009, ApJ, in press, (arXiv:0907.2386) 

\bibitem[\protect\citeauthoryear{B{\"o}ttcher, Dermer, \& Finke}{2008}]{2008ApJ...679L...9B} B{\"o}ttcher M., Dermer C.~D., Finke J.~D., 2008, ApJ, 679, L9 

\bibitem[\protect\citeauthoryear{Burrows et al.}{2005}]{2005SSRv..120..165B} Burrows D.~N., et al., 2005, SSRv, 120, 165 

\bibitem[\protect\citeauthoryear{Costamante \& Ghisellini}{2002}]{2002A&A...384...56C} Costamante L., Ghisellini G., 2002, A\&A, 384, 56 (CG02)

\bibitem[]{} De Angelis A., Mansutti O., Persic M., 2008, La Rivista del Nuovo Cimento, 31, n.4, 187 (arXiv:0712.0315)

\bibitem[\protect\citeauthoryear{Donato et al.}{2001}]{2001A&A...375..739D} Donato D., Ghisellini G., Tagliaferri G., Fossati G., 2001, A\&A, 375, 739 

\bibitem[\protect\citeauthoryear{Fichtel et al.}{1994}]{1994ApJS...94..551F} Fichtel C.~E., et al., 1994, ApJS, 94, 551 

\bibitem[]{} Foschini, L. et al. 2009,  Proc. of "Accretion and ejection in AGN: a global view", Como, June 2009 (arXiv:0908.3313)

\bibitem[]{} Fossati G., Maraschi L., Celotti A., Comastri A. \& Ghisellini G., 1998, MNRAS, 299, 433
             
\bibitem[\protect\citeauthoryear{Georganopoulos \& Kazanas}{2003}]{2003ApJ...594L..27G} Georganopoulos M., Kazanas D., 2003, ApJ, 594, L27 

\bibitem[]{} Ghisellini G., Celotti A., Fossati G., Maraschi L. \& Comastri A., 1998, MNRAS, 301, 451

\bibitem[Ghisellini et al.(2005)]{2005A&A...432..401G} Ghisellini G., Tavecchio F., \& Chiaberge M.\ 2005, A\&A , 432, 401 

\bibitem[\protect\citeauthoryear{Ghisellini et al.}{2007}]{2007MNRAS.382L..82G} Ghisellini G., Foschini L., Tavecchio F., Pian E., 2007, MNRAS, 382, L82 

\bibitem[\protect\citeauthoryear{Ghisellini \& Tavecchio}{2008}]{2008MNRAS.387.1669G} Ghisellini G., Tavecchio F., 2008a, MNRAS, 387, 1669 

\bibitem[\protect\citeauthoryear{Ghisellini \& Tavecchio}{2008}]{2008MNRAS.386L..28G} Ghisellini G., Tavecchio F., 2008b, MNRAS, 386, L28 

\bibitem[]{} Ghisellini G. \& Tavecchio F., 2009, MNRAS, 397, 985  

\bibitem[\protect\citeauthoryear{Ghisellini et al.}{2009}]{2009arXiv0909.0932G} Ghisellini G., Tavecchio F., Foschini L., 
Ghirlanda G., Maraschi L., Celotti A., 2009a, MNRAS, submitted  (arXiv:0909.0932) 

\bibitem[\protect\citeauthoryear{Ghisellini et al.}{2009}]{2009MNRAS.393L..16G} Ghisellini G., Tavecchio F., Bodo G., 
Celotti A., 2009b, MNRAS, 393, L16 

\bibitem[]{} Ghisellini G., Tavecchio F. \& Ghirlanda G., 2009, MNRAS, in press (arXiv:0906.2195)
             
\bibitem[\protect\citeauthoryear{Giannios, Uzdensky, \& Begelman}{2009}]{2009MNRAS.395L..29G} Giannios D., Uzdensky D.~A., Begelman M.~C., 2009, MNRAS, 395, L29 

\bibitem[\protect\citeauthoryear{Giommi \& Padovani}{1994}]{1994MNRAS.268L..51G} Giommi P., Padovani P., 1994, MNRAS, 268, L51 

\bibitem[\protect\citeauthoryear{Hartman et al.}{1999}]{1999ApJS..123...79H} Hartman R.~C., et al., 1999, ApJS, 123, 79 

\bibitem[\protect\citeauthoryear{Kalberla et al.}{2005}]{2005A&A...440..775K} Kalberla P.~M.~W., Burton W.~B., Hartmann D., Arnal E.~M., Bajaja E., Morras R., P{\"o}ppel W.~G.~L., 2005, A\&A, 440, 775 

\bibitem[\protect\citeauthoryear{Katarzy{\'n}ski et al.}{2006}]{2006MNRAS.368L..52K} Katarzy{\'n}ski K., Ghisellini G., Tavecchio F., Gracia J., Maraschi L., 2006, MNRAS, 368, L52 

\bibitem[\protect\citeauthoryear{Kneiske et al.}{2004}]{2004A&A...413..807K} Kneiske T.~M., Bretz T., Mannheim K., Hartmann D.~H., 2004, A\&A, 413, 807 

\bibitem[\protect\citeauthoryear{Leonardo et  al.}{2009}]{2009arXiv0907.0959L} Leonardo E., et al., 2009, to appear in the proc. of the  ICRC 2009, Lodz, Poland (arXiv:0907.0959) 

\bibitem[\protect\citeauthoryear{Maraschi et al.}{1999}]{1999ApJ...526L..81M} Maraschi L., et al., 1999, ApJ, 526, L81 

\bibitem[\protect\citeauthoryear{Maraschi \& Tavecchio}{2003}]{2003ApJ...593..667M} Maraschi L., Tavecchio F., 2003, ApJ, 593, 667 

\bibitem[\protect\citeauthoryear{Mazin \& Raue}{2007}]{2007A&A...471..439M} Mazin D., Raue M., 2007, A\&A, 471, 439 

\bibitem[\protect\citeauthoryear{M{\"u}cke et al.}{2003}]{2003APh....18..593M} M{\"u}cke A., Protheroe R.~J., Engel R., 
Rachen J.~P., Stanev T., 2003, APh, 18, 593 

\bibitem[\protect\citeauthoryear{Nandikotkur et al.}{2007}]{2007ApJ...657..706N} Nandikotkur G., Jahoda K.~M., Hartman R.~C., Mukherjee R., Sreekumar P., B{\"o}ttcher M., Sambruna R.~M., Swank  J.~H., 2007, ApJ, 657, 706 

\bibitem[\protect\citeauthoryear{Neronov, Semikoz, \& Sibiryakov}{2008}]{2008MNRAS.391..949N} Neronov A., Semikoz D., Sibiryakov S., 2008, MNRAS, 391, 949 

\bibitem[\protect\citeauthoryear{Ong}{2009}]{2009ATel.2084....1O} Ong R.~A., 2009, ATel, 2084, 1 

\bibitem[\protect\citeauthoryear{Pei}{1992}]{1992ApJ...395..130P} Pei Y.~C., 1992, ApJ, 395, 130 

\bibitem[\protect\citeauthoryear{Poole et al.}{2008}]{2008MNRAS.383..627P} Poole T.~S., et al., 2008, MNRAS, 383, 627

\bibitem[\protect\citeauthoryear{Roming et al.}{2005}]{2005SSRv..120...95R} Roming P.~W.~A., et al., 2005, SSRv, 120, 95 

\bibitem[\protect\citeauthoryear{Sbarufatti, Treves, \& Falomo}{2005}]{2005ApJ...635..173S} Sbarufatti B., Treves A., Falomo R., 2005, ApJ, 635, 173 

\bibitem[\protect\citeauthoryear{Schlegel, Finkbeiner, 
\& Davis}{1998}]{1998ApJ...500..525S} Schlegel D.~J., Finkbeiner D.~P., Davis M., 1998, ApJ, 500, 525 

\bibitem[\protect\citeauthoryear{Stanev \& Franceschini}{1998}]{1998ApJ...494L.159S} Stanev T., Franceschini A., 1998, ApJ, 494, L159 

\bibitem[\protect\citeauthoryear{Stecker, de Jager, \& Salamon}{1992}]{1992ApJ...390L..49S} Stecker F.W., de Jager O.~C., Salamon M.~H., 1992, ApJ, 390, L49 

\bibitem[\protect\citeauthoryear{Stecker, de Jager, \& Salamon}{1996}]{1996ApJ...473L..75S} Stecker F.~W., de Jager O.~C., Salamon M.~H., 1996, ApJ, 473, L75 

\bibitem[\protect\citeauthoryear{Stecker, Baring, \& Summerlin}{2007}]{2007ApJ...667L..29S} Stecker F.W., Baring M.~G., Summerlin E.~J., 2007, ApJ, 667, L29 

\bibitem[\protect\citeauthoryear{Tagliaferri et al.}{2008}]{2008ApJ...679.1029T} Tagliaferri G., et al., 2008, ApJ, 679, 1029 

\bibitem[\protect\citeauthoryear{Tavecchio, Maraschi, \& Ghisellini}{1998}]{1998ApJ...509..608T} Tavecchio F., Maraschi L., Ghisellini G., 1998, ApJ, 509, 608 

\bibitem[\protect\citeauthoryear{Tavecchio \& Ghisellini}{2008}]{2008MNRAS.385L..98T} Tavecchio F., Ghisellini G., 2008, MNRAS, 385, L98 

\bibitem[\protect\citeauthoryear{Tavecchio \& Ghisellini}{2009}]{2009MNRAS.394L.131T} Tavecchio F., Ghisellini G., 2009, MNRAS, 394, L131 

\bibitem[\protect\citeauthoryear{Tavecchio et al.}{2009}]{2009MNRAS.tmpL.292T} Tavecchio F., Ghisellini G., Ghirlanda G., Costamante L., Franceschini A., 2009, MNRAS, in press (arXiv:0905.0899) 

\bibitem[\protect\citeauthoryear{Teshima}{2009}]{2009ATel.2098....1T} Teshima M., 2009, ATel, 2098, 1 

\bibitem[\protect\citeauthoryear{Zweerink et al.}{1997}]{1997ApJ...490L.141Z} Zweerink J.~A., et al., 1997, ApJ, 490, 
L141 



\end{thebibliography}
\end{document}